\documentclass[onecolumn,peerreview]{IEEEtran}
\ifCLASSINFOpdf
\else
    \usepackage[dvips]{graphicx}
\fi
\usepackage{amsmath,amssymb}
\newcommand{\bysame}{%
    \leavevmode\hbox to 3em{\hrulefill}\,}

\begin{document}
%
\title{Derivation of Mutual Information and Linear Minimum Mean-Square Error for Viterbi Decoding of Convolutional Codes Using the Innovations Method}
%
%
%

\author{Masato~Tajima,~\IEEEmembership{Senior~Member,~IEEE}
\thanks{M. Tajima is with University of Toyama, 3190 Gofuku,
Toyama 930-8555, Japan (e-mail: masatotjm@kind.ocn.ne.jp).}
\thanks{Manuscript received April 19, 2005; revised August 26, 2015.}}

%
%

\markboth{Journal of \LaTeX\ Class Files,~Vol.~14, No.~8, August~2015}%
{Tajima \MakeLowercase{\textit{}}: Derivation of Mutual Information and Linear Minimum Mean-Square Error for Viterbi Decoding of Convolutional Codes Using the Innovations Method}
%



\maketitle

\begin{abstract}
We see that convolutional coding/Viterbi decoding has the structure of the Kalman filter (or the linear minimum variance filter). First, we calculate the covariance matrix of the innovation (i.e., the soft-decision input to the main decoder in a Scarce-State-Transition (SST) Viterbi decoder). Then a covariance matrix corresponding to that of the one-step prediction error in the Kalman filter is obtained. Furthermore, from that matrix, a covariance matrix corresponding to that of the filtering error in the Kalman filter is derived using the formula in the Kalman filter. As a result, the average mutual information per branch for Viterbi decoding of convolutional codes is given using these covariance matrices. Also, the trace of the latter matrix represents the linear minimum mean-square error (LMMSE). We show that an approximate value of the average mutual information is sandwiched between half the SNR times the average filtering and one-step prediction LMMSEs. In the case of QLI codes, from the covariance matrix of the soft-decision input to the main decoder, we can get a matrix. We show that the trace of this matrix has some connection with the linear smoothing error.
\end{abstract}


%
\IEEEpeerreviewmaketitle

\section{Introduction and Scenario}
%
%
%
%
In~\cite{taji 19}, by comparing with the results in linear filtering theory~\cite{kai 74}, we introduced the notion of innovations~\cite{kai 681,kai 682,kai 70,kuni 76} for Viterbi decoding of convolutional codes. More precisely, we have shown that the hard-decision input to the main decoder in an SST Viterbi decoder (see Fig.1)~\cite{kubo 93,pin 91,taji 03} can be regarded as the innovation corresponding to the input to the decoder. Here note that since the framework of coding theory is determined by hard-decision data, the soft-decision input to the main decoder in an SST Viterbi decoder is equally regarded as the innovation. Our motivation for the above paper has come from the works of Kailath and Frost~\cite{kai 681,kai 682}. They proposed the {\it innovations method} in order to solve linear filtering problems~\cite{kai 681} and showed the effectiveness of the method. They also extended the method to linear smoothing problems~\cite{kai 682}. Moreover, they applied the innovations method to nonlinear estimation problems~\cite{fro 71}.
\par
Suppose that the observations are given by
\begin{equation}
\mbox{\boldmath $z$}_k=\mbox{\boldmath $s$}_k+\mbox{\boldmath $w$}_k,~k=0, 1, \cdots ,
\end{equation}
where $\{\mbox{\boldmath $s$}_k\}$ is a zero-mean finite-variance signal process and $\{\mbox{\boldmath $w$}_k\}$ is a zero-mean white noise process. It is assumed that $\mbox{\boldmath $w$}_k$ has a covariance matrix $E[\mbox{\boldmath $w$}_k \mbox{\boldmath $w$}_l^T]=W_k\delta_{kl}$, where $E[\,\cdot \,]$ denotes the expectation, ``$^T$'' means transpose, and $\delta_{kl}$ is Kronecker's delta. The innovation (denoted $\mbox{\boldmath $\nu$}_k$) is defined by $\mbox{\boldmath $\nu$}_k\stackrel{\triangle}{=}\mbox{\boldmath $z$}_k-\mbox{\boldmath $\hat s$}_{k \vert k-1}$, where $\mbox{\boldmath $\hat s$}_{k \vert k-1}$ is the ``linear'' least-squares estimate of $\mbox{\boldmath  $s$}_k$ given $\{\mbox{\boldmath $z$}_l,~0 \leq l \leq k-1\}$. Kailath~\cite[Section III-B]{kai 681} showed the following.
\newtheorem{pro}{Proposition}
\begin{pro}[Kailath~\cite{kai 681}]
The covariance matrix of the innovation $\mbox{\boldmath $\nu$}_k$ is given by
\begin{equation}
E[\mbox{\boldmath $\nu$}_k\mbox{\boldmath $\nu$}_l^T]=(S_k+W_k)\delta_{kl} ,
\end{equation}
where $S_k$ is the covariance matrix of the error in the estimate, i.e.,  $\mbox{\boldmath $s$}_k-\mbox{\boldmath $\hat s$}_{k \vert k-1}$.
\end{pro}
\par
Now suppose that the observations are given by
\begin{equation}
\mbox{\boldmath $z$}_k=\sqrt{\rho}\,\mbox{\boldmath $x$}_k+\mbox{\boldmath $w$}_k ,~k=0, 1, \cdots ,
\end{equation}
where $\rho$ is a parameter related to the {\it signal-to-noise ratio} (SNR), where $\{\mbox{\boldmath $x$}_k\}$ is a signal process independent of $\rho$, and where $\mbox{\boldmath $w$}_k$ are standard Gaussian random vectors with $E[\mbox{\boldmath $w$}_k\mbox{\boldmath $w$}_l^T]=I\delta_{kl}$ ($I$ is the identity matrix). $\{\mbox{\boldmath $x$}_k\}$ and $\{\mbox{\boldmath $w$}_k\}$ are assumed to be independent. Note that the covariance matrix (denoted $R_k$) of the innovation $\mbox{\boldmath $\nu$}_k$  is given by
\begin{equation}
R_k=E[\mbox{\boldmath $\nu$}_k\mbox{\boldmath $\nu$}_k^T]=I+\rho M_k ,
\end{equation}
where $M_k$ is the covariance matrix of the estimation error $\mbox{\boldmath $x$}_k-\mbox{\boldmath $\hat x$}_{k \vert k-1}$ ($\mbox{\boldmath $\hat x$}_{k \vert k-1}$ is the linear least-squares estimate of $\mbox{\boldmath $x$}_k$ given $\{\mbox{\boldmath $z$}_l,~0 \leq l \leq k-1\}$). This fact implies that if the innovations corresponding to the observations are given and if the associated covariance matrix $R_k$ is calculated, then the covariance matrix $M_k$ can be obtained. Here suppose that $\mbox{\boldmath $z$}_k=\sqrt{\rho}\,\mbox{\boldmath $x$}_k+\mbox{\boldmath $w$}_k$ corresponds to the observation equation in the Kalman filter (or linear minimum variance filter)~\cite{ari 77,jaz 70,kai 681}. Then we see that the covariance matrix (denoted $P_k$) of the filtering error $\mbox{\boldmath $x$}_k-\mbox{\boldmath $\hat x$}_{k \vert k}$ ($\mbox{\boldmath $\hat x$}_{k \vert k}$ is the linear least-squares estimate of $\mbox{\boldmath $x$}_k$ given $\{\mbox{\boldmath $z$}_l,~0 \leq l \leq k\}$) can be calculated from $M_k$ by applying the formula in the Kalman filter, where $\mbox{tr}(P_k)$ ($\mbox{tr}(\cdot)$ denotes the trace) represents the mean-square error in the estimate $\mbox{\boldmath $\hat x$}_{k \vert k}$, i.e., the linear minimum mean-square error (LMMSE).
\par 
On the other hand, the input-output mutual information for the discrete-time Kalman filter with Gaussian signals has been derived~\cite{ari 77}. Suppose again that $\mbox{\boldmath $z$}_k=\sqrt{\rho}\,\mbox{\boldmath $x$}_k+\mbox{\boldmath $w$}_k$ corresponds to the observation equation in the Kalman filter, where $\mbox{\boldmath $x$}_k$ is assumed to be Gaussian. Let $\mbox{\boldmath $x$}^k=\{\mbox{\boldmath $x$}_0, \mbox{\boldmath $x$}_1, \cdots, \mbox{\boldmath $x$}_k\}$ and $\mbox{\boldmath $z$}^k=\{\mbox{\boldmath $z$}_0, \mbox{\boldmath $z$}_1, \cdots, \mbox{\boldmath $z$}_k\}$. Then the mutual information $I[\mbox{\boldmath $x$}^k; \mbox{\boldmath $z$}^k]$ between $\mbox{\boldmath $x$}^k$ and $\mbox{\boldmath $z$}^k$ is given by
\begin{eqnarray}
I[\mbox{\boldmath $x$}^k; \mbox{\boldmath $z$}^k] &=& \frac{1}{2}\sum_{j=0}^k\log \vert R_j \vert \\
&=& \frac{1}{2}\sum_{j=0}^k\log\frac{\vert M_j \vert}{\vert P_j \vert} ,
\end{eqnarray}
where ``$\log$'' is the natural logarithm and $\vert A \vert$ denotes the determinant of a square matrix $A$.
(If $\mbox{\boldmath $x$}_j$ are not Gaussian, then the above equality becomes an inequality.) Note that $I[\mbox{\boldmath $x$}^k; \mbox{\boldmath $z$}^k]$ is represented using $R_j$, or using $M_j$ and $P_j$. This fact again implies that if the innovations are obtained and if the associated covariance matrices are calculated, then the input-output mutual information can be derived. Thus we see that the derivation of the input-output mutual information and the LMMSE is reduced to the calculation of the covariance matrix of the innovation. We have noticed that the above argument also applies to Viterbi decoding of convolutional codes and hence the associated input-output mutual information and LMMSE shall be derived. In this paper, we show that this is really possible. In our case, the signal vector $\mbox{\boldmath $x$}_k$ is obtained from the encoded block $\mbox{\boldmath $y$}_k$ using a one-to-one mapping. Since the sequence $\{\mbox{\boldmath $y$}_k\}$ constitutes a discrete dynamical system, $\{\mbox{\boldmath $x$}_k\}$ also constitutes a discrete dynamical system. Hence we see that convolutional coding corresponds to the signal process. Also, from the model of the received data for code symbols, the observation equation is given by (3). Then we can assume the structure of the Kalman filter and hence the innovations method can be applied. In addition, we already know that the soft-decision input to the main decoder (denoted $\mbox{\boldmath $r$}_k=\left(r_k^{(1)}, \cdots, r_k^{(n_0)}\right)$) in an SST Viterbi decoder is regarded as the innovation associated with the received data. Then our goal comes down to the calculation of the covariance matrix of $\mbox{\boldmath $r$}_k$. In~\cite{taji 19}, we obtained the distribution function of the single component $r_k^{(l)}$. After that, we derived the joint distribution function of $\mbox{\boldmath $r$}_k$ for a two-dimensional case~\cite{taji 21}. In this paper, we derive the joint distribution function of $\mbox{\boldmath $r$}_k$ for a general multidimensional case and then obtain the associated covariance matrix. Thus we can accomplish the purpose. The above is our basic scenario and constitutes the main part of the paper.
\par
Now in this paper, we are mainly concerned with the mutual information and mean-square error. Including these notions, the mutual information, the {\it likelihood ratio} (LR), and the mean-square error are central concerns in information theory, detection theory, and estimation theory. From the late 1960s to the early 1970s, the relation between the LR and the mean-square error was actively discussed. Kailath~\cite{kai 69} showed that the LR is expressed in terms of the causal least-squares estimate. In~\cite{kai 683}, he also showed that conversely the causal least-squares estimate can be obtained from the LR. Esposito~\cite{espo 68} showed that the noncausal least-squares estimate is given by the gradient of the log LR. This result was also discussed in~\cite{kai 683}. Note that its extended version was presented by Zakai~\cite{zak 05}. Subsequently, Duncan~\cite{dun 70}, and Kadota, Zakai, and Ziv~\cite{kado 71} derived the relation between the mutual information and the causal mean-square error. Their works are regarded as extensions of the work of Gelfand and Yaglom (see~\cite{ari 77,dun 70}), who discussed for the first time the relation between the mutual information and the filtering error. Later, Guo, Shamai, and Verd\'{u}~\cite{guo 05} derived new formulas regarding the relation between the mutual information and the noncausal mean-square error. Moreover, Zakai~\cite{zak 05} extended the results of Guo et al.~\cite{guo 05}. He showed that the principal relation between the mutual information and the estimation error holds also in the abstract Wiener Space~\cite{ito 78,zak 05}. We remark that the additive Gaussian noise channel is assumed for all above works.
\par
Among those works, we have noted the work of Guo et al.~\cite{guo 05}. It deals with the relation between the mutual information and the minimum mean-square error (MMSE) associated with the least-squares estimate (i.e., the best estimate measured in mean-square sense\footnote{It is given by the conditional expectation with respect to the past observations, which is in general a highly ``nonlinear'' functional of the observations~\cite{kai 98}.}). On the other hand, we are concerned with the ``linear'' least-squares estimate based on the past observations. In connection with the subject, the linear minimum variance filter (the Kalman filter in a broad sense)~\cite{ari 77,jaz 70,kai 681} plays an essential role in this paper. In addition, this paper is much influenced by the work of Guo et al.~\cite{guo 05} In particular, Corollary 3 and Theorem 9 for discrete-time Gaussian channels~\cite[Section IV]{guo 05} are closely related to the results in our paper. Another reason why we have noted the paper of Guo et al.~\cite{guo 05} is that their channel model (i.e., input-output relation) is equal to our observation equation. Actually, it has provided a strong motivation for our work.
\par
The rest of the paper is organized as follows. In Section II, we briefly review the Kalman filter, whose structure is essentially used in this paper. In Section III, we discuss the mutual information associated with the Kalman filter. In Section IV, we examine the relationship between the mutual information and the LMMSE. In our argument, a condition related to the eigenvalues of $P_k$ is assumed. It is shown that (an approximate value of) the average mutual information is sandwiched between half $\rho$ times the average filtering and one-step prediction LMMSEs. In Section V, we focus on the innovations associated with Viterbi decoding of convolutional codes. First, we calculate the covariance matrix of the innovation (i.e., $\mbox{\boldmath $r$}_k$). As a result, a covariance matrix corresponding to $M_k$ in the Kalman filter is obtained. Furthermore, a covariance matrix corresponding to $P_k$ is derived from that matrix using the formula in the Kalman filter. Note that the calculation of the latter matrix is rather difficult in a general case. Hence this is done for the case of $n_0=2$. Next, we apply the obtained results to concrete rate $1/2$ QLI codes~\cite{joha 99}, where a QLI code is regarded as a general code. In Section VI, we examine the relation between Viterbi decoding of QLI codes and the smoothing error. For QLI codes, the precise innovation has not been extracted~\cite{taji 19}. However, from the covariance matrix of the soft-decision input to the main decoder, we can get a matrix. We show that the trace of this matrix has some connection with the linear smoothing error.
\par
Let us close this section by introducing the basic notions needed for this paper. Notations in this paper are same as those in~\cite{taji 19} in principle. We always assume that the underlying field is $\mbox{GF}(2)$. Let $G(D)$ be a generator matrix for an $(n_0, k_0)$ time-invariant convolutional code, where $G(D)$ is assumed to be {\it minimal}~\cite{forn 70}. A corresponding check matrix $H(D)$ is also assumed to be minimal. Hence, they have the same constraint length, denoted $\nu$. Denote by $\mbox{\boldmath $i$}=\{\mbox{\boldmath $i$}_k\}$ and $\mbox{\boldmath $y$}=\{\mbox{\boldmath $y$}_k\}$ an information sequence and the corresponding code sequence, respectively, where $\mbox{\boldmath $i$}_k=\left(i_k^{(1)}, \cdots, i_k^{(k_0)}\right)$ is the information block at $t=k$ and $\mbox{\boldmath $y$}_k=\left(y_k^{(1)}, \cdots, y_k^{(n_0)}\right)$ is the encoded block at $t=k$. It is assumed that a code sequence {\boldmath $y$} is transmitted symbol by symbol over a memoryless AWGN channel using BPSK modulation~\cite{hell 71}. Let $\mbox{\boldmath $z$}=\{\mbox{\boldmath $z$}_k\}$ be a received sequence, where $\mbox{\boldmath $z$}_k=\left(z_k^{(1)}, \cdots, z_k^{(n_0)}\right)$ is the received block at $t=k$. Each component $z_j$ of {\boldmath $z$} is represented as
\begin{eqnarray}
z_j &=& x_j\sqrt{2E_s/N_0}+w_j \\
&=& cx_j+w_j ,
\end{eqnarray}
where $c=\sqrt{2E_s/N_0}$. Here, $x_j$ takes $\pm 1$ depending on whether the code symbol $y_j$ is $0$ or $1$. This is, $x_j$ is the equiprobable binary signal. $E_s$ and $N_0$ denote the energy per channel symbol and the single-sided noise spectral density, respectively. (Let $E_b$ be the energy per information bit. Then the relationship between $E_b$ and $E_s$ is defined by $E_s=RE_b$, where $R$ is the code rate.) $w_j$ is a zero-mean unit variance Gaussian random variable with probability density function
\begin{equation}
q(y)=\frac{1}{\sqrt{2\pi}}e^{-\frac{y^2}{2}} .
\end{equation}
Each $w_j$ is independent of all others. The hard-decision (denoted ``$^h$'') data of $z_j$ is defined by
\begin{equation}
z_j^h=\left\{
\begin{array}{rl}
0,& \quad z_j \geq 0 \\
1,& \quad z_j < 0 .
\end{array} \right.
\end{equation}
In this case, the channel error probability (denoted $\epsilon$) is given by
\begin{equation}
\epsilon=\frac{1}{\sqrt{2\pi}} \int_{\sqrt{2E_s/N_0}}^{\infty}e^{-\frac{y^2}{2}}dy \stackrel{\triangle}{=}Q\bigl(\sqrt{2E_s/N_0}\bigr) .
\end{equation}
Note that the above signal model can be seen as the block at $t=k$. Let $\rho=c^2=2E_s/N_0$. Then we can rewrite as
\begin{equation}
\mbox{\boldmath $z$}_k=\sqrt{\rho}\,\mbox{\boldmath $x$}_k+\mbox{\boldmath $w$}_k ,
\end{equation}
where $\mbox{\boldmath $x$}_k=\left(x_k^{(1)}, \cdots, x_k^{(n_0)}\right)$ and $\mbox{\boldmath $w$}_k=\left(w_k^{(1)}, \cdots, w_k^{(n_0)}\right)$.
\par
In this paper, we consider an SST Viterbi decoder~\cite{kubo 93,pin 91,taji 03} which consists of a pre-decoder and a main decoder (see Fig.1). In the case of a general convolutional code, the inverse encoder $G^{-1}(D)$ is used as a pre-decoder. Let $\mbox{\boldmath $r$}_k=\left(r_k^{(1)}, \cdots, r_k^{(n_0)}\right)$ be the soft-decision input to the main decoder. $r_k^{(l)}~(1 \leq l \leq n_0)$ is given by
\begin{equation}
r_k^{(l)}=\left\{
\begin{array}{rl}
\vert z_k^{(l)} \vert,& \quad r_k^{(l)h}=0 \\
- \vert z_k^{(l)} \vert,& \quad r_k^{(l)h}=1 .
\end{array} \right.
\end{equation}
\par
Let $\mbox{\boldmath $v$}_k=(v_k^1, \cdots, v_k^n)$ be an $n$-tuple of variables. Also, let $\mbox{\boldmath $p$}(D)=(p_1(D), \cdots, p_n(D))$ be an $n$-tuple of polynomials in $D$. Since each $p_l(D)$ is a delay operator with respect to $k$, $\sum_{l=1}^np_l(D)v_k^l$ is well defined, where $D^jv_k^l=v_{k-j}^l$. In this paper, we express the above variable as $\mbox{\boldmath $v$}_k\mbox{\boldmath $p$}^T(D)$. Using this notation, we have
\begin{equation}
\mbox{\boldmath $y$}_k=\mbox{\boldmath $i$}_kG(D) .
\end{equation}
Also, the syndrome at $t=k$ is defined by
\begin{equation}
\mbox{\boldmath $\zeta$}_k=\mbox{\boldmath $z$}_k^hH^T(D) .
\end{equation}
Note that $\mbox{\boldmath $\zeta$}_k=\mbox{\boldmath $e$}_kH^T(D)$ holds, where $\mbox{\boldmath $e$}_k=\left(e_k^{(1)}, \cdots, e_k^{(n_0)}\right)$ is the error at $t=k$.

\section{Kalman Filter}
In this paper, the linear minimum variance filter (i.e., the Kalman filter in a broad sense)~\cite{ari 77,jaz 70,kai 681} plays an essential role. Hence we briefly review the Kalman filter with assumption of Gaussianness and then the linear minimum variance filter (without assumption of Gaussianness). In this section, column vectors are assumed.

\subsection{The Discrete-Time Kalman Filter}
Suppose that the $n$-vector signal process $\{\mbox{\boldmath $x$}_k\}$ and $m$-vector observation process $\{\mbox{\boldmath $z$}_k\}$ are given by the following equations:
\begin{eqnarray}
\mbox{\boldmath $x$}_{k+1} &=& F_k\mbox{\boldmath $x$}_k+\mbox{\boldmath $u$}_k \\
\mbox{\boldmath $z$}_k &=& H_k\mbox{\boldmath $x$}_k+\mbox{\boldmath $w$}_k,~k=0, 1, \cdots ,
\end{eqnarray}
where $F_k$ and $H_k$ are $n \times n$ and $m \times n$ matrices, respectively, and where $\mbox{\boldmath $u$}_k$ and $\mbox{\boldmath $w$}_k$ are $n$-vector and $m$-vector white Gaussian noises, respectively. It is assumed that
\begin{equation}
\left.
\begin{array}{ll}
E[\mbox{\boldmath $u$}_k]=\mbox{\boldmath $0$},& E[\mbox{\boldmath $u$}_k\mbox{\boldmath $u$}_l^T]=U_k \delta_{kl} \\
E[\mbox{\boldmath $w$}_k]=\mbox{\boldmath $0$},& E[\mbox{\boldmath $w$}_k\mbox{\boldmath $w$}_l^T]=W_k \delta_{kl} \\
E[\mbox{\boldmath $u$}_k\mbox{\boldmath $w$}_l^T]=0. & 
\end{array} \right\}
\end{equation}
Moreover, we assume that
\begin{equation}
E[\mbox{\boldmath $u$}_k\mbox{\boldmath $x$}_l^T]=0~(l \leq k),~~E[\mbox{\boldmath $w$}_k\mbox{\boldmath $x$}_l^T]=0~(l \leq k) .
\end{equation}
(This means that the future noise is uncorrelated with the past signal~\cite{kai 681,kai 69}.) The initial state $\mbox{\boldmath $x$}_0$ is assumed to be a Gaussian random vector with mean $\mbox{\boldmath $\bar x$}_0$ and covariance matrix $X_0$.
\par
{\it Remark 1:} Since $\mbox{\boldmath $u$}_k$ is Gaussian, $E[\mbox{\boldmath $u$}_k\mbox{\boldmath $u$}_l^T]=U_k \delta_{kl}$ implies that $\mbox{\boldmath $u$}_k$ and $\mbox{\boldmath $u$}_l$ are independent for $k \neq l$. Similarly, $\mbox{\boldmath $w$}_k$ and $\mbox{\boldmath $w$}_l$ are independent for $k \neq l$. Also, $\mbox{\boldmath $u$}_k$ and $\mbox{\boldmath $w$}_l$ are independent for any $k$ and $l$. Moreover, since $E[\mbox{\boldmath $u$}_k\mbox{\boldmath $x$}_0^T]=0$ and $E[\mbox{\boldmath $w$}_k\mbox{\boldmath $x$}_0^T]=0$ hold, $\mbox{\boldmath $x$}_0$ is independent of $\mbox{\boldmath $u$}_k$ and $\mbox{\boldmath $w$}_k$. We see that $\{\mbox{\boldmath $x$}_k\}$ and $\{\mbox{\boldmath $w$}_k\}$ are independent as a result.
\par
Under the above conditions, the discrete-time Kalman filter is described as follows.
\begin{pro}[Arimoto~\cite{ari 77}]
The least-squares estimate of $\mbox{\boldmath $x$}_k$ given the observations $\{\mbox{\boldmath $z$}_0, \mbox{\boldmath $z$}_1, \cdots, \mbox{\boldmath $z$}_k\}$ (denoted $\mbox{\boldmath $\hat x$}_{k \vert k}$) is expressed as
\begin{equation}
\mbox{\boldmath $\hat x$}_{k \vert k}=\mbox{\boldmath $\hat x$}_{k \vert k-1}+P_kH_k^TW_k^{-1}(\mbox{\boldmath $z$}_k-H_k\mbox{\boldmath $\hat x$}_{k \vert k-1}),~k=0, 1, \cdots ,
\end{equation}
where $\mbox{\boldmath $\hat x$}_{k \vert k-1}$ is the least-squares estimate of $\mbox{\boldmath $x$}_k$ given $\{\mbox{\boldmath $z$}_0, \mbox{\boldmath $z$}_1, \cdots, \mbox{\boldmath $z$}_{k-1}\}$. Here, we have the following relations:
\begin{eqnarray}
\mbox{\boldmath $\hat x$}_{k+1 \vert k} &=& F_k \mbox{\boldmath $\hat x$}_{k \vert k} \\
P_k &=& M_k-M_kH_k^T(W_k+H_k M_k H_k^T)^{-1}H_k M_k \\
M_{k+1} &=& F_kP_kF_k^T+U_k ,
\end{eqnarray}
where $P_k$ and $M_k$ are defined by
\begin{eqnarray}
P_k &=& E[(\mbox{\boldmath $x$}_k-\mbox{\boldmath $\hat x$}_{k \vert k})(\mbox{\boldmath $x$}_k-\mbox{\boldmath $\hat x$}_{k \vert k})^T] \\
M_k &=& E[(\mbox{\boldmath $x$}_k-\mbox{\boldmath $\hat x$}_{k \vert k-1})(\mbox{\boldmath $x$}_k-\mbox{\boldmath $\hat x$}_{k \vert k-1})^T] .
\end{eqnarray}
\end{pro}
\par
{\it Remark 2:} $\mbox{\boldmath $\hat x$}_{k \vert k}$ is equal to the conditional expectation of $\mbox{\boldmath $x$}_k$ with respect to the given observations $\{\mbox{\boldmath $z$}_0, \mbox{\boldmath $z$}_1, \cdots, \mbox{\boldmath $z$}_k\}$. In this case, since $\{\mbox{\boldmath $x$}_k, \mbox{\boldmath $z$}_k\}$ are jointly Gaussian, it is a linear functional of $\mbox{\boldmath $z$}_0, \mbox{\boldmath $z$}_1, \cdots, \mbox{\boldmath $z$}_k$.
\par
{\it Remark 3:} The quantity
\begin{equation}
\mbox{\boldmath $\nu$}_k\stackrel{\triangle}{=}\mbox{\boldmath $z$}_k-H_k\mbox{\boldmath $\hat x$}_{k \vert k-1}
\end{equation}
represents the innovation~\cite{kai 681,kai 682,kai 70,kuni 76,oksen 98} corresponding to $\mbox{\boldmath $z$}_k$. The associated covariance matrix (denoted $R_k$) is given by 
\begin{equation}
R_k=W_k+H_k M_k H_k^T . 
\end{equation}
\par
{\it Remark 4:} It follows from (22) that if $M_k$ is given, then $P_k$ is obtained based only on the observation equation.
\par
We remark that the above four equations constitute a recursion structure.
\begin{itemize}
\item [1)] Suppose that $\mbox{\boldmath $\hat x$}_{0 \vert -1}=\mbox{\boldmath $\bar x$}_0$ and $M_0=X_0>0$ have been given as initial values. (Let $A$ and $B$ be symmetric matrices. $A>0~(A \geq 0)$ means that $A$ is positive definite (positive semi-definite). $A>B~(A \geq B)$ shows that $A-B>0~(A-B \geq 0)$.)
\item [2)] $P_0$ is determined from (22). Then $\mbox{\boldmath $\hat x$}_{0 \vert 0}$ is obtained from (20) on the assumption that $\mbox{\boldmath $z$}_0$ is given. Subsequently, $\mbox{\boldmath $\hat x$}_{1 \vert 0}$ is determined from (21) and $M_1$ is obtained from (23). That is, we have $\mbox{\boldmath $\hat x$}_{1 \vert 0}$ and $M_1$.
\item [3)] In general, suppose that $\mbox{\boldmath $\hat x$}_{k \vert k-1}$ and $M_k$ have been obtained. $P_k$ is determined from (22). Then $\mbox{\boldmath $\hat x$}_{k \vert k}$ is obtained from (20) on the assumption that $\mbox{\boldmath $z$}_k$ is given. Subsequently, $\mbox{\boldmath $\hat x$}_{k+1 \vert k}$ is determined from (21) and $M_{k+1}$ is obtained from (23). That is, when $\mbox{\boldmath $z$}_k$ is given, $\mbox{\boldmath $\hat x$}_{k+1 \vert k}$ and $M_{k+1}$ are obtained by way of $\mbox{\boldmath $\hat x$}_{k \vert k}$ and $P_k$.
\end{itemize}
\par
Note that $P_k$ has an alternative expression. Let $A$, $B$, and $C$ be matrices with appropriate sizes. Then the identical equation
\begin{equation}
(A^{-1}+C^TB^{-1}C)^{-1}=A-AC^T(CAC^T+B)^{-1}CA
\end{equation}
holds, where $A>0$ and $B>0$~\cite{ari 77,jaz 70}. By applying this formula, we have
\begin{equation}
P_k=(M_k^{-1}+H_k^TW_k^{-1}H_k)^{-1} .
\end{equation}
If $M_k>0$, then $M_k^{-1}>0$ holds~\cite{horn 13}. Similarly, since $M_k^{-1}+H_k^TW_k^{-1}H_k>0$, we have $(M_k^{-1}+H_k^TW_k^{-1}H_k)^{-1}>0$, i.e., $P_k>0$. Here, we can assume that $F_k$ is non-singular (see~\cite{ari 77,jaz 70}). Then it follows from (23) that $M_{k+1}>0$. Hence it can be assumed that $M_k>0$ and $P_k>0$. Moreover, note that $P_k \leq M_k$ holds from (22). This means that the mean-square error does not increase given a newly obtained observation.

\subsection{Linear Minimum Variance Filter}
In Proposition 2, it is assumed that $\mbox{\boldmath $x$}_0$, $\mbox{\boldmath $u$}_k$, and $\mbox{\boldmath $w$}_k$ are Gaussian. On the other hand, consider the case where this assumption does not hold. Even so, we can construct the Kalman filter, which is optimal in the sense that the mean-square error is minimized based on a ``linear'' estimation method. This filter is called the linear minimum variance filter. In this case, ``conditional expectation'' is replaced with ``orthogonal projection'' and ``independence'' between random variables is replaced with ``orthogonality'' (uncorrelated) between them. Suppose that the signal and observation processes are defined by (16) and (17), respectively, where noises $\mbox{\boldmath $u$}_k$ and $\mbox{\boldmath $w$}_k$ are assumed to satisfy the orthogonal conditions given by (18) and (19). Also, $\mbox{\boldmath $x$}_0$ is a random vector with mean $\mbox{\boldmath $\bar x$}_0$ and covariance matrix $X_0$. Under these conditions, we have the following. (It is slightly modified.)
\begin{pro}[Kailath~\cite{kai 681}]
Let $\mbox{\boldmath $\hat x$}_{k \vert k-1}$ be the linear least-squares estimate of $\mbox{\boldmath $x$}_k$ given the observations $\{\mbox{\boldmath $z$}_0, \mbox{\boldmath $z$}_1, \cdots, \mbox{\boldmath $z$}_{k-1}\}$. Also, let  $\mbox{\boldmath $\nu$}_k=\mbox{\boldmath $z$}_k-H_k\mbox{\boldmath $\hat x$}_{k \vert k-1}$ be the innovation associated with $\mbox{\boldmath $z$}_k$. Then we obtain the expression
\begin{equation}
\mbox{\boldmath $\hat x$}_{k+1 \vert k}=F_k\mbox{\boldmath $\hat x$}_{k \vert k-1}+F_kK_k\mbox{\boldmath $\nu$}_k,~k=0, 1, \cdots ,
\end{equation}
where
\begin{eqnarray}
K_k &\stackrel{\triangle}{=}& P_kH_k^TW_k^{-1} \\
&=& M_kH_k^T(W_k+H_k M_k H_k^T)^{-1} .
\end{eqnarray}
We also have a recursion relation
\begin{equation}
M_{k+1}=F_k \left(M_k-K_kR_kK_k^T \right)F_k^T+U_k ,
\end{equation}
where $R_k=W_k+H_kM_kH_k^T$ is the covariance matrix of the innovation $\mbox{\boldmath $\nu$}_k$.
\end{pro}
\par
{\it Remark 1:} Let $Z^k$ be the space spanned by $\mbox{\boldmath $z$}_0, \mbox{\boldmath $z$}_1, \cdots, \mbox{\boldmath $z$}_k$. Then $\mbox{\boldmath $\hat x$}_{k \vert k}$ is the orthogonal projection of $\mbox{\boldmath $x$}_k$ onto $Z^k$. Similarly, $\mbox{\boldmath $\hat x$}_{k \vert k-1}$ is the orthogonal projection of $\mbox{\boldmath $x$}_k$ onto $Z^{k-1}$.
\par
{\it Remark 2:} $\mbox{\boldmath $z$}_k$ is not Gaussian in general. Hence strictly speaking, $\mbox{\boldmath $\nu$}_k$ is not the innovation. We may call it a quasi-innovation. However, following Kailath~\cite{kai 681}, we call it the innovation in this paper.
\par
Note that (30) is the combined form of (20) and (21). Similarly, (33) is the combined form of (22) and (23). Hence the obtained estimator has the same structure as the discrete-time Kalman filter given by Proposition 2. Thus we see that the Kalman filter can be constructed even in the case where $\mbox{\boldmath $x$}_0$, $\mbox{\boldmath $u$}_k$, and $\mbox{\boldmath $w$}_k$ are not Gaussian. For that reason, the linear minimum variance filter is also called the Kalman filter~\cite{kai 681}. In the following, we call the linear minimum variance filter simply the Kalman filter.

\section{Mutual Information Associated with the Discrete-Time Kalman Filter}
\subsection{Gaussian Signals}
Consider the discrete-time Kalman filter described in Section II-A, where $\mbox{\boldmath $x$}_0$, $\mbox{\boldmath $u$}_k$, and $\mbox{\boldmath $w$}_k$ are assumed to be Gaussian. We assume the same conditions as those given in Section II-A. In this case, the input-output mutual information $I[\mbox{\boldmath $x$}^k; \mbox{\boldmath $z$}^k]$ has been derived.
\begin{pro}[Arimoto~\cite{ari 77}]
Suppose that the same conditions as those of Proposition 2 are satisfied. Then the mutual information $I[\mbox{\boldmath $x$}^k; \mbox{\boldmath $z$}^k]$ is given by
\begin{eqnarray}
I[\mbox{\boldmath $x$}^k; \mbox{\boldmath $z$}^k] &=& \frac{1}{2}\sum_{j=0}^k \log \frac{\vert W_j+H_jM_jH_j^T \vert}{\vert W_j \vert } \\
&=& \frac{1}{2}\sum_{j=0}^k \log \frac{\vert R_j \vert}{\vert W_j \vert } \\
&=& \frac{1}{2}\sum_{j=0}^k \log  \frac{\vert M_j \vert}{\vert P_j \vert} .
\end{eqnarray}
\end{pro}
\par
{\it Remark:} $I[\mbox{\boldmath $x$}^k; \mbox{\boldmath $z$}^k]$ is expressed using $R_j~(0 \leq j \leq k)$ and $W_j~(0 \leq j \leq k)$, or using $M_j~(0 \leq j \leq k)$ and $P_j~(0 \leq j \leq k)$. Hence if $M_j~(0 \leq j \leq k)$ are given, then $I[\mbox{\boldmath $x$}^k; \mbox{\boldmath $z$}^k]$ is obtained based only on the observation equation.
\par
The outline of the proof in~\cite{ari 77} is as follows. Let
\begin{equation}
\mbox{\boldmath $z$}^k=\left(
\begin{array}{c}
\mbox{\boldmath $z$}_0 \\
\mbox{\boldmath $z$}_1 \\
\vdots \\
\mbox{\boldmath $z$}_k
\end{array}
\right),~~\mbox{\boldmath $x$}^k=\left(
\begin{array}{c}
\mbox{\boldmath $x$}_0 \\
\mbox{\boldmath $x$}_1 \\
\vdots \\
\mbox{\boldmath $x$}_k
\end{array}
\right),~~\mbox{\boldmath $w$}^k=\left(
\begin{array}{c}
\mbox{\boldmath $w$}_0 \\
\mbox{\boldmath $w$}_1 \\
\vdots \\
\mbox{\boldmath $w$}_k
\end{array}
\right) .
\end{equation}
Also, set
\begin{equation}
\mbox{\boldmath $H$}_k=\left(
\begin{array}{cccc}
H_0 & 0 & \cdots & 0 \\
0 & H_1 & \cdots & 0 \\
& & \ddots & \\
0 & 0 & \cdots & H_k
\end{array}
\right) .
\end{equation}
Then we have
\begin{equation}
\mbox{\boldmath $z$}^k=\mbox{\boldmath $H$}_k\mbox{\boldmath $x$}^k+\mbox{\boldmath $w$}^k .
\end{equation}
Let us define as
\begin{equation}
X_{ij}=E[(\mbox{\boldmath $x$}_i-\mbox{\boldmath $\bar x$}_i)(\mbox{\boldmath $x$}_j-\mbox{\boldmath $\bar x$}_j)^T],~~\mbox{\boldmath $\bar x$}_i=E[\mbox{\boldmath $x$}_i] .
\end{equation}
Then the covariance matrix of $\mbox{\boldmath $z$}^k$ (denoted $\mbox{\boldmath $Z$}_k$) is given by
\begin{equation}
\mbox{\boldmath $Z$}_k=\mbox{\boldmath $H$}_k\mbox{\boldmath $X$}_k\mbox{\boldmath $H$}_k^T+\mbox{\boldmath $W$}_k,
\end{equation}
where
\begin{equation}
\mbox{\boldmath $X$}_k=\left(
\begin{array}{cccc}
X_{00} & X_{01} & \cdots & X_{0k} \\
X_{10} & X_{11} & \cdots & X_{1k} \\
& & \ddots & \\
X_{k0} & X_{k1} & \cdots & X_{kk}
\end{array}
\right)
\end{equation}
and
\begin{equation}
\mbox{\boldmath $W$}_k=\left(
\begin{array}{cccc}
W_0 & 0 & \cdots & 0 \\
0 & W_1 & \cdots & 0 \\
& & \ddots & \\
0 & 0 & \cdots & W_k
\end{array}
\right) .
\end{equation}
Since $\mbox{\boldmath $x$}^k$ and $\mbox{\boldmath $w$}^k$ are Gaussian and are independent, the mutual information $I[\mbox{\boldmath $x$}^k; \mbox{\boldmath $z$}^k]$ between $\mbox{\boldmath $x$}^k$ and $\mbox{\boldmath $z$}^k$ is given by
\begin{eqnarray}
I[\mbox{\boldmath $x$}^k; \mbox{\boldmath $z$}^k] &=& \frac{1}{2}\log \frac{\vert \mbox{\boldmath $W$}_k+\mbox{\boldmath $H$}_k\mbox{\boldmath $X$}_k\mbox{\boldmath $H$}_k^T \vert}{\vert \mbox{\boldmath $W$}_k \vert} \nonumber \\
&=&  \frac{1}{2}\log \frac{\vert \mbox{\boldmath $Z$}_k \vert}{\vert \mbox{\boldmath $W$}_k \vert} .
\end{eqnarray}
Here it is shown that
\begin{eqnarray}
\vert \mbox{\boldmath $Z$}_k \vert &=& \vert \mbox{\boldmath $Z$}_{k-1} \vert \cdot \vert W_k+H_kM_kH_k^T \vert \\
\vert \mbox{\boldmath $W$}_k \vert &=& \vert \mbox{\boldmath $W$}_{k-1} \vert \cdot \vert W_k \vert .
\end{eqnarray}
Using this recursion relations, we finally have
\begin{eqnarray}
I[\mbox{\boldmath $x$}^k; \mbox{\boldmath $z$}^k] &=& \frac{1}{2}\sum_{j=0}^k \log \frac{\vert W_j+H_jM_jH_j^T \vert}{\vert W_j \vert } \\
&=& \frac{1}{2}\sum_{j=0}^k \log \frac{\vert R_j \vert}{\vert W_j \vert } .
\end{eqnarray}
\par
Note that the right-hand side has another expression. From the equality
\begin{displaymath}
P_j=(M_j^{-1}+H_j^TW_j^{-1}H_j)^{-1} ,
\end{displaymath}
we have
\begin{displaymath}
M_jP_j^{-1}=I_n+M_jH_j^TW_j^{-1}H_j ,
\end{displaymath}
where $I_n$ is the identity matrix of size $n \times n$. Then it follows that
\begin{eqnarray}
\vert M_jP_j^{-1} \vert &=& \vert I_n+M_jH_j^TW_j^{-1}H_j \vert \nonumber \\
&=& \vert I_m+W_j^{-1}H_j M_jH_j^T \vert \nonumber \\
&=& \vert W_j^{-1}(W_j+H_j M_jH_j^T) \vert \nonumber \\
&=& \frac{\vert W_j+H_j M_jH_j^T \vert}{\vert W_j \vert} \nonumber \\
&=& \frac{\vert R_j \vert}{\vert W_j \vert} .
\end{eqnarray}
In the above modifications, we have used the relation
\begin{equation}
 \vert I_n+AB \vert =\vert I_m+BA \vert ,
\end{equation}
where $A$ is an $n \times m$ matrix and $B$ is an $m \times n$ matrix. Hence we have
\begin{equation}
I[\mbox{\boldmath $x$}^k; \mbox{\boldmath $z$}^k]=\frac{1}{2}\sum_{j=0}^k \log \frac{\vert M_j \vert}{\vert P_j \vert } .
\end{equation}

\subsection{Non-Gaussian Signals}
In this section, we consider non-Gaussian signals $\mbox{\boldmath $x$}_k$. Hence $\mbox{\boldmath $x$}_0$ and $\mbox{\boldmath $u$}_k$ are not necessarily Gaussian. On the other hand, $\mbox{\boldmath $w$}_k$ is assumed to be Gaussian. Moreover, we assume the stronger conditions with respect to $\mbox{\boldmath $u$}_k$ and $\mbox{\boldmath $w$}_k$ compared with those in Section II-B. In addition to (18), it is assumed that
\begin{itemize}
\item $\mbox{\boldmath $u$}_k$ and $\mbox{\boldmath $u$}_l$ are independent for $k \neq l$. $\mbox{\boldmath $u$}_k$ and $\mbox{\boldmath $w$}_l$ are independent for any $k$ and $l$.
\end{itemize}
({\it Remark:} Since $\mbox{\boldmath $w$}_k$ is Gaussian, $E[\mbox{\boldmath $w$}_k\mbox{\boldmath $w$}_l^T]=W_k \delta_{kl}$ implies that $\mbox{\boldmath $w$}_k$ and $\mbox{\boldmath $w$}_l$ are independent for $k \neq l$.) Also, (19) is replaced with
\begin{itemize}
\item $\mbox{\boldmath $u$}_k$ and $\mbox{\boldmath $x$}_l$ are independent for $l \leq k$. $\mbox{\boldmath $w$}_k$ and $\mbox{\boldmath $x$}_l$ are independent for $l \leq k$.
\end{itemize}
Note that $\mbox{\boldmath $w$}^k$ is independent of $\mbox{\boldmath $x$}^k$ as a result. Then we have the following.
\begin{pro}
Suppose that the above conditions are satisfied. Then the mutual information $I[\mbox{\boldmath $x$}^k; \mbox{\boldmath $z$}^k]$ satisfies the inequality
\begin{eqnarray}
I[\mbox{\boldmath $x$}^k; \mbox{\boldmath $z$}^k] &\leq& \frac{1}{2}\sum_{j=0}^k \log \frac{\vert W_j+H_jM_jH_j^T \vert}{\vert W_j \vert } \\
&=& \frac{1}{2}\sum_{j=0}^k \log \frac{\vert R_j \vert}{\vert W_j \vert } \\
&=& \frac{1}{2}\sum_{j=0}^k \log  \frac{\vert M_j \vert}{\vert P_j \vert} .
\end{eqnarray}
\end{pro}
\begin{IEEEproof}
Note the relation
\begin{displaymath}
\mbox{\boldmath $z$}^k=\mbox{\boldmath $H$}_k\mbox{\boldmath $x$}^k+\mbox{\boldmath $w$}^k .
\end{displaymath}
Since $\mbox{\boldmath $w$}^k$ is Gaussian and is independent of $\mbox{\boldmath $x$}^k$, we can assume an additive Gaussian noise channel. Here the covariance matrix $\mbox{\boldmath $X$}_k$ is regarded as an energy constraint on the input $\mbox{\boldmath $x$}^k$ (see~\cite[Section 7.4]{gall 68}). Then the covariance matrix of the channel output $\mbox{\boldmath $z$}^k$ is limited to
\begin{displaymath}
\mbox{\boldmath $Z$}_k=\mbox{\boldmath $H$}_k\mbox{\boldmath $X$}_k\mbox{\boldmath $H$}_k^T+\mbox{\boldmath $W$}_k .
\end{displaymath}
Consider the problem of maximizing the entropy $H(\mbox{\boldmath $z$}^k)$ subject to a constraint on $\mbox{\boldmath $Z$}_k$. It is known that $H(\mbox{\boldmath $z$}^k)$ is maximized when $\mbox{\boldmath $z$}^k$ is Gaussian. Then we have
\begin{equation}
I[\mbox{\boldmath $x$}^k; \mbox{\boldmath $z$}^k]\leq \frac{1}{2}\log \frac{\vert \mbox{\boldmath $W$}_k+\mbox{\boldmath $H$}_k\mbox{\boldmath $X$}_k\mbox{\boldmath $H$}_k^T \vert}{\vert \mbox{\boldmath $W$}_k \vert} .
\end{equation}
Modifications of the right-hand side are the same as those in Section III-A.
\end{IEEEproof}


\section{Relationship Between Mutual Information and LMMSE}
In this section, we discuss the relationship between the mutual information and LMMSE for the discrete-time Kalman filter. Though the structure of the Kalman filter is assumed in this paper, we are mainly concerned with the observation process. As was stated in the previous sections, when $M_j~(0 \leq j \leq k)$ are given, $P_j~(0 \leq j \leq k)$ and the mutual information $I[\mbox{\boldmath $x$}^k; \mbox{\boldmath $z$}^k]$ are obtained based only on the observation equation. That is, the precise description of how the signal process is generated is not used in the derivation (see~\cite{kai 681}). This fact is essentially important in this paper. In the following, we assume that $m=n=n_0$. Also, taking into account the model of the received data for code symbols (see Section I), we assume that the observations are given by
\begin{equation}
\mbox{\boldmath $z$}_k=\sqrt{\rho}\,\mbox{\boldmath $x$}_k+\mbox{\boldmath $w$}_k,~k=0, 1, \cdots ,
\end{equation}
where $\rho=c^2=2E_s/N_0$, where $\{\mbox{\boldmath $x$}_k\}$ is a  finite-variance signal process independent of $\rho$, and where $\mbox{\boldmath $w$}_k$ are standard Gaussian random vectors with $E[\mbox{\boldmath $w$}_k\mbox{\boldmath $w$}_l^T]=I_{n_0}\delta_{kl}$. We assume the same conditions as those in Section III-B except the above. That is, $\mbox{\boldmath $x$}_k$ is not, in general, Gaussian and the noise $\{\mbox{\boldmath $w$}_k\}$ is independent of $\{\mbox{\boldmath $x$}_k\}$. Since $H_k=\sqrt{\rho}\,I_{n_0}$ and $W_k=I_{n_0}$, we have
\begin{equation}
R_k=I_{n_0}+\rho M_k .
\end{equation}
We also have
\begin{equation}
P_k=M_k-\rho M_k(I_{n_0}+\rho M_k)^{-1}M_k .
\end{equation}
Note that Proposition 5 in Section III-B is restated as follows.
\begin{pro}
Under the above conditions, the mutual information $I[\mbox{\boldmath $x$}^k; \mbox{\boldmath $z$}^k]$ satisfies the inequality
\begin{eqnarray}
I[\mbox{\boldmath $x$}^k; \mbox{\boldmath $z$}^k] &\leq& \frac{1}{2}\sum_{j=0}^k \log  \vert I_{n_0}+\rho M_j \vert \\
&=& \frac{1}{2}\sum_{j=0}^k \log  \vert R_j \vert \\
&=& \frac{1}{2}\sum_{j=0}^k \log \frac{\vert M_j \vert}{\vert P_j \vert} .
\end{eqnarray}
\end{pro}
\par
{\it Remark:} Let $\rho>0$ be sufficiently small. Then we have $\mbox{\boldmath $z$}_k \approx \mbox{\boldmath $w$}_k$. Since $\mbox{\boldmath $x$}^k$ and $\mbox{\boldmath $w$}^k$ are independent, this means that $I[\mbox{\boldmath $x$}^k; \mbox{\boldmath $z$}^k] \approx 0$. On the other hand, we have $\log \vert I_{n_0}+\rho M_j \vert \approx \log \vert I_{n_0}\vert=0$. Hence the bound in (59) is tight for sufficiently small $\rho>0$.
\par
Notice that Guo et al.~\cite{guo 05} deal with a discrete-time Gaussian noise channel given by $\mbox{\boldmath $z$}_k=\sqrt{\rho}\,\mbox{\boldmath $x$}_k+\mbox{\boldmath $w$}_k$. Hence we think of the results (Corollary 3 and Theorem 9) in~\cite[Section IV]{guo 05}. In what follows, we always assume that the same conditions as those of Proposition 6 are satisfied. First, consider~\cite[Corollay 3]{guo 05}. In our case, it is modified as follows.
\begin{pro}
We have
\begin{equation}
\frac{d}{d \rho}I[\mbox{\boldmath $x$}^k; \mbox{\boldmath $z$}^k] \leq \frac{1}{2}\sum_{j=0}^k \mbox{tr}(P_j) .
\end{equation}
\end{pro}
\begin{IEEEproof}
For~\cite[Corollary 3]{guo 05}, it suffices to note the following inequalities:
\par
(the smoothing MMSE at time $j$ given the entire observations $\mbox{\boldmath $z$}^k$)$\leq$(the filtering MMSE at time $j$)$\leq$(the filtering LMMSE at time $j$ ($=\mbox{tr}(P_j)$)).
\end{IEEEproof}
\par
Next, let us examine the relationship between the mutual information and the one-step prediction LMMSE (i.e., $\mbox{tr}(M_k)$). We have the following.
\newtheorem{lem}{Lemma}
\begin{lem}
\begin{equation}
\log \vert I_{n_0}+\rho M_j \vert \leq \rho\,\mbox{tr}(M_j) .
\end{equation}
\end{lem}
\begin{IEEEproof}
Let $A>0$ be a matrix of size $n_0 \times n_0$. Then the inequality
\begin{equation}
\vert A \vert ^{\frac{1}{n_0}} \leq \frac{1}{n_0}\mbox{tr}(A) 
\end{equation}
holds~\cite{cov 06,hardy 52}. Letting $A=I_{n_0}+\rho M_j $, we have
\begin{equation}
\frac{1}{n_0}\log \vert I_{n_0}+\rho M_j  \vert \leq \log \mbox{tr}(I_{n_0}+\rho M_j )-\log n_0 .
\end{equation}
Here, $\log \mbox{tr}(I_{n_0}+\rho M_j )$ is modified as follows:
\begin{eqnarray}
\log \mbox{tr}(I_{n_0}+\rho M_j) &=& \log (n_0+\rho\,\mbox{tr}(M_j)) \nonumber \\
&=& \log \left(n_0(1+\frac{\rho}{n_0}\mbox{tr}(M_j))\right) \nonumber \\
&=& \log n_0+\log \left(1+\frac{\rho}{n_0}\mbox{tr}(M_j)\right) \nonumber \\
&\leq& \log n_0+\frac{\rho}{n_0}\mbox{tr}(M_j) ,
\end{eqnarray}
where we have used the inequality $\log x \leq x-1~(x>0)$. Then
\begin{eqnarray}
\frac{1}{n_0} \log \vert I_{n_0}+\rho M_j \vert &\leq& \log \mbox{tr}(I_{n_0}+\rho M_j)-\log n_0 \nonumber \\
&\leq& \log n_0+\frac{\rho}{n_0}\mbox{tr}(M_j)-\log n_0 \nonumber \\
&=& \frac{\rho}{n_0}\mbox{tr}(M_j)
\end{eqnarray}
or $\log \vert I_{n_0}+\rho M_j \vert \leq \rho\,\mbox{tr}(M_j)$ has been derived.
\end{IEEEproof}
\par
From the above lemma, we directly have the following.
\begin{pro}
The mutual information $I[\mbox{\boldmath $x$}^k; \mbox{\boldmath $z$}^k]$ satisfies
\begin{eqnarray}
I[\mbox{\boldmath $x$}^k; \mbox{\boldmath $z$}^k] &\leq& \frac{1}{2}\sum_{j=0}^k \log  \vert I_{n_0}+\rho M_j \vert \nonumber \\
&\leq& \frac{\rho}{2}\sum_{j=0}^k\mbox{tr}(M_j) .
\end{eqnarray}
\end{pro}
\begin{IEEEproof}
Direct consequence of Proposition 6 and Lemma 1.
\end{IEEEproof}
\par
Note that the above corresponds to the right-hand inequality in~\cite[Theorem 9]{guo 05}. Then we would like to know whether the left-hand inequality in~\cite[Theorem 9]{guo 05} holds in our case. For that purpose, let us examine the relationship between the mutual information and the filtering LMMSE (i.e., $\mbox{tr}(P_k)$). We need the following.
\begin{lem}
Let $R_j$ be the covariance matrix of the innovation $\mbox{\boldmath $\nu$}_j$. Then we have
\begin{equation}
M_j=P_jR_j=R_jP_j .
\end{equation}
\end{lem}
\begin{IEEEproof}
From the relation
\begin{displaymath}
P_j=(M_j^{-1}+H_j^TW_j^{-1}H_j)^{-1} ,
\end{displaymath}
we have
\begin{displaymath}
M_jP_j^{-1}=I_n+M_jH_j^TW_j^{-1}H_j .
\end{displaymath}
Noting that $n=n_0$, $H_j=\sqrt{\rho}\,I_{n_0}$, and $W_j=I_{n_0}$, the above is reduced to
\begin{displaymath}
M_jP_j^{-1}=I_{n_0}+\rho M_j=R_j .
\end{displaymath}
Hence we have $M_j=R_jP_j$. Similarly, we have $M_j=P_jR_j$.
\end{IEEEproof}
\par
Since $P_j(\rho)>0$ is a symmetric matrix, we can find an orthogonal matrix $Q(\rho)$ such that $Q(\rho)^{-1}P_j(\rho)Q(\rho)$ is a diagonal matrix (denoted $\Gamma(\rho)$). That is,
\begin{equation}
Q^{-1}(\rho)P_j(\rho)Q(\rho)=\Gamma(\rho) ,
\end{equation}
where the diagonal entries of $\Gamma(\rho)$ are the eigenvalues of $P_j(\rho)$ (denoted $\lambda_l(\rho),~1 \leq l \leq n_0)$. Note that $\lambda_l(\rho)>0~(1 \leq l \leq n_0)$. In the following, $P_j(\rho)$, $Q(\rho)$, $\Gamma(\rho)$, and $\lambda_l(\rho)$ are written simply as $P_j$, $Q$, $\Gamma$, and $\lambda_l$, respectively. Let $\lambda_{max}$ be the maximum value of $\lambda_l~(1 \leq l \leq n_0)$. Moreover, define
\begin{equation}
\Gamma+\rho \Phi \stackrel{\triangle}{=}Q^{-1}P_jQ+\rho Q^{-1}\frac{d P_j}{d \rho}Q .
\end{equation}
\begin{lem}
Let $(\Gamma+\rho \Phi)_{ll}~(1 \leq l \leq n_0)$ be the diagonal entries of $\Gamma+\rho \Phi$. Then we have
\begin{eqnarray}
(\Gamma+\rho \Phi)_{ll} &=& \lambda_l+\rho \frac{d \lambda_l}{d \rho} \nonumber \\
&=& \frac{d}{d \rho}(\rho \lambda_l)~(1 \leq l \leq n_0) .
\end{eqnarray}
\end{lem}
\begin{IEEEproof}
We have
\begin{eqnarray}
\frac{d}{d\rho}\Gamma &=& \frac{d}{d\rho}(Q^{-1}P_jQ) \nonumber \\
&=& \left(-Q^{-1}\frac{dQ}{d\rho}Q^{-1}\right)P_jQ+Q^{-1}\frac{dP_j}{d\rho}Q+Q^{-1}P_jQ\left(Q^{-1}\frac{dQ}{d\rho}\right) \nonumber \\
&=& -\left(Q^{-1}\frac{dQ}{d\rho}\right)\Gamma+Q^{-1}\frac{dP_j}{d\rho}Q+\Gamma\left(Q^{-1}\frac{dQ}{d\rho}\right) .
\end{eqnarray}
Note the $(l, l)$ entries. Since $((Q^{-1}\frac{dQ}{d\rho})\Gamma)_{ll}=(\Gamma(Q^{-1}\frac{dQ}{d\rho}))_{ll}$, we have $\frac{d \lambda_l}{d\rho}=\Phi_{ll}$. Then we obtain
\begin{eqnarray}
(\Gamma+\rho \Phi)_{ll} &=& \Gamma_{ll}+\rho \Phi_{ll} \nonumber \\
&=& \lambda_l+\rho \frac{d \lambda_l}{d\rho} \nonumber \\
&=& \frac{d}{d\rho}(\rho \lambda_l) . \nonumber
\end{eqnarray}
\end{IEEEproof}
\par
Under the above, we have the following.
\begin{lem}
Suppose that $\lambda_{max}<\frac{1}{\rho}$. If $\frac{d}{d \rho}(\rho \lambda_l)>0~(1 \leq l \leq n_0)$, then we have
\begin{equation}
\frac{d}{d \rho}\log \vert I_{n_0}+\rho M_j \vert>\frac{d}{d \rho}(\rho \,\mbox{tr}(P_j)) .
\end{equation}
Moreover, if $\frac{d}{d \gamma}(\gamma \lambda_l)>0~(1 \leq l \leq n_0)$ holds for $\gamma \in (0,~\rho]$, then we have
\begin{equation}
\log \vert I_{n_0}+\rho M_j \vert>\rho\,\mbox{tr}(P_j).
\end{equation}
\end{lem}
\begin{IEEEproof}
See Appendix A.
\end{IEEEproof}
\par
{\it Remark 1:} Note that
\begin{eqnarray}
\frac{d}{d \rho}\log \vert I_{n_0}+\rho M_j \vert &=& \frac{1}{\vert I_{n_0}+\rho M_j \vert}\frac{d}{d \rho} \vert I_{n_0}+\rho M_j \vert \nonumber \\
&=& \mbox{tr} \left((I_{n_0}+\rho M_j)^{-1} \frac{d}{d \rho}(I_{n_0}+\rho M_j) \right) .
\end{eqnarray}
Here, let $\rho>0$ be sufficiently small. In this case, $M_j=M_j(\rho)$ can be regarded as a constant matrix. (Note that $M_j \approx X_j$, where $X_j$ is the covariance matrix of $\mbox{\boldmath $x$}_j$, which is independent of $\rho$.) Then we have
\begin{eqnarray}
\frac{d}{d \rho}\log \vert I_{n_0}+\rho M_j \vert &=& \mbox{tr} \left((I_{n_0}+\rho M_j)^{-1} \frac{d}{d \rho}(I_{n_0}+\rho M_j) \right)  \nonumber \\
&\approx& \mbox{tr}(R_j^{-1}M_j) \nonumber \\
&=& \mbox{tr}(P_j) .
\end{eqnarray}
Taking the limit as $\rho \rightarrow 0$, we obtain
\begin{eqnarray}
\left. \frac{d}{d \rho}\log \vert I_{n_0}+\rho M_j \vert \right|_{\rho=0} &=& \mbox{tr}(P_j(0)) \nonumber \\
&=& \left. \frac{d}{d \rho}(\rho \,\mbox{tr}(P_j)) \right|_{\rho=0} ,
\end{eqnarray}
where we have used the fact that $P_j$ is also regarded as a constant matrix for sufficiently small $\rho>0$.
\par
{\it Remark 2:} Note the following equalities:
\begin{eqnarray}
\mbox{tr}(\Gamma+\rho \Phi) &=& \sum_{l=1}^{n_0}(\Gamma+\rho \Phi)_{ll} \nonumber \\
&=& \sum_{l=1}^{n_0}\frac{d}{d \rho}(\rho \lambda_l) \nonumber \\
&=& \frac{d}{d \rho}\left(\rho \sum_{l=1}^{n_0}\lambda_l \right) \nonumber \\
&=&  \frac{d}{d \rho}(\rho \,\mbox{tr}(P_j)) ,
\end{eqnarray}
where we have used the relation $\mbox{tr}(P_j)=\sum_{l=1}^{n_0}\lambda_l$~\cite{horn 13}. Hence if
\begin{displaymath}
\mbox{tr}(\Gamma+\rho \Phi)=\frac{d}{d \rho}(\rho\,\mbox{tr}(P_j))>0
\end{displaymath}
holds (i.e., $\rho\,\mbox{tr}(P_j)$ increases as $\rho$ increases), then the assumption $\frac{d}{d \rho}(\rho \lambda_l)>0~(1 \leq l \leq n_0)$ is satisfied with high probability. On the other hand, consider the case that $\frac{d}{d \rho}(\rho\,\mbox{tr}(P_j))<0$ holds (i.e., $\rho\,\mbox{tr}(P_j)$ decreases as $\rho$ increases). If this is the case, since $\rho\,\mbox{tr}(P_j)$ becomes smaller, we are likely to have the inequality $\log \vert I_{n_0}+\rho M_j \vert >\rho\,\mbox{tr}(P_j)$. Hence, even if the assumption of Lemma 4 is not satisfied, there is a possibility that $\log \vert I_{n_0}+\rho M_j \vert >\rho\,\mbox{tr}(P_j)$ holds.
\par
From Lemma 4, we directly have the following.
\begin{pro}
Suppose that $\lambda_{max}<\frac{1}{\rho}$. If $\frac{d}{d \gamma}(\gamma \lambda_l)>0~(1 \leq l \leq n_0)$ holds for $\gamma \in (0,~\rho]$. Then we have
\begin{equation}
\frac{\rho}{2}\sum_{j=0}^k \mbox{tr}(P_j)<\frac{1}{2}\sum_{j=0}^k\log \vert I_{n_0}+\rho M_j \vert .
\end{equation}
\end{pro}
\begin{IEEEproof}
Direct consequence of Lemma 4.
\end{IEEEproof}
\par
Since $\mbox{\boldmath $x$}_j$ is not Gaussian, $\frac{1}{2}\sum_{j=0}^k\log \vert I_{n_0}+\rho M_j \vert$ does not coincide with the mutual information $I[\mbox{\boldmath $x$}^k; \mbox{\boldmath $z$}^k]$. However, it is considered that the above corresponds to the left-hand inequality in~\cite[Theorem 9]{guo 05}. Thus combining Proposition 8 and Proposition 9, a result similar to that of Guo et al.~\cite[Theorem 9]{guo 05} has been derived.


\section{Innovations for Viterbi Decoding of Convolutional Codes}
\begin{figure}[tb]
\begin{center}
\includegraphics[width=10.0cm,clip]{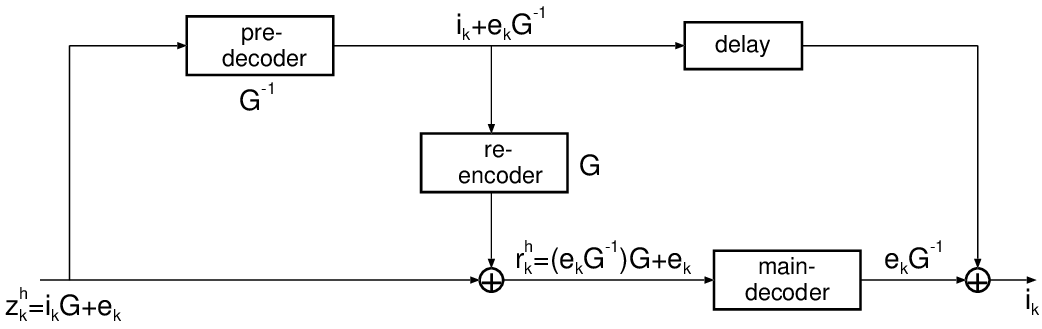}
\end{center}
\caption{The structure of an SST Viterbi decoder (pre-decoder: $G^{-1}$).}
\label{Fig.1}
\end{figure}
The basic structure of an SST Viterbi decoder is shown in Fig.1. In~\cite{taji 19}, we showed that the hard-decision input $\mbox{\boldmath $r$}_k^h$ to the main decoder in an SST Viterbi decoder is considered as the innovation associated with the received data $\mbox{\boldmath $z$}_k^h$ for the Viterbi decoder. Hence the soft-decision input $\mbox{\boldmath $r$}_k$ to the main decoder is equally regarded as the innovation. Also, we see that the structure of the Kalman filter can be assumed for convolutional coding/Viterbi decoding. In fact, recall the model of the received data relevant to code symbols (see Section I). The observations are given by
\begin{displaymath}
\mbox{\boldmath $z$}_k=\sqrt{\rho}\,\mbox{\boldmath $x$}_k+\mbox{\boldmath $w$}_k,~k=0, 1, \cdots ,
\end{displaymath} 
where $\rho=c^2=2E_s/N_0$, where $\mbox{\boldmath $x$}_k$ is a vector whose components are equiprobable binary signals, not dependent on $\rho$, and where $\mbox{\boldmath $w$}_k$ are standard Gaussian random vectors with $E[\mbox{\boldmath $w$}_k\mbox{\boldmath $w$}_l^T]=I_{n_0}\delta_{kl}$. Here, $\{\mbox{\boldmath $x$}_k\}$ and $\{\mbox{\boldmath $w$}_k\}$ are independent. On the other hand, the sequence of code symbols $\{y_j\}$ constitutes a discrete dynamical system and the correspondence between $y_j$ and $x_j~(=\pm 1)$ is one-to-one. Then we see that $\{x_j\}$ also constitutes a discrete dynamical system. This fact implies that convolutional coding corresponds to the signal process. Thus we can actually assume the structure of the Kalman filter for convolutional coding/Viterbi decoding. Hence we can use the results in Section IV.

\subsection{Covariance Matrix of the Innovation}
As stated in Section I, if the covariance matrix of $\mbox{\boldmath $r$}_k$ (denoted $\Sigma_r$) is calculated, then a matrix corresponding to $M_k$ (denoted $\Sigma_x$) in the Kalman filter is obtained. Moreover, the covariance matrix of the filtering error (denoted $\Sigma_c$) is derived from $\Sigma_x$ using the formula in the Kalman filter. Then our main purpose in this section is to obtain the covariance matrix $\Sigma_r$. Note that $\Sigma_r$ is calculated using the joint probability density function $p_r(x_1, \cdots, x_{n_0})$ of $\mbox{\boldmath $r$}_k=\left(r_k^{(1)}, \cdots, r_k^{(n_0)}\right)$. Let $\mbox{\boldmath $v$}_k=\left(v_k^{(1)}, \cdots, v_k^{(n_0)}\right)$ be the encoded block for the main decoder in an SST Viterbi decoder. In~\cite{taji 19}, we showed that the distribution of $r_k^{(l)}\,(1 \leq l \leq n_0)$ has the form
\begin{equation}
p_r(y)=(1-\alpha_l)q(y-c)+\alpha_lq(y+c) ,
\end{equation}
where $c=\sqrt{\rho}$. Moreover, in~\cite{taji 21}, we showed the following important relation:
\begin{equation}
\alpha_l=P\left(v_k^{(l)}=1\right),~l=1, \cdots, n_0 ,
\end{equation}
($P(\cdot)$ denotes the probability). Since time-invariant convolutional codes are considered in this paper, $\alpha_l$ is not dependent on $k$ (i.e., time or depth). From this relation, we have noticed that $p_r(y)$ can be extended to a multidimensional case. In the case of $n_0=2$, this was done in~\cite{taji 21}. In a general case, $p_r(x_1, \cdots, x_{n_0})$ is given by
\begin{eqnarray}
p_r(x_1, \cdots, x_{n_0}) &=& \alpha_{00 \cdots 0}q(x_1-c)q(x_2-c)\cdots q(x_{n_0}-c) \nonumber \\
&& +\alpha_{00 \cdots 1}q(x_1-c)q(x_2-c)\cdots q(x_{n_0}+c) \nonumber \\
&& +\cdots+ \alpha_{11 \cdots 1}q(x_1+c)q(x_2+c)\cdots q(x_{n_0}+c) ,
\end{eqnarray}
where $\alpha_{ij \cdots l}=P(v_k^{(1)}=i, v_k^{(2)}=j, \cdots, v_k^{(n_0)}=l)$. $p_r(x_1, \cdots, x_{n_0})$ is justified by the marginal relation
\begin{equation}
\int_{-\infty}^{\infty}p_r(x_1, \cdots, x_i, \cdots, x_{n_0})dx_i=p_r(x_1, \cdots, x_{i-1}, x_{i+1}, \cdots, x_{n_0}) .
\end{equation}
Note that since $\alpha_{ij \cdots l}$ is not dependent on $k$, $p_r(x_1, \cdots, x_{n_0})$ is also not dependent on $k$.
\par
Let us calculate the associated covariance matrix $\Sigma_r=(\gamma_{ij})$. Since~\cite[Lemma 1]{taji 21} shall be used in the derivation, we write it down in a slightly modified form. A proof is almost the same as that given there and hence omitted.
\begin{lem}
Let $i \neq j$. Then the following quantities have the same value:
\begin{equation}
P(v_k^{(i)}=0, v_k^{(j)}=0)-P(v_k^{(i)}=0)P(v_k^{(j)}=0)
\end{equation}
\begin{equation}
P(v_k^{(i)}=0)P(v_k^{(j)}=1)-P(v_k^{(i)}=0, v_k^{(j)}=1)
\end{equation}
\begin{equation}
P(v_k^{(i)}=1)P(v_k^{(j)}=0)-P(v_k^{(i)}=1, v_k^{(j)}=0)
\end{equation}
\begin{equation}
P(v_k^{(i)}=1, v_k^{(j)}=1)-P(v_k^{(i)}=1)P(v_k^{(j)}=1) .
\end{equation}
The common value is denoted by $\theta_{ij}$.
\end{lem}
\par
In the above lemma, define $\tilde \alpha_{lm}=P(v_k^{(i)}=l, v_k^{(j)}=m)$. Then $\tilde \alpha_{lm}$ are expressed as
\begin{eqnarray}
\tilde \alpha_{00} &=& 1-\alpha_i-\alpha_j+u \\
\tilde \alpha_{01} &=& \alpha_j-u \\
\tilde \alpha_{10} &=& \alpha_i-u \\
\tilde \alpha_{11}  &=& u ,
\end{eqnarray}
where $ u~(0 \leq u \leq \min(\alpha_i, \alpha_j))$ is an arbitrary constant.
\par
Using Lemma 5, we have the following.
\begin{pro}
The covariance matrix $\Sigma_r$ of $\mbox{\boldmath $r$}_k$ is given by
\begin{equation}
\Sigma_r=\left(
\begin{array}{cccc}
1+4 \rho \alpha_1(1-\alpha_1) & 4 \rho \theta_{12} & \cdots & 4 \rho \theta_{1n_0} \\
4 \rho \theta_{12} & 1+4 \rho \alpha_2(1-\alpha_2) & \cdots & 4 \rho \theta_{2n_0} \\
& & \ddots & \\
4 \rho \theta_{1n_0} & 4 \rho \theta_{2n_0} & \cdots & 1+4 \rho \alpha_{n_0}(1-\alpha_{n_0})
\end{array}
\right) .
\end{equation}
\end{pro}
\begin{IEEEproof}
See Appendix B.
\end{IEEEproof}
\par
Note that $\Sigma_r$ is rewritten as
\begin{eqnarray}
\Sigma_r &=& \left(
\begin{array}{cccc}
1 & 0 & \cdots & 0 \\
0 & 1 & \cdots & 0 \\
& & \ddots & \\
0 & 0 & \cdots & 1
\end{array}
\right)+\rho \left(
\begin{array}{cccc}
4 \alpha_1(1-\alpha_1) & 4 \theta_{12} & \cdots & 4 \theta_{1n_0} \\
4 \theta_{12} & 4 \alpha_2(1-\alpha_2) & \cdots & 4 \theta_{2n_0} \\
& & \ddots & \\
4 \theta_{1n_0} & 4 \theta_{2n_0} & \cdots & 4 \alpha_{n_0}(1-\alpha_{n_0})
\end{array}
\right) \nonumber \\
&=& I_{n_0}+\rho \Sigma_x ,
\end{eqnarray}
where $I_{n_0}$ is the covariance matrix of the observation noise $\mbox{\boldmath $w$}_k$ and where
\begin{equation}
\Sigma_x=\left(
\begin{array}{cccc}
4 \alpha_1(1-\alpha_1) & 4 \theta_{12} & \cdots & 4 \theta_{1n_0} \\
4 \theta_{12} & 4 \alpha_2(1-\alpha_2) & \cdots & 4 \theta_{2n_0} \\
& & \ddots & \\
4 \theta_{1n_0} & 4 \theta_{2n_0} & \cdots & 4 \alpha_{n_0}(1-\alpha_{n_0})
\end{array}
\right) .
\end{equation}
Notice that $\Sigma_r$ and $\Sigma_x$ are not dependent on $k$. Here recall that the covariance matrix $R_k$ of the innovation $\mbox{\boldmath $\nu$}_k$ is expressed as $R_k=I_{n_0}+\rho M_k$. Then by comparing with the above result, it is reasonable to consider that $\Sigma_x$ is a matrix corresponding to $M_k$ in the Kalman filter. We will show that this expectation is justified.
\par
Set $\Sigma_x=4\Sigma_v$. Then we have the following.
\begin{pro}
$\Sigma_v$ is the covariance matrix of $\mbox{\boldmath $v$}_k=\left(v_k^{(1)}, \cdots, v_k^{(n_0)}\right)$ (i.e., the encoded block for the main decoder).
\end{pro}
\begin{IEEEproof}
See Appendix C.
\end{IEEEproof}
\par
Now set
\begin{equation}
\mbox{\boldmath $\tilde x$}_k=\left(
\begin{array}{c}
\tilde x_k^{(1)} \\
\tilde x_k^{(2)} \\
\vdots \\
\tilde x_k^{(n_0)}
\end{array}
\right) ,
\end{equation}
where
\begin{equation}
\tilde x_k^{(l)}=1-2v_k^{(l)}~(1 \leq l \leq n_0) .
\end{equation}
Note that $\tilde x_k^{(l)}$ takes $\pm 1$ depending on whether $v_k^{(l)}$ is $0$ or $1$. Then we have the following.
\begin{pro}
$\Sigma_x$ is the covariance matrix of $\mbox{\boldmath $\tilde x$}_k$.
\end{pro}
\begin{IEEEproof}
See Appendix D.
\end{IEEEproof}
\par
In addition, a very important property of $\Sigma_x$ has been derived.
\newtheorem{cor}{Corollary}
\begin{cor}
$\Sigma_x \geq 0$, i.e., $\Sigma_x$ is positive semi-definite.
\end{cor}
\begin{IEEEproof}
A covariance matrix is positive semi-definite~\cite{jaz 70}.
\end{IEEEproof}
\par
{\it Remark 1:} Let $\Sigma_v=(v_{ij})$. We can assume that $v_{ij}>0~(1 \leq i, j \leq n_0)$. Suppose that $i \neq j$. Then $v_{ij}$ is the covariance of $v_k^{(i)}$ and $v_k^{(j)}$. Since $i \neq j$, it is expected that $v_k^{(i)}$ and $v_k^{(j)}$ are weakly correlated and hence $v_{ij}$ has a small positive value. In this case, there is a possibility that the inequality
\begin{equation}
v_{ii}>\sum_{j \neq i}v_{ij}~(1 \leq i \leq n_0)
\end{equation}
holds. That is, $\Sigma_v$ is a strictly diagonally dominant matrix~\cite{horn 13}. It is known that a symmetric strictly diagonally dominant matrix with positive diagonal entries is positive definite~\cite[Theorem 6.1.10]{horn 13}. Hence, if $\Sigma_v$ is strictly diagonally dominant, then we have $\Sigma_v>0$. As a result, we also have $\Sigma_x>0$.
\par
Now note the hard-decision input $\mbox{\boldmath $r$}_k^h$~\cite{taji 19} to the main decoder in an SST Viterbi decoder. It is expressed as
\begin{eqnarray}
\mbox{\boldmath $r$}_k^h &=& (\mbox{\boldmath $i$}_k-\mbox{\boldmath $\hat i$}(k \vert k))G(D)+\mbox{\boldmath $e$}_k \nonumber \\
&=& \mbox{\boldmath $v$}_k+\mbox{\boldmath $e$}_k .
\end{eqnarray}
Hence we have
\begin{eqnarray}
\mbox{\boldmath $v$}_k &=& (\mbox{\boldmath $i$}_k-\mbox{\boldmath $\hat i$}(k \vert k))G(D) \nonumber \\
&=& \mbox{\boldmath $i$}_kG(D)-\mbox{\boldmath $\hat i$}(k \vert k)G(D) .
\end{eqnarray}
This means that $\mbox{\boldmath $v$}_k$ is regarded as an estimation error of the encoded block. Also, noting that $\tilde x_k^{(l)}$ is defined by $\tilde x_k^{(l)}=1-2v_k^{(l)}$, we see that $\mbox{\boldmath $\tilde x$}_k$ has a similar meaning. Then taking into account the obtained results, we can conclude that $\Sigma_x$ is a matrix corresponding to $M_k$ in the Kalman filter.
\par
{\it Remark 2:} Note the equality $\mbox{\boldmath $v$}_k=\mbox{\boldmath $i$}_kG(D)-\mbox{\boldmath $\hat i$}(k \vert k)G(D)$. As an estimate of $\mbox{\boldmath $i$}_k$, not $\mbox{\boldmath $\hat i$}(k \vert k-1)$ but $\mbox{\boldmath $\hat i$}(k \vert k)$ is used~\cite{taji 19}. However, $\mbox{\boldmath $i$}_kG(D)-\mbox{\boldmath $\hat i$}(k \vert k)G(D)$ is a formal expression. On the other hand, $\mbox{\boldmath $v$}_k$ has a concrete meaning, i.e., the encoded block for the main decoder in an SST Viterbi decoder. Then $\mbox{\boldmath $v$}_k$ should be given preference over $\mbox{\boldmath $i$}_kG(D)-\mbox{\boldmath $\hat i$}(k \vert k)G(D)$.
\par
Moreover, we already know the equation connecting $P_k$ and $M_k$ (see Section II). Then the covariance matrix $\Sigma_c$ of the filtering error can be obtained by replacing $M_k$ with $\Sigma_x$ in the equation. That is, $\Sigma_c$ is given by
\begin{equation}
\Sigma_c\stackrel{\triangle}{=}\Sigma_x-\rho \Sigma_x(I_{n_0}+\rho \Sigma_x)^{-1}\Sigma_x .
\end{equation}
As a result, the mutual information and the LMMSE for Viterbi decoding of convolutional codes are derived based on the procedure described in Section I. Thus we can accomplish our scenario stated in Section I.

\subsection{Mutual Information and LMMSE for Viterbi Decoding of Convolutional Codes}
Since $\Sigma_r$, $\Sigma_x$, and $\Sigma_c$ have been calculated, the results in the previous section can be transferred to the case of convolutional coding/Viterbi decoding. That is, we can evaluate the mutual information (strictly speaking, the average mutual information per branch) using Proposition 6. Also, we can compare $\frac{\rho}{2}\mbox{tr}(\Sigma_c)$, $\frac{1}{2} \log \vert I_{n_0}+\rho \Sigma_x \vert$, and $\frac{\rho}{2}\mbox{tr}(\Sigma_x)$ using Propositions 8 and 9. For that purpose, let us note Propositions 6, 7, 8, and 9 again. (For simplicity, assumptions needed to derive the results are omitted.) The associated results are modified as follows:
\begin{eqnarray}
\frac{1}{\rho}\left(\frac{I[\mbox{\boldmath $x$}^k; \mbox{\boldmath $z$}^k]}{k+1}\right) &\leq& \frac{1}{2 \rho}\left(\frac{1}{k+1}\sum_{j=0}^k \log \vert I_{n_0}+\rho M_j \vert \right) \\
\frac{d}{d \rho}\left(\frac{I[\mbox{\boldmath $x$}^k; \mbox{\boldmath $z$}^k]}{k+1}\right) &\leq& \frac{1}{2}\left(\frac{1}{k+1}\sum_{j=0}^k \mbox{tr}(P_j) \right) \\
\frac{1}{2 \rho}\left(\frac{1}{k+1}\sum_{j=0}^k \log \vert I_{n_0}+\rho M_j \vert \right) &\leq& \frac{1}{2}\left(\frac{1}{k+1}\sum_{j=0}^k \mbox{tr}(M_j) \right) \\
\frac{1}{2}\left(\frac{1}{k+1}\sum_{j=0}^k \mbox{tr}(P_j) \right) &<& \frac{1}{2 \rho}\left(\frac{1}{k+1}\sum_{j=0}^k \log \vert I_{n_0}+\rho M_j \vert \right) .
\end{eqnarray}
Noting that $\Sigma_r$, $\Sigma_x$, and $\Sigma_c$ are not dependent on $k$, it follows from the above inequalities that
\begin{eqnarray}
\frac{1}{\rho}\left(\frac{I[\mbox{\boldmath $x$}^k; \mbox{\boldmath $z$}^k]}{k+1}\right) &\leq& \frac{1}{2 \rho} \log \vert I_{n_0}+\rho \Sigma_x \vert \\
\frac{d}{d \rho}\left(\frac{I[\mbox{\boldmath $x$}^k; \mbox{\boldmath $z$}^k]}{k+1}\right) &\leq& \frac{1}{2}\mbox{tr}(\Sigma_c) \\
\frac{1}{2 \rho} \log \vert I_{n_0}+\rho \Sigma_x \vert &\leq& \frac{1}{2}\mbox{tr}(\Sigma_x) \\
\frac{1}{2}\mbox{tr}(\Sigma_c) &<& \frac{1}{2 \rho} \log \vert I_{n_0}+\rho \Sigma_x \vert .
\end{eqnarray}
(We remark that in the fourth inequality, the conditions corresponding to those of Lemma 4 are implicitly assumed.) Note that $\frac{1}{2} \log \vert I_{n_0}+\rho \Sigma_x \vert$ is an approximate value of the average mutual information per branch. Also, $\mbox{tr}(\Sigma_c)$ represents the LMMSE per branch. The above are the fundamental relations for Viterbi decoding of convolutional codes.
\par
In the following, we will discuss the case of $n_0=2$ in more detail.

\subsection{Mutual Information and LMMSE in the Case of $n_0=2$}
In the case of $n_0=2$~\cite{taji 21}, $\Sigma_r$ is given by
\begin{eqnarray}
\Sigma_r &=& \left(
\begin{array}{cc}
1+4 \rho \alpha_1(1-\alpha_1) & 4 \rho \theta_{12} \\
4 \rho \theta_{12} & 1+4 \rho \alpha_2(1-\alpha_2)
\end{array}
\right) \nonumber \\
&=& I_2+\rho \Sigma_x ,
\end{eqnarray}
where
\begin{equation}
\Sigma_x=\left(
\begin{array}{cc}
4\alpha_1(1-\alpha_1) & 4 \theta_{12} \\
4 \theta_{12} & 4\alpha_2(1-\alpha_2)
\end{array}
\right) .
\end{equation}
Also, we can derive $\Sigma_c$ from $\Sigma_x$ using the formula $\Sigma_c=\Sigma_x-\rho \Sigma_x(I_{n_0}+\rho \Sigma_x)^{-1}\Sigma_x$. We have the following.
\begin{pro}
Let
\begin{equation}
\Sigma_x=\left(
\begin{array}{cc}
\sigma_1^2 & \sigma_{12} \\
\sigma_{12} & \sigma_2^2
\end{array}
\right) .
\end{equation}
Then we have
\begin{eqnarray}
\Sigma_c &\stackrel{\triangle}{=}& \left(
\begin{array}{cc}
c_{11} & c_{12} \\
c_{12} & c_{22}
\end{array}
\right) \nonumber \\
&=& \frac{1}{\Delta_r}\left(
\begin{array}{cc}
\sigma_1^2+\rho \Delta_x & \sigma_{12} \\
\sigma_{12} & \sigma_2^2+\rho \Delta_x
\end{array}
\right) ,
\end{eqnarray}
where 
\begin{eqnarray}
\Delta_x &=& \vert \Sigma_x \vert \nonumber \\
&=& \sigma_1^2\sigma_2^2-(\sigma_{12})^2 \\
\Delta_r &=& \vert \Sigma_r \vert \nonumber \\
&=& 1+\rho(\sigma_1^2+\sigma_2^2+\rho \Delta_x) .
\end{eqnarray}
\end{pro}
\begin{IEEEproof}
See Appendix E.
\end{IEEEproof}
\par
With respect to the relationship between $\Sigma_c$ and $\Sigma_x$, we have the following.
\begin{pro}
$0 \leq \Sigma_c \leq \Sigma_x$.
\end{pro}
\begin{IEEEproof}
The right-hand inequality follows from the relation
\begin{displaymath}
\Sigma_x-\Sigma_c=\rho \Sigma_x (I_2+\rho \Sigma_x)^{-1} \Sigma_x .
\end{displaymath}
Consider the left-hand inequality. It follows from Proposition 13 that
\begin{eqnarray}
\vert \Sigma_c \vert &=& \frac{1}{(\Delta_r)^2}\bigl((\sigma_1^2+\rho \Delta_x)(\sigma_2^2+\rho \Delta_x)-(\sigma_{12})^2 \bigr) \nonumber \\
&=& \frac{\Delta_x \bigl(1+\rho(\sigma_1^2+\sigma_2^2+\rho \Delta_x)\bigr)}{(\Delta_r)^2} \nonumber \\
&=& \frac{\Delta_x}{\Delta_r}\geq 0 .
\end{eqnarray}
Then noting that $c_{11} \geq 0$ and $c_{22} \geq 0$, we have $\Sigma_c \geq 0$~\cite[Theorem 7.2.5]{horn 13}.
\end{IEEEproof}
\par
Recall that $0 \leq P_k \leq M_k$ holds in general. Hence the above inequalities also justify our expectation that $\Sigma_x$ is corresponding to $M_k$.
\par
\begin{cor}
$0 \leq \mbox{tr}(\Sigma_c) \leq \mbox{tr}(\Sigma_x)$.
\end{cor}
\begin{IEEEproof}
Direct consequence of Proposition 14.
\end{IEEEproof}
\par
Now recall the results in the previous subsection. In the case of $n_0=2$, we have
\begin{eqnarray}
\frac{1}{\rho}\left(\frac{I[\mbox{\boldmath $x$}^k; \mbox{\boldmath $z$}^k]}{k+1}\right) &\leq& \frac{1}{2 \rho} \log \vert I_2+\rho \Sigma_x \vert \\
\frac{d}{d \rho}\left(\frac{I[\mbox{\boldmath $x$}^k; \mbox{\boldmath $z$}^k]}{k+1}\right) &\leq& \frac{1}{2}\mbox{tr}(\Sigma_c) \\
\frac{1}{2 \rho} \log \vert I_2+\rho \Sigma_x \vert &\leq& \frac{1}{2}\mbox{tr}(\Sigma_x) \\
\frac{1}{2}\mbox{tr}(\Sigma_c) &<& \frac{1}{2 \rho} \log \vert I_2+\rho \Sigma_x \vert ,
\end{eqnarray}
where (cf. Proposition 13)
\begin{eqnarray}
\frac{1}{2 \rho} \log \vert I_2+\rho \Sigma_x \vert &=& \frac{1}{2 \rho}\log \left(1+\rho(\sigma_1^2+\sigma_2^2+\rho \Delta_x)\right) \\
\frac{1}{2}\mbox{tr}(\Sigma_x) &=& \frac{1}{2}(\sigma_1^2+\sigma_2^2) \\
\frac{1}{2}\mbox{tr}(\Sigma_c) &=& \frac{\frac{1}{2}(\sigma_1^2+\sigma_2^2)+\rho \Delta_x}{1+\rho(\sigma_1^2+\sigma_2^2+\rho \Delta_x)} .
\end{eqnarray}
The variables on the right-hand sides are defined as follows:
\begin{eqnarray}
\sigma_1^2 &=& 4\alpha_1(1-\alpha_1) \\
\sigma_2^2 &=& 4\alpha_2(1-\alpha_2) \\
\sigma_{12} &=& 4\theta_{12} \\
\theta_{12} &=& \alpha_{11}-\alpha_1 \alpha_2 \\
&=& P(v_k^{(1)}=1, v_k^{(2)}=1)-\alpha_1 \alpha_2 \\
\Delta_x &=& \sigma_1^2 \sigma_2^2-(\sigma_{12})^2 .
\end{eqnarray}
Hence our calculation reduces to the evaluation of the values of $\alpha_{11}$, $\alpha_1$ and $\alpha_2$. Note that these are functions of the channel error probability $\epsilon=Q\bigl(\sqrt{2E_s/N_0}\bigr)=Q\bigl(\sqrt{E_b/N_0}\bigr)$.

\subsection{Numerical Results}
In this section, we examine the relationship between $\frac{1}{2}\mbox{tr}(\Sigma_c)$, $\frac{1}{2 \rho}\log \vert I_2+\rho \Sigma_x \vert$, and $\frac{1}{2}\mbox{tr}(\Sigma_x)$ using concrete convolutional codes. Note that the structure of an SST Viterbi decoder for a general convolutional code and that for a QLI code are slightly different (Fig.1 and Fig.6). Also, in~\cite{taji 19}, we showed that a QLI code is related to a filtered estimate in linear estimation theory when it is regarded as a general code, whereas a QLI code has some connection with a smoothed estimate when it is regarded as an inherent QLI code. That is, we can consider the two estimates for a QLI code. Taking into account this important fact, we take QLI codes and then regard them as general codes. We have taken the following two QLI codes~\cite{joha 99}:
\begin{itemize}
\item [1)] $C_1~(\nu=2):~G_1(D)=(1+D+D^2, 1+D^2)$.
\item [2)] $C_2~(\nu=6):~G_2(D)=(1+D^3+D^5+D^6, 1+D+D^3+D^5+D^6)$.
\end{itemize}
\par
Note that the encoded block for the main decoder in an SST Viterbi decoder becomes $\mbox{\boldmath $v$}_k=\mbox{\boldmath $e$}_kG^{-1}G=\mbox{\boldmath $e$}_k(G^{-1}G)$. The $G^{-1}G$'s for the above codes are given as follows:
\begin{equation}
G_1^{-1}G_1=\left(
\begin{array}{cc}
D+D^2+D^3 & D+D^3 \\
1+D^3 & 1+D+D^2+D^3
\end{array}
\right)
\end{equation}
\begin{equation}
G_2^{-1}G_2=\left(
\begin{array}{cc}
1+D^2+D^3+D^4+D^5+D^6+D^9+D^{11} & 1+D+D^2+D^4+D^9+D^{11} \\
D^2+D^4+D^9+D^{11} & D^2+D^3+D^4+D^5+D^6+D^9+D^{11}
\end{array}
\right) .
\end{equation}
Also, $\alpha_{11}$, $\alpha_1$, and $\alpha_2$ are calculated as follows:
\par
1) $C_1$: 
\begin{eqnarray}
\alpha_{11} &=& 4\epsilon-22\epsilon^2+58\epsilon^3-80\epsilon^4+56\epsilon^5-16\epsilon^6 \\
\alpha_1 &=& 5\epsilon-20\epsilon^2+40\epsilon^3-40\epsilon^4+16\epsilon^5 \\
\alpha_2 &=& 6\epsilon-30\epsilon^2+80\epsilon^3-120\epsilon^4+96\epsilon^5-32\epsilon^6 .
\end{eqnarray}
\par
2) $C_2$:
\begin{eqnarray}
\alpha_{11} &=& 9\epsilon-123\epsilon^2+942\epsilon^3-4700\epsilon^4+16464\epsilon^5 \nonumber \\
&& -42128\epsilon^6+80224\epsilon^7-114048\epsilon^8+119680\epsilon^9-90112\epsilon^{10} \nonumber \\
&& +46080\epsilon^{11}-14336\epsilon^{12}+2048\epsilon^{13} \\
\alpha_1 &=& 12\epsilon-132\epsilon^2+880\epsilon^3-3960\epsilon^4+12672\epsilon^5 \nonumber \\
&& -29568\epsilon^6+50688\epsilon^7-63360\epsilon^8+56320\epsilon^9-33792\epsilon^{10} \nonumber \\
&& +12288\epsilon^{11}-2048\epsilon^{12} \\
\alpha_2 &=& 13\epsilon-156\epsilon^2+1144\epsilon^3-5720\epsilon^4+20592\epsilon^5 \nonumber \\
&& -54912\epsilon^6+109824\epsilon^7-164736\epsilon^8+183040\epsilon^9-146432\epsilon^{10} \nonumber \\
&& +79872\epsilon^{11}-26624\epsilon^{12}+4096\epsilon^{13} .
\end{eqnarray}
\par
{\it Remark:} $\alpha_{11}$, $\alpha_1$, and $\alpha_2$ are calculated directly based on the relations $\alpha_{11}=P(v_k^{(1)}=1, v_k^{(2)}=1)$, $\alpha_1=P(v_k^{(1)}=1)$, and $\alpha_2=P(v_k^{(2)}=1)$, respectively. In the case of $\nu=6$, $v_k^{(l)}\,(l=1, 2)$ consists of many error terms and hence calculations are rather complicated. We see that $\alpha_{11}(\epsilon=\frac{1}{2})=\frac{1}{4}$, $\alpha_1(\epsilon=\frac{1}{2})=\frac{1}{2}$, and $\alpha_2(\epsilon=\frac{1}{2})=\frac{1}{2}$, which imply that the obtained results are reasonable.
\par
When the SNR ($E_b/N_0$) is given, $\epsilon=Q\bigl(\sqrt{E_b/N_0}\bigr)$ is determined and hence the value of each variable is obtained. Table I and Table II show the variables related to $\Sigma_x$ for $C_1$ and $C_2$, respectively.
\begin{table}[tb]
\caption{Variables related to $\Sigma_x$ for $C_1$ (as a general code)}
\label{Table 1}
\begin{center}
\begin{tabular}{c*{6}{|c}}
$E_b/N_0~(\mbox{dB})$ & $\alpha_1$ & $4\alpha_1(1-\alpha_1)$ & $\alpha_2$ & $4\alpha_2(1-\alpha_2)$ & $\theta_{12}$ & $\frac{1}{2}\mbox{tr}(\Sigma_x)$ \\
\hline
$-10$ & $0.4995$ & $1.0000$ & $0.4999$ & $1.0000$ & $0.0038$ & $1.0000$ \\
$-9$ & $0.4992$ & $1.0000$ & $0.4998$ & $1.0000$ & $0.0053$ & $1.0000$ \\
$-8$ & $0.4986$ & $1.0000$ & $0.4996$ & $1.0000$ & $0.0074$ & $1.0000$ \\
$-7$ & $0.4976$ & $1.0000$ & $0.4992$ & $1.0000$ & $0.0102$ & $1.0000$ \\
$-6$ & $0.4958$ & $0.9999$ & $0.4984$ & $1.0000$ & $0.0142$ & $1.0000$ \\
$-5$ & $0.4930$ & $0.9998$ & $0.4970$ & $1.0000$ & $0.0193$ & $0.9999$ \\
$-4$ & $0.4883$ & $0.9995$ & $0.4945$ & $0.9999$ & $0.0262$ & $0.9997$ \\
$-3$ & $0.4808$ & $0.9985$ & $0.4900$ & $0.9996$ & $0.0352$ & $0.9991$ \\
$-2$ & $0.4691$ & $0.9962$ & $0.4823$ & $0.9987$ & $0.0465$ & $0.9975$ \\
$-1$ & $0.4515$ & $0.9906$ & $0.4696$ & $0.9963$ & $0.0602$ & $0.9935$ \\
$0$ & $0.4259$ & $0.9780$ & $0.4494$ & $0.9898$ & $0.0758$ & $0.9839$ \\
$1$ & $0.3904$ & $0.9520$ & $0.4191$ & $0.9738$ & $0.0917$ & $0.9629$ \\
$2$ & $0.3442$ & $0.9029$ & $0.3766$ & $0.9391$ & $0.1050$ & $0.9210$ \\
$3$ & $0.2879$ & $0.8201$ & $0.3213$ & $0.8723$ & $0.1116$ & $0.8462$ \\
$4$ & $0.2255$ & $0.6986$ & $0.2565$ & $0.7628$ & $0.1076$ & $0.7307$ \\
$5$ & $0.1621$ & $0.5433$ & $0.1876$ & $0.6096$ & $0.0921$ & $0.5765$ \\
$6$ & $0.1049$ & $0.3756$ & $0.1231$ & $0.4318$ & $0.0681$ & $0.4037$ \\
$7$ & $0.0599$ & $0.2252$ & $0.0710$ & $0.2638$ & $0.0427$ & $0.2445$ \\
$8$ & $0.0293$ & $0.1138$ & $0.0349$ & $0.1347$ & $0.0222$ & $0.1243$ \\
$9$ & $0.0120$ & $0.0474$ & $0.0143$ & $0.0564$ & $0.0094$ & $0.0519$ \\
$10$ & $0.0039$ & $0.0155$ & $0.0047$ & $0.0187$ & $0.0031$ & $0.0171$
\end{tabular}
\end{center}
\end{table}
\begin{table}[tb]
\caption{Variables related to $\Sigma_x$ for $C_2$ (as a general code)}
\label{Table 2}
\begin{center}
\begin{tabular}{c*{6}{|c}}
$E_b/N_0~(\mbox{dB})$ & $\alpha_1$ & $4\alpha_1(1-\alpha_1)$ & $\alpha_2$ & $4\alpha_2(1-\alpha_2)$ & $\theta_{12}$ & $\frac{1}{2}\mbox{tr}(\Sigma_x)$ \\
\hline
$-10$ & $0.5000$ & $1.0000$ & $0.5000$ & $1.0000$ & $0.0000$ & $1.0000$ \\
$-9$ & $0.5000$ & $1.0000$ & $0.5000$ & $1.0000$ & $0.0000$ & $1.0000$ \\
$-8$ & $0.5000$ & $1.0000$ & $0.5000$ & $1.0000$ & $0.0001$ & $1.0000$ \\
$-7$ & $0.5000$ & $1.0000$ & $0.5000$ & $1.0000$ & $0.0001$ & $1.0000$ \\
$-6$ & $0.5000$ & $1.0000$ & $0.5000$ & $1.0000$ & $0.0003$ & $1.0000$ \\
$-5$ & $0.5000$ & $1.0000$ & $0.5000$ & $1.0000$ & $0.0006$ & $1.0000$ \\
$-4$ & $0.4999$ & $1.0000$ & $0.5000$ & $1.0000$ & $0.0014$ & $1.0000$ \\
$-3$ & $0.4998$ & $1.0000$ & $0.4999$ & $1.0000$ & $0.0026$ & $1.0000$ \\
$-2$ & $0.4994$ & $1.0000$ & $0.4996$ & $1.0000$ & $0.0051$ & $1.0000$ \\
$-1$ & $0.4981$ & $1.0000$ & $0.4988$ & $1.0000$ & $0.0095$ & $1.0000$ \\
$0$ & $0.4949$ & $0.9999$ & $0.4965$ & $1.0000$ & $0.0173$ & $1.0000$ \\
$1$ & $0.4869$ & $0.9993$ & $0.4903$ & $0.9996$ & $0.0298$ & $0.9995$ \\
$2$ & $0.4695$ & $0.9963$ & $0.4759$ & $0.9977$ & $0.0482$ & $0.9970$ \\
$3$ & $0.4362$ & $0.9837$ & $0.4462$ & $0.9884$ & $0.0718$ & $0.9861$ \\
$4$ & $0.3814$ & $0.9437$ & $0.3948$ & $0.9557$ & $0.0955$ & $0.9497$ \\
$5$ & $0.3048$ & $0.8476$ & $0.3195$ & $0.8697$ & $0.1092$ & $0.8587$ \\
$6$ & $0.2159$ & $0.6771$ & $0.2289$ & $0.7060$ & $0.1028$ & $0.6916$ \\
$7$ & $0.1319$ & $0.4580$ & $0.1412$ & $0.4851$ & $0.0770$ & $0.4716$ \\
$8$ & $0.0674$ & $0.2514$ & $0.0726$ & $0.2693$ & $0.0449$ & $0.2604$ \\
$9$ & $0.0283$ & $0.1100$ & $0.0306$ & $0.1187$ & $0.0202$ & $0.1144$ \\
$10$ & $0.0093$ & $0.0369$ & $0.0100$ & $0.0396$ & $0.0069$ & $0.0383$
\end{tabular}
\end{center}
\end{table}
\par
Now recall that $\rho \lambda_{max}<1$ is assumed in the argument for Lemma 4, where $\lambda_{max}$ is the maximum eigenvalue of $P_j$ (see Section IV). Since $\Sigma_c$ is corresponding to $P_j$, we have examined whether this condition is satisfied with respect to $\Sigma_c$. From Proposition 13, the two eigenvalues of $\Sigma_c$ are given by
\begin{eqnarray}
\lambda_1 &=& \frac{\frac{1}{2}(\sigma_1^2+\sigma_2^2)+\rho \Delta_x}{\Delta_r}-\frac{\sqrt{\frac{1}{4}(\sigma_1^2-\sigma_2^2)^2+(\sigma_{12})^2}}{\Delta_r} \nonumber \\
&=& \frac{1}{2}\mbox{tr}(\Sigma_c)-\frac{\sqrt{\frac{1}{4}(\sigma_1^2-\sigma_2^2)^2+(\sigma_{12})^2}}{\Delta_r} \\
\lambda_2 &=& \frac{\frac{1}{2}(\sigma_1^2+\sigma_2^2)+\rho \Delta_x}{\Delta_r}+\frac{\sqrt{\frac{1}{4}(\sigma_1^2-\sigma_2^2)^2+(\sigma_{12})^2}}{\Delta_r} \nonumber \\
&=& \frac{1}{2}\mbox{tr}(\Sigma_c)+\frac{\sqrt{\frac{1}{4}(\sigma_1^2-\sigma_2^2)^2+(\sigma_{12})^2}}{\Delta_r}~(\stackrel{\triangle}{=}\lambda_{max}) .
\end{eqnarray}
From Tables I and II, we observe that $\sigma_1^2 \approx \sigma_2^2$ holds. Then $\lambda_1$ and $\lambda_2$ are approximated by
\begin{eqnarray}
\tilde \lambda_1 &=& \frac{1}{2}\mbox{tr}(\Sigma_c)-\frac{\sigma_{12}}{\Delta_r} \nonumber \\
&=& \frac{1}{2}\mbox{tr}(\Sigma_c)-\frac{4\theta_{12}}{\Delta_r} \\
\tilde \lambda_2 &=& \frac{1}{2}\mbox{tr}(\Sigma_c)+\frac{\sigma_{12}}{\Delta_r} \nonumber \\
&=& \frac{1}{2}\mbox{tr}(\Sigma_c)+\frac{4\theta_{12}}{\Delta_r}~(\stackrel{\triangle}{=}\tilde \lambda_{max}) ,
\end{eqnarray}
respectively. The values of $\tilde \lambda_1$, $\tilde \lambda_2$, $\rho \tilde \lambda_1$ and $\rho \tilde \lambda_{max}$ for $C_1$ and $C_2$ are shown in Tables III and IV, respectively.
\begin{table}[tb]
\caption{$\tilde \lambda_1$, $\tilde \lambda_2$, $\rho \tilde \lambda_1$ and $\rho \tilde \lambda_{max}$ for $C_1$ (as a general code)}
\label{Table 3}
\begin{center}
\begin{tabular}{c*{5}{|c}}
$E_b/N_0~(\mbox{dB})$ & $\rho$ & $\tilde \lambda_1$ & $\tilde \lambda_2$ & $\rho \tilde \lambda_1$ & $\rho \tilde \lambda_{max}$ \\
\hline
$-10$ & $0.1000$ & $0.8965$ & $0.9217$ & $0.0897$ & $0.0922$ \\
$-9$ & $0.1259$ & $0.8715$ & $0.9049$ & $0.1097$ & $0.1139$ \\
$-8$ & $0.1585$ & $0.8410$ & $0.8852$ & $0.1333$ & $0.1403$ \\
$-7$ & $0.1995$ & $0.8051$ & $0.8619$ & $0.1606$ & $0.1719$ \\
$-6$ & $0.2512$ & $0.7625$ & $0.8351$ & $0.1915$ & $0.2098$ \\
$-5$ & $0.3162$ & $0.7143$ & $0.8035$ & $0.2259$ & $0.2541$ \\
$-4$ & $0.3981$ & $0.6598$ & $0.7672$ & $0.2627$ & $0.3054$ \\
$-3$ & $0.5012$ & $0.6001$ & $0.7255$ & $0.3008$ & $0.3636$ \\
$-2$ & $0.6310$ & $0.5367$ & $0.6775$ & $0.3387$ & $0.4275$ \\
$-1$ & $0.7943$ & $0.4711$ & $0.6233$ & $0.3742$ & $0.4951$ \\
$0$ & $1.0000$ & $0.4050$ & $0.5628$ & $0.4050$ & $0.5628$ \\
$1$ & $1.2589$ & $0.3405$ & $0.4973$ & $0.4287$ & $0.6261$ \\
$2$ & $1.5849$ & $0.2792$ & $0.4290$ & $0.4425$ & $0.6799$ \\
$3$ & $1.9953$ & $0.2223$ & $0.3611$ & $0.4436$ & $0.7205$ \\
$4$ & $2.5119$ & $0.1710$ & $0.2964$ & $0.4295$ & $0.7445$ \\
$5$ & $3.1623$ & $0.1252$ & $0.2368$ & $0.3959$ & $0.7488$ \\
$6$ & $3.9811$ & $0.0858$ & $0.1830$ & $0.3416$ & $0.7285$ \\
$7$ & $5.0119$ & $0.0534$ & $0.1346$ & $0.2676$ & $0.6746$ \\
$8$ & $6.3096$ & $0.0288$ & $0.0908$ & $0.1817$ & $0.5729$ \\
$9$ & $7.9433$ & $0.0128$ & $0.0522$ & $0.1017$ & $0.4146$ \\
$10$ & $10.0000$ & $0.0045$ & $0.0227$ & $0.0450$ & $0.2270$
\end{tabular}
\end{center}
\end{table}
\begin{table}[tb]
\caption{$\tilde \lambda_1$, $\tilde \lambda_2$, $\rho \tilde \lambda_1$ and $\rho \tilde \lambda_{max}$ for $C_2$ (as a general code)}
\label{Table 4}
\begin{center}
\begin{tabular}{c*{5}{|c}}
$E_b/N_0~(\mbox{dB})$ & $\rho$ & $\tilde \lambda_1$ & $\tilde \lambda_2$ & $\rho \tilde \lambda_1$ & $\rho \tilde \lambda_{max}$ \\
\hline
$-10$ & $0.1000$ & $0.9091$ & $0.9091$ & $0.0909$ & $0.0909$ \\
$-9$ & $0.1259$ & $0.8881$ & $0.8881$ & $0.1118$ & $0.1118$ \\
$-8$ & $0.1585$ & $0.8629$ & $0.8635$ & $0.1368$ & $0.1369$ \\
$-7$ & $0.1995$ & $0.8334$ & $0.8340$ & $0.1663$ & $0.1664$ \\
$-6$ & $0.2512$ & $0.7984$ & $0.8000$ & $0.2006$ & $0.2010$ \\
$-5$ & $0.3162$ & $0.7584$ & $0.7612$ & $0.2398$ & $0.2407$ \\
$-4$ & $0.3981$ & $0.7124$ & $0.7182$ & $0.2836$ & $0.2859$ \\
$-3$ & $0.5012$ & $0.6615$ & $0.6707$ & $0.3315$ & $0.3362$ \\
$-2$ & $0.6310$ & $0.6054$ & $0.6208$ & $0.3820$ & $0.3917$ \\
$-1$ & $0.7943$ & $0.5453$ & $0.5689$ & $0.4331$ & $0.4519$ \\
$0$ & $1.0000$ & $0.4821$ & $0.5167$ & $0.4821$ & $0.5167$ \\
$1$ & $1.2589$ & $0.4175$ & $0.4645$ & $0.5256$ & $0.5848$ \\
$2$ & $1.5849$ & $0.3535$ & $0.4123$ & $0.5603$ & $0.6535$ \\
$3$ & $1.9953$ & $0.2919$ & $0.3597$ & $0.5824$ & $0.7177$ \\
$4$ & $2.5119$ & $0.2340$ & $0.3064$ & $0.5878$ & $0.7696$ \\
$5$ & $3.1623$ & $0.1808$ & $0.2542$ & $0.5717$ & $0.8039$ \\
$6$ & $3.9811$ & $0.1324$ & $0.2046$ & $0.5271$ & $0.8145$ \\
$7$ & $5.0119$ & $0.0898$ & $0.1588$ & $0.4501$ & $0.7959$ \\
$8$ & $6.3096$ & $0.0534$ & $0.1164$ & $0.3369$ & $0.7344$ \\
$9$ & $7.9433$ & $0.0265$ & $0.0765$ & $0.2105$ & $0.6077$ \\
$10$ & $10.0000$ & $0.0097$ & $0.0397$ & $0.0970$ & $0.3970$
\end{tabular}
\end{center}
\end{table}
We see that $\rho \tilde \lambda_{max}<1$ actually holds for both $C_1$ and $C_2$. Hence our argument is justified.
\par
We have calculated the values of $\frac{1}{2}\mbox{tr}(\Sigma_c)$, $\frac{1}{2 \rho}\log \vert I_2+\rho \Sigma_x \vert$, and $\frac{1}{2}\mbox{tr}(\Sigma_x)$. Results are shown in Tables V and VI. They are shown in Figs. 2 and 3 as well.
\begin{table}[tb]
\caption{$\frac{1}{2}\mbox{tr}(\Sigma_c)$, $\frac{1}{2 \rho}\log \vert I_2+\rho \Sigma_x \vert $, and $\frac{1}{2}\mbox{tr}(\Sigma_x)$ for $C_1$ (as a general code)}
\label{Table 5}
\begin{center}
\begin{tabular}{c*{6}{|c}}
$E_b/N_0~(\mbox{dB})$ & $\frac{\rho}{2}\mbox{tr}(\Sigma_c)$ & $\frac{1}{2}\mbox{tr}(\Sigma_c)$ & $\frac{1}{1+\rho}$ & $\frac{1}{2 \rho}\log \vert I_2+\rho \Sigma_x \vert $ & $\frac{\log(1+\rho)}{\rho}$ & $\frac{1}{2}\mbox{tr}(\Sigma_x)$ \\
\hline
$-10$ & $0.0909$ & $0.9091$ & $0.9091$ & $0.9531$ & $0.9531$ & $1.0000$ \\
$-9$ & $0.1118$ &$0.8882$ & $0.8882$ & $0.9419$ & $0.9419$ & $1.0000$ \\
$-8$ & $0.1368$ & $0.8631$ & $0.8632$ & $0.9282$ & $0.9282$ & $1.0000$ \\
$-7$ & $0.1663$ & $0.8335$ & $0.8337$ & $0.9117$ & $0.9118$ & $1.0000$ \\
$-6$ & $0.2007$ & $0.7988$ & $0.7992$ & $0.8918$ & $0.8921$ & $1.0000$ \\
$-5$ & $0.2400$ & $0.7589$ & $0.7597$ & $0.8683$ & $0.8689$ & $0.9999$ \\
$-4$ & $0.2840$ & $0.7135$ & $0.7153$ & $0.8404$ & $0.8418$ & $0.9997$ \\
$-3$ & $0.3322$ & $0.6628$ & $0.6661$ & $0.8077$ & $0.8106$ & $0.9991$ \\
$-2$ & $0.3831$ & $0.6071$ & $0.6131$ & $0.7696$ & $0.7753$ & $0.9975$ \\
$-1$ & $0.4346$ & $0.5472$ & $0.5573$ & $0.7251$ & $0.7360$ & $0.9935$ \\
$0$ & $0.4839$ & $0.4839$ & $0.5000$ & $0.6732$ & $0.6931$ & $0.9839$ \\
$1$ & $0.5274$ & $0.4189$ & $0.4427$ & $0.6130$ & $0.6473$ & $0.9629$ \\
$2$ & $0.5612$ & $0.3541$ & $0.3869$ & $0.5438$ & $0.5992$ & $0.9210$ \\
$3$ & $0.5820$ & $0.2917$ & $0.3339$ & $0.4664$ & $0.5498$ & $0.8462$ \\
$4$ & $0.5870$ & $0.2337$ & $0.2847$ & $0.3834$ & $0.5001$ & $0.7307$ \\
$5$ & $0.5724$ & $0.1810$ & $0.2403$ & $0.2984$ & $0.4510$ & $0.5765$ \\
$6$ & $0.5351$ & $0.1344$ & $0.2008$ & $0.2166$ & $0.4033$ & $0.4037$ \\
$7$ & $0.4711$ & $0.0940$ & $0.1663$ & $0.1434$ & $0.3579$ & $0.2445$ \\
$8$ & $0.3773$ & $0.0598$ & $0.1368$ & $0.0834$ & $0.3153$ & $0.1243$ \\
$9$ & $0.2582$ & $0.0325$ & $0.1118$ & $0.0405$ & $0.2758$ & $0.0519$ \\
$10$ & $0.1360$ & $0.0136$ & $0.0909$ & $0.0152$ & $0.2398$ & $0.0171$
\end{tabular}
\end{center}
\end{table}
\begin{table}[tb]
\caption{$\frac{1}{2}\mbox{tr}(\Sigma_c)$, $\frac{1}{2 \rho}\log \vert I_2+\rho \Sigma_x \vert $, and $\frac{1}{2}\mbox{tr}(\Sigma_x)$ for $C_2$ (as a general code)}
\label{Table 6}
\begin{center}
\begin{tabular}{c*{6}{|c}}
$E_b/N_0~(\mbox{dB})$ & $\frac{\rho}{2}\mbox{tr}(\Sigma_c)$ & $\frac{1}{2}\mbox{tr}(\Sigma_c)$ & $\frac{1}{1+\rho}$ & $\frac{1}{2 \rho}\log \vert I_2+\rho \Sigma_x \vert $ & $\frac{\log(1+\rho)}{\rho}$ & $\frac{1}{2}\mbox{tr}(\Sigma_x)$ \\
\hline
$-10$ & $0.0909$ & $0.9091$ & $0.9091$ & $0.9531$ & $0.9531$ & $1.0000$ \\
$-9$ & $0.1118$ & $0.8881$ & $0.8882$ & $0.9419$ & $0.9419$ & $1.0000$ \\
$-8$ & $0.1368$ & $0.8632$ & $0.8632$ & $0.9282$ & $0.9282$ & $1.0000$ \\
$-7$ & $0.1663$ & $0.8337$ & $0.8337$ & $0.9118$ & $0.9118$ & $1.0000$ \\
$-6$ & $0.2008$ & $0.7992$ & $0.7992$ & $0.8921$ & $0.8921$ & $1.0000$ \\
$-5$ & $0.2403$ & $0.7598$ & $0.7597$ & $0.8689$ & $0.8689$ & $1.0000$ \\
$-4$ & $0.2848$ & $0.7153$ & $0.7153$ & $0.8418$ & $0.8418$ & $1.0000$ \\
$-3$ & $0.3339$ & $0.6661$ & $0.6661$ & $0.8106$ & $0.8106$ & $1.0000$ \\
$-2$ & $0.3869$ & $0.6131$ & $0.6131$ & $0.7752$ & $0.7753$ & $1.0000$ \\
$-1$ & $0.4425$ & $0.5571$ & $0.5573$ & $0.7358$ & $0.7360$ & $1.0000$ \\
$0$ & $0.4994$ & $0.4994$ & $0.5000$ & $0.6925$ & $0.6931$ & $1.0000$ \\
$1$ & $0.5552$ & $0.4410$ & $0.4427$ & $0.6453$ & $0.6473$ & $0.9995$ \\
$2$ & $0.6069$ & $0.3829$ & $0.3869$ & $0.5936$ & $0.5992$ & $0.9970$ \\
$3$ & $0.6501$ & $0.3258$ & $0.3339$ & $0.5356$ & $0.5498$ & $0.9861$ \\
$4$ & $0.6787$ & $0.2702$ & $0.2847$ & $0.4688$ & $0.5001$ & $0.9497$ \\
$5$ & $0.6878$ & $0.2175$ & $0.2403$ & $0.3915$ & $0.4510$ & $0.8587$ \\
$6$ & $0.6708$ & $0.1685$ & $0.2008$ & $0.3057$ & $0.4033$ & $0.6916$ \\
$7$ & $0.6230$ & $0.1243$ & $0.1663$ & $0.2184$ & $0.3579$ & $0.4716$ \\
$8$ & $0.5357$ & $0.0849$ & $0.1368$ & $0.1378$ & $0.3153$ & $0.2604$ \\
$9$ & $0.4091$ & $0.0515$ & $0.1118$ & $0.0737$ & $0.2758$ & $0.1144$ \\
$10$ & $0.2470$ & $0.0247$ & $0.0909$ & $0.0304$ & $0.2398$ & $0.0383$
\end{tabular}
\end{center}
\end{table}
\begin{figure}[tb]
\begin{center}
\includegraphics[width=10.0cm,clip]{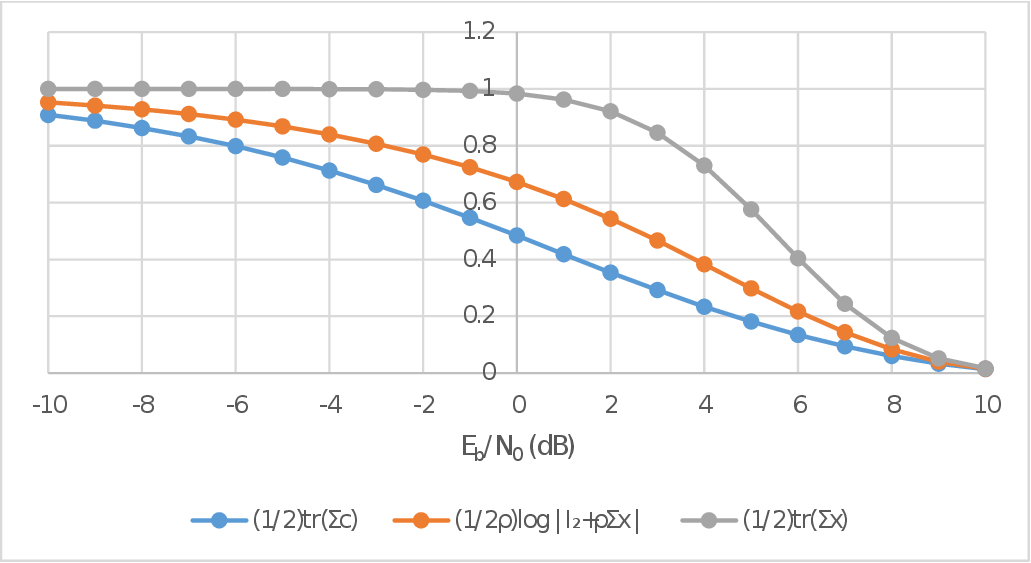}
\end{center}
\caption{Relationship between $\frac{1}{2}\mbox{tr}(\Sigma_c)$, $\frac{1}{2 \rho}\log \vert I_2+\rho \Sigma_x \vert $, and $\frac{1}{2}\mbox{tr}(\Sigma_x)$ ($C_1$).}
\label{Fig.2}
\end{figure}
\begin{figure}[tb]
\begin{center}
\includegraphics[width=10.0cm,clip]{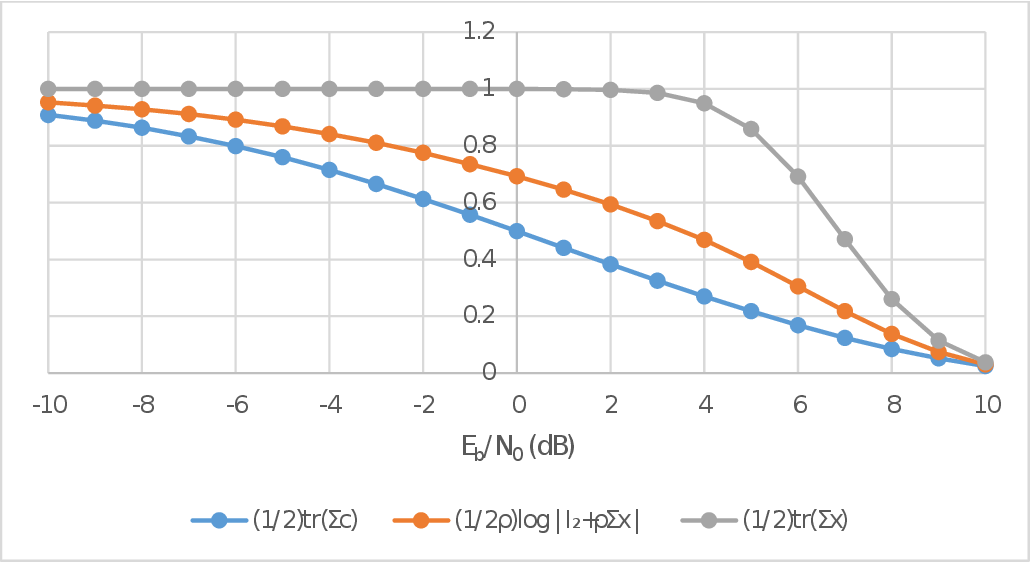}
\end{center}
\caption{Relationship between $\frac{1}{2}\mbox{tr}(\Sigma_c)$, $\frac{1}{2 \rho}\log \vert I_2+\rho \Sigma_x \vert $, and $\frac{1}{2}\mbox{tr}(\Sigma_x)$ ($C_2$).}
\label{Fig.3}
\end{figure}
From Figs. 2 and 3, we observe that the inequalities
\begin{equation}
\frac{1}{2}\mbox{tr}(\Sigma_c)<\frac{1}{2 \rho}\log \vert I_2+\rho \Sigma_x \vert<\frac{1}{2}\mbox{tr}(\Sigma_x)
\end{equation}
hold for both $C_1$ and $C_2$.
\par
The right-hand inequality $\frac{1}{2 \rho}\log \vert I_2+\rho \Sigma_x \vert<\frac{1}{2}\mbox{tr}(\Sigma_x)$ is understandable from Lemma 1. Then consider the left-hand inequality $\frac{1}{2}\mbox{tr}(\Sigma_c)<\frac{1}{2 \rho}\log \vert I_2+\rho \Sigma_x \vert$. This is related to Lemma 4, where the condition
\begin{displaymath}
(\Gamma+\rho \Phi)_{ll}=\frac{d}{d\rho}(\rho \lambda_l)>0~(1 \leq l \leq n_0)
\end{displaymath}
is assumed. Since
\begin{eqnarray}
\mbox{tr}(\Gamma+\rho \Phi) &=& \sum_{l=1}^{n_0}(\Gamma+\rho \Phi)_{ll} \nonumber \\
&=& \frac{d}{d\rho}(\rho\,\mbox{tr}(P_j)) , \nonumber
\end{eqnarray}
if $\frac{d}{d\rho}(\rho\,\mbox{tr}(P_j))>0$, then it is expected that $\frac{d}{d\rho}(\rho \lambda_l)>0~(1 \leq l \leq n_0)$ holds with high probability. Now consider $\Sigma_c$, where $n_0=2$. Let $\lambda_1$ and $\lambda_2$ be the eigenvalues of $\Sigma_c$. Also, let $\tilde \lambda_1$ and $\tilde \lambda_2$ be approximate values of them. In this case, if $\frac{d}{d\rho}(\rho\,\mbox{tr}(\Sigma_c))>0$, then it is expected that $\frac{d}{d\rho}(\rho \lambda_l)>0~(l=1, 2)$ holds with high probability. In order to examine it, let us look at the variable $\frac{\rho}{2}\mbox{tr}(\Sigma_c)$ in Tables V and VI. In the case of $C_1$, we observe that $\frac{\rho}{2}\mbox{tr}(\Sigma_c)$ is an increasing function of $\rho$ for SNR values smaller than $4\mbox{dB}$. That is, $\frac{d}{d\rho}(\rho\,\mbox{tr}(\Sigma_c))>0$ holds. Here we see from Table III that the behaviors of $\rho \tilde \lambda_1$ and $\rho \tilde \lambda_2$ are similar to that of $\frac{\rho}{2}\mbox{tr}(\Sigma_c)$. That is, $\frac{d}{d\rho}(\rho \tilde \lambda_l)>0~(l=1, 2)$ holds approximately in the same range of SNR values. Similarly, in the case of $C_2$, $\frac{\rho}{2}\mbox{tr}(\Sigma_c)$ is an increasing function of $\rho$ for SNR values smaller than $5\mbox{dB}$ and hence $\frac{d}{d\rho}(\rho\,\mbox{tr}(\Sigma_c))>0$ holds. As in the case of $C_1$, we see from Table IV that the behaviors of $\rho \tilde \lambda_1$ and $\rho \tilde \lambda_2$ are similar to that of $\frac{\rho}{2}\mbox{tr}(\Sigma_c)$. That is, $\frac{d}{d\rho}(\rho \tilde \lambda_l)>0~(l=1, 2)$ holds approximately in the same range of SNR values. Hence we can say that $\frac{d}{d\rho}(\rho \lambda_l)>0~(l=1, 2)$ holds approximately in the same range of SNR values as that in which $\frac{d}{d\rho}(\rho\,\mbox{tr}(\Sigma_c))>0$ holds. Thus it has been shown that our expectation is valid. As a result, we have $\frac{1}{2}\mbox{tr}(\Sigma_c)<\frac{1}{2 \rho}\log \vert I_2+\rho \Sigma_x \vert$.
\par
On the other hand, for higher SNRs (i.e., the SNR exceeds $4\mbox{dB}$ for $C_1$ and the SNR exceeds $5\mbox{dB}$ for $C_2$), we observe that $\frac{\rho}{2}\mbox{tr}(\Sigma_c)$ is a decreasing function of $\rho$. That is, $\rho\,\mbox{tr}(\Sigma_c)$ becomes smaller as $\rho$ increases. This means that we are likely to have the inequality $\log \vert I_2+\rho \Sigma_x \vert>\rho\,\mbox{tr}(\Sigma_c)$. Then there is a possibility that $\frac{1}{2}\mbox{tr}(\Sigma_c)<\frac{1}{2 \rho}\log \vert I_2+\rho \Sigma_x \vert$ holds. We think that the inequality $\frac{1}{2}\mbox{tr}(\Sigma_c)<\frac{1}{2 \rho}\log \vert I_2+\rho \Sigma_x \vert$ is justified in this manner.

\subsection{Discussion}
The values of $\frac{1}{1+\rho}$ and $\frac{\log(1+\rho)}{\rho}$ are shown in Tables V and VI as well. We observe that
\begin{eqnarray}
\frac{1}{2 \rho}\log \vert I_2+\rho \Sigma_x \vert &\leq& \frac{\log(1+\rho)}{\rho} \\
\frac{1}{2}\mbox{tr}(\Sigma_c) &\leq& \frac{1}{1+\rho} .
\end{eqnarray}
Let
\begin{equation}
\mbox{\boldmath $x$}_j=\left(
\begin{array}{c}
x_j^{(1)} \\
x_j^{(2)}
\end{array}
\right) ,
\end{equation}
where $\mbox{\boldmath $x$}_j$ is standard Gaussian, i.e., the associated covariance matrix is $I_2$. Then we have
\begin{equation}
I[\mbox{\boldmath $x$}_j; \mbox{\boldmath $z$}_j]=\frac{1}{2}\log \vert I_2+\rho I_2 \vert=\log(1+\rho) .
\end{equation}
That is, $\log(1+\rho)$ represents the mutual information per branch for a Gaussian case. Note that the above inequalities show the effect of coding. We also observe that the bounds in (143) and (144) are tight for small $\rho>0$. These are explained as follows.

\subsubsection{$\frac{1}{2 \rho}\log \vert I_2+\rho \Sigma_x \vert \leq \frac{\log(1+\rho)}{\rho}$}
Note that $P_j \leq M_j \leq X_j$ hold in general, where $X_j$ is the covariance matrix of $\mbox{\boldmath $x$}_j$. We see that $X_j=I_{n_0}$ for the signal model in this paper. Then noting that
\begin{displaymath}
I_{n_0}+\rho M_j \leq I_{n_0}+\rho I_{n_0}=(1+\rho)I_{n_0} ,
\end{displaymath}
we have
\begin{displaymath}
\vert I_{n_0}+\rho M_j \vert \leq \vert (1+\rho)I_{n_0} \vert=(1+\rho)^{n_0} .
\end{displaymath}
Hence we have
\begin{displaymath}
\frac{1}{k+1}\sum_{j=0}^k \log \vert I_{n_0}+\rho M_j \vert \leq n_0 \log (1+\rho) .
\end{displaymath}
By replacing the left-hand side with $\log \vert I_{n_0}+\rho \Sigma_x \vert=\log \vert I_2+\rho \Sigma_x \vert$, we obtain
\begin{displaymath}
\log \vert I_2+\rho \Sigma_x \vert \leq 2 \log (1+\rho) ,
\end{displaymath}
or equivalently
\begin{equation}
\frac{1}{2 \rho}\log \vert I_2+\rho \Sigma_x \vert \leq\frac{\log (1+\rho)}{\rho} .
\end{equation}
\par
Let us show that the bound in (147) is tight for sufficiently small $\rho>0$. When $\rho>0$ is sufficiently small, we have $\Sigma_x \approx I_2$ (cf. $M_j \approx X_j=I_{n_0}$). Then it follows from
\begin{displaymath}
\vert I_2+\rho\Sigma_x \vert \approx \vert (1+\rho)I_2 \vert=(1+\rho)^2
\end{displaymath}
that
\begin{equation}
\frac{1}{2 \rho}\log \vert I_2+\rho \Sigma_x \vert \approx \frac{1}{2 \rho}2\log(1+\rho)=\frac{\log (1+\rho)}{\rho} .
\end{equation}

\subsubsection{$\frac{1}{2}\mbox{tr}(\Sigma_c) \leq \frac{1}{1+\rho}$}
First, we show the inequality $\frac{1}{2}\mbox{tr}(\Sigma_c) \leq \frac{1}{1+\rho}$ by direct computation. Note that
\begin{equation}
\frac{1}{2}\mbox{tr}(\Sigma_c)=\frac{\frac{1}{2}(\sigma_1^2+\sigma_2^2)+\rho \Delta_x}{1+\rho(\sigma_1^2+\sigma_2^2+\rho \Delta_x)} .
\end{equation}
Consider the difference
\begin{eqnarray}
\lefteqn{\frac{1}{1+\rho}-\frac{\frac{1}{2}(\sigma_1^2+\sigma_2^2)+\rho \Delta_x}{1+\rho(\sigma_1^2+\sigma_2^2+\rho \Delta_x)}} \nonumber \\
&& =\frac{1+\rho(\sigma_1^2+\sigma_2^2+\rho \Delta_x)-(1+\rho)(\frac{1}{2}(\sigma_1^2+\sigma_2^2)+\rho \Delta_x)}{(1+\rho)(1+\rho(\sigma_1^2+\sigma_2^2+\rho \Delta_x))} \nonumber \\
&& \stackrel{\triangle}{=}\frac{\phi}{\psi} .
\end{eqnarray}
By direct computation, $\phi$ is reduced to
\begin{equation}
\phi=1-\frac{1}{2}(\sigma_1^2+\sigma_2^2)+\frac{\rho}{2}\left((\sigma_1^2+\sigma_2^2)-2\sigma_1^2\sigma_2^2+2(\sigma_{12})^2\right) .
\end{equation}
Here, note that
\begin{displaymath}
0 \leq \sigma_1^2=4\alpha_1(1-\alpha_1) \leq 1
\end{displaymath}
\begin{displaymath}
0 \leq \sigma_2^2=4\alpha_2(1-\alpha_2) \leq 1
\end{displaymath}
for $0 \leq \epsilon \leq \frac{1}{2}$ (cf.~\cite[Lemma 13]{taji 19}). Then we have
\begin{displaymath}
1-\frac{1}{2}(\sigma_1^2+\sigma_2^2) \geq 0 .
\end{displaymath}
Also, since
\begin{displaymath}
\frac{\sigma_1^2+\sigma_2^2}{2} \geq \sqrt{\sigma_1^2\sigma_2^2} \geq \sigma_1^2\sigma_2^2 ,
\end{displaymath}
we have
\begin{displaymath}
(\sigma_1^2+\sigma_2^2)-2\sigma_1^2\sigma_2^2 \geq 0 .
\end{displaymath}
Thus we obtain $\phi \geq 0$. That is, the inequality $\frac{1}{2}\mbox{tr}(\Sigma_c) \leq \frac{1}{1+\rho}$ has been proved.
\par
Next, let $\rho>0$ be sufficiently small. From the definition of $M_j$, $M_j \approx X_j=I_{n_0}$ holds. This means that $M_j$ can be regarded as a constant matrix. In this case, noting that $M_j \leq I_{n_0}$, it is shown that
\begin{equation}
\frac{d}{d \rho}\log \vert I_{n_0}+\rho M_j \vert \leq \frac{d}{d \rho}\log \vert I_{n_0}+\rho I_{n_0} \vert .
\end{equation}
(For the relation
\begin{eqnarray}
\lefteqn{\frac{d}{d \rho}\log \vert I_{n_0}+\rho I_{n_0} \vert -\frac{d}{d \rho}\log \vert I_{n_0}+\rho M_j \vert} \nonumber \\
&& =\frac{d}{d \rho}\log \frac{\vert I_{n_0}+\rho I_{n_0} \vert}{\vert I_{n_0}+\rho M_j \vert} , \nonumber
\end{eqnarray}
it suffices to note that
\begin{displaymath}
\frac{\vert I_{n_0}+\rho I_{n_0} \vert}{\vert I_{n_0}+\rho M_j \vert}
\end{displaymath}
is a non-decreasing function of $\rho$.) Note that since $M_j \approx I_{n_0}$, the bound in (152) is tight.
\par
If $M_j$ is independent of $\rho$, then we have
\begin{eqnarray}
\frac{d}{d \rho}\log \vert I_{n_0}+\rho M_j \vert &=& \frac{1}{\vert I_{n_0}+\rho M_j \vert}\frac{d}{d \rho} \vert I_{n_0}+\rho M_j \vert \nonumber \\
&=& \mbox{tr} \left((I_{n_0}+\rho M_j)^{-1} \frac{d}{d \rho}(I_{n_0}+\rho M_j) \right) \nonumber \\
&=& \mbox{tr}(R_j^{-1}M_j) \nonumber \\
&=& \mbox{tr}(P_j) ,
\end{eqnarray}
where the last equality is justified by the relation $M_j=R_jP_j$.
\par
On the other hand, noting that
\begin{displaymath}
\vert I_{n_0}+\rho I_{n_0} \vert=\vert (1+\rho)I_{n_0} \vert=(1+\rho)^{n_0} ,
\end{displaymath}
we have
\begin{displaymath}
\frac{d}{d \rho}\log \vert I_{n_0}+\rho I_{n_0} \vert=\frac{n_0}{1+\rho} .
\end{displaymath}
Thus we obtain the inequality
\begin{equation}
\mbox{tr}(P_j) \leq \frac{n_0}{1+\rho} .
\end{equation}
We remark that $M_j$ is not a constant matrix in the strict sense ($M_j \approx I_{n_0}$). Hence the expression
\begin{equation}
\mbox{tr}(P_j)\stackrel{<}{\sim}\frac{n_0}{1+\rho}
\end{equation}
may be more appropriate. Then we have
\begin{displaymath}
\frac{1}{k+1}\sum_{j=0}^k\mbox{tr}(P_j)\stackrel{<}{\sim}\frac{n_0}{1+\rho} .
\end{displaymath}
By replacing the left-hand side with $\mbox{tr}(\Sigma_c)$ (let $n_0=2$), we finally have
\begin{equation}
\frac{1}{2}\mbox{tr}(\Sigma_c)\stackrel{<}{\sim}\frac{1}{1+\rho} .
\end{equation}
The above shows that the bound in (144) is tight for sufficiently small $\rho>0$.

\subsubsection{Influence of the constraint length $\nu$}
Suppose that the code rate is $\frac{1}{2}$ and let $\nu$ be the constraint length. Consider a portion of the associated code trellis with section length $\ell$ (from depth $j$ to depth $j+\ell$ and is denoted $T_c^{\ell}$). Note that the number of states in $T_c^{\ell}$ is $2^{\nu}$ and $2^{\ell}$ sub-paths diverge from each state at depth $j$. Then there are $2^{\nu} \times 2^{\ell}=2^{\nu+\ell}$ sub-paths on $T_c^{\ell}$. Here note any sub-path
\begin{equation}
\mbox{\boldmath $y$}_{j+1}=\mbox{\boldmath $c$}_{j+1}, \mbox{\boldmath $y$}_{j+2}=\mbox{\boldmath $c$}_{j+2}, \cdots, \mbox{\boldmath $y$}_{j+\ell}=\mbox{\boldmath $c$}_{j+\ell} ,
\end{equation}
where $\mbox{\boldmath $c$}_{j+l}~(1 \leq l \leq \ell)$ represents two code bits with which the branch is labeled. Since the number of sub-paths is $2^{\nu+\ell}$ and they are equally likely, we have
\begin{equation}
P(\mbox{\boldmath $y$}_{j+1}=\mbox{\boldmath $c$}_{j+1}, \cdots, \mbox{\boldmath $y$}_{j+1}=\mbox{\boldmath $c$}_{j+\ell})=\frac{1}{2^{\nu+\ell}} .
\end{equation}
On the other hand, noting that
\begin{equation}
P(\mbox{\boldmath $y$}_{j+l}=\mbox{\boldmath $c$}_{j+l})=\frac{1}{4}~(1 \leq l \leq \ell) ,
\end{equation}
we have
\begin{equation}
P(\mbox{\boldmath $y$}_{j+1}=\mbox{\boldmath $c$}_{j+1}) \times \cdots \times P(\mbox{\boldmath $y$}_{j+\ell}=\mbox{\boldmath $c$}_{j+\ell})=(1/4)^{\ell}=\frac{1}{2^{2\ell}} .
\end{equation}
Since $\frac{1}{2^{\nu+\ell}}\neq \frac{1}{2^{2\ell}}$ for $\ell>\nu$, the above means that if $\ell>\nu$, then $\mbox{\boldmath $y$}_{j+1}, \cdots, \mbox{\boldmath $y$}_{j+\ell}$ are not mutually independent.
\par
Consider the above result from the coding viewpoint. We see that $2\ell$ code bits are related to $T_c^{\ell}$. Hence the number of possible patterns generated by these $2\ell$ bits is $2^{2\ell}$. However, if $\ell>\nu$, then there are only $2^{\nu+\ell}~(<2^{2\ell})$ sub-paths on $T_c^{\ell}$. That is, the patterns on $T_c^{\ell}$ are restricted compared with all possible patterns which can be generated by $2\ell$ bits. Note that this restriction corresponds to coding. Thus two facts: $\mbox{\boldmath $y$}_{j+1}, \cdots, \mbox{\boldmath $y$}_{j+\ell}$ are not mutually independent for $\ell>\nu$ and coding has an influence on the trellis for $\ell>\nu$ are equivalent.
\par
Now let us look at the values of $\frac{1}{2 \rho}\log \vert I_2+\rho \Sigma_x \vert$ for $C_1~(\nu=2)$ and $C_2~(\nu=6)$ (see Tables V and VI). We observe that $\frac{1}{2 \rho}\log \vert I_2+\rho \Sigma_x \vert$ for $C_2$ is larger than that for $C_1$. From the above argument, in the case of $\nu=6$, encoded blocks $\mbox{\boldmath $y$}_j$ are likely to be mutually independent compared with the case of $\nu=2$. That is, $C_2$ is less influenced by coding compared with $C_1$. Hence the derived results are reasonable. A similar argument applies for the values of $\frac{1}{2}\mbox{tr}(\Sigma_x)$ and $\frac{1}{2}\mbox{tr}(\Sigma_c)$ (see Tables V and VI).

\subsubsection{Mutual information for equiprobable binary inputs}
Consider the observations:
\begin{displaymath}
\mbox{\boldmath $z$}_j=\sqrt{\rho}\,\mbox{\boldmath $x$}_j+\mbox{\boldmath $w$}_j,~j=0, 1, \cdots, k .
\end{displaymath}
We see that these equations can be regarded as parallel additive Gaussian noise channels~\cite{gall 68}. In fact, we have the following:
\begin{itemize}
\item Since $\mbox{\boldmath $w$}_j$ and $\mbox{\boldmath $x$}_l$ are independent for $l \leq j$, $\mbox{\boldmath $w$}_j$ and $\mbox{\boldmath $x$}_j$ are independent.
\item $\mbox{\boldmath $w$}_j$ is Gaussian.
\item $\mbox{\boldmath $w$}_j$ and $\mbox{\boldmath $w$}_l$ are independent for $j \neq l$.
\end{itemize}
Then~\cite[Section 7.5]{gall 68} noting that $\mbox{\boldmath $x$}_j~(0 \leq j \leq k)$ are not, in general, mutually independent (see 3)), we have
\begin{equation}
I[\mbox{\boldmath $x$}^k; \mbox{\boldmath $z$}^k] \leq \sum_{j=0}^kI[\mbox{\boldmath $x$}_j; \mbox{\boldmath $z$}_j] ,
\end{equation}
where $I[\mbox{\boldmath $x$}_j; \mbox{\boldmath $z$}_j]$ denotes the mutual information associated with a branch. Note that $I[\mbox{\boldmath $x$}_j; \mbox{\boldmath $z$}_j]$ is given by
\begin{equation}
I[\mbox{\boldmath $x$}_j; \mbox{\boldmath $z$}_j]=n_0I(\rho) ,
\end{equation}
where $I(\rho)$ is the mutual information associated with a single code symbol $y_j^{(l)}~(1 \leq l \leq n_0)$ of $\mbox{\boldmath $y$}_j$ for an equiprobable binary input~\cite{guo 05}. In the above, we have used the fact that $y_j^{(l)}~(1 \leq l \leq n_0)$ are mutually independent. In the following, let $n_0=2$. Then we have
\begin{equation}
\frac{1}{\rho}\left(\frac{I[\mbox{\boldmath $x$}^k; \mbox{\boldmath $z$}^k]}{k+1}\right) \leq \frac{2I(\rho)}{\rho} \leq \frac{\log(1+\rho)}{\rho} ,
\end{equation}
where $\log(1+\rho)=2 \times \frac{\log(1+\rho)}{2}$ is the mutual information associated with a branch for a standard Gaussian input. On the other hand, we already know the inequalities
\begin{equation}
\frac{1}{\rho}\left(\frac{I[\mbox{\boldmath $x$}^k; \mbox{\boldmath $z$}^k]}{k+1}\right) \leq \frac{1}{2 \rho} \log \vert I_2+\rho \Sigma_x \vert \leq \frac{\log(1+\rho)}{\rho} .
\end{equation}
Then the question arises: Is there some quantitative relation between $\frac{1}{2 \rho} \log \vert I_2+\rho \Sigma_x \vert$ and $\frac{2I(\rho)}{\rho}$? We remark that $\frac{1}{2} \log \vert I_2+\rho \Sigma_x \vert$ is directly related to the mutual information for Gaussian inputs. However, it is not clear how $\frac{1}{2} \log \vert I_2+\rho \Sigma_x \vert$ is related to the mutual information for equiprobable binary inputs.
\par
In order to answer the above question, we have calculated the values of $\frac{1}{2 \rho} \log \vert I_2+\rho \Sigma_x \vert$ and $\frac{2I(\rho)}{\rho}$ for $C_1$ and $C_2$.
\par
{\it Remark:} See Guo et al.~\cite{guo 05}. $I(\rho)$ is given by
\begin{eqnarray}
I(\rho) &=& \rho-\int_{-\infty}^{\infty}\frac{e^{-\frac{y^2}{2}}}{\sqrt{2 \pi}}\log \cosh(\rho-\sqrt{\rho}y)dy \\
&=& \rho-\int_{-\infty}^{\infty}\frac{e^{-\frac{(y-\rho)^2}{2\rho}}}{\sqrt{2 \pi \rho}}\log \cosh(y)dy .
\end{eqnarray}
For the integration on the right-hand side, we have used a piecewise linear approximation of $\log \cosh(y)$. Hence the obtained values may be slightly different from the exact values. Note that when $\rho>0$ is sufficiently small, we have
\begin{equation}
I(\rho) \approx \frac{\rho}{2}-\frac{\rho^2}{2}.
\end{equation}
\par
Results for $C_1$ and $C_2$ are shown in Fig.4 and Fig.5, respectively. In these figures, the values of $\frac{\log(1+\rho)}{\rho}$ are shown as well.
\begin{figure}[tb]
\begin{center}
\includegraphics[width=10.0cm,clip]{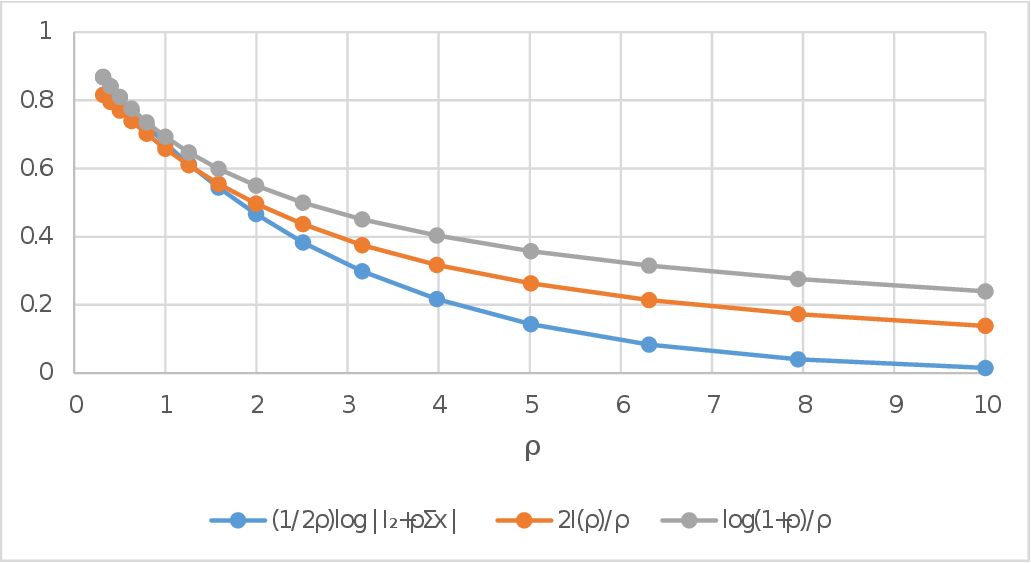}
\end{center}
\caption{Relationship between $\frac{1}{2 \rho}\log \vert I_2+\rho \Sigma_x \vert $, $\frac{2I(\rho)}{\rho}$, and $\frac{\log(1+\rho)}{\rho}$ ($C_1$).}
\label{Fig.4}
\end{figure}
\begin{figure}[tb]
\begin{center}
\includegraphics[width=10.0cm,clip]{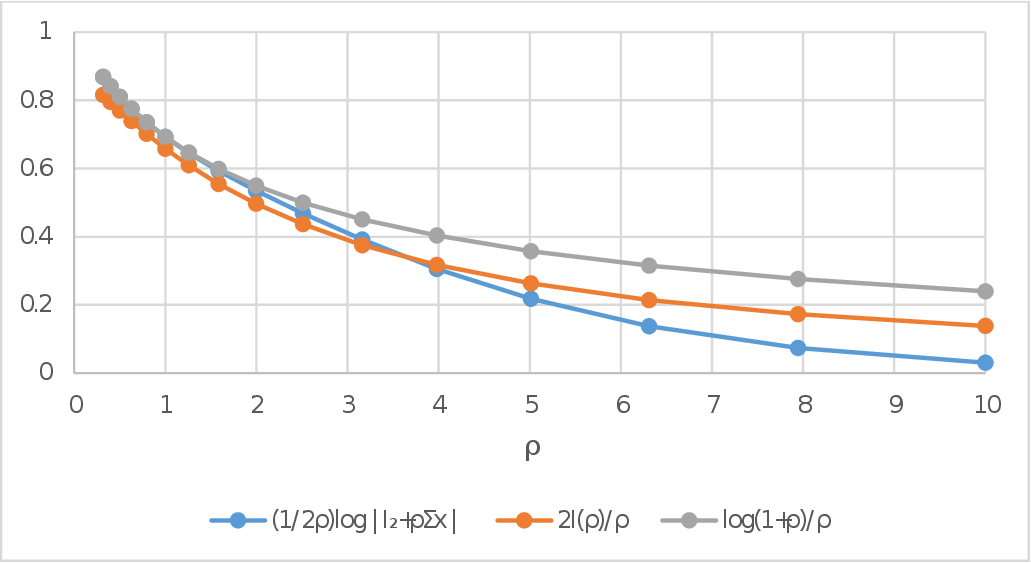}
\end{center}
\caption{Relationship between $\frac{1}{2 \rho}\log \vert I_2+\rho \Sigma_x \vert $, $\frac{2I(\rho)}{\rho}$, and $\frac{\log(1+\rho)}{\rho}$ ($C_2$).}
\label{Fig.5}
\end{figure}
From Figs. 4 and 5, we observe that $\frac{1}{2 \rho}\log \vert I_2+\rho \Sigma_x \vert < \frac{2I(\rho)}{\rho}$ holds for large $\rho$. However, we have a reverse inequality for small $\rho$. More precisely, in the case of $C_1$, a reversion occurs at $\rho=1.3 \sim 1.6$ ($E_b/N_0=1 \mbox{dB} \sim 2 \mbox{dB}$), whereas, in the case of $C_2$, a reversion occurs at $\rho=3.2 \sim 4.0$ ($E_b/N_0=5 \mbox{dB} \sim 6 \mbox{dB}$). We see that a reason for having a reverse inequality comes from the fact that $\frac{1}{2 \rho}\log \vert I_2+\rho \Sigma_x \vert$ is almost equal to $\frac{\log(1+\rho)}{\rho}$ for small $\rho>0$ (see Tables V and VI).
\par
As a result, with respect to the bound for the average mutual information per branch, we have
\begin{equation}
\frac{1}{\rho}\left(\frac{I[\mbox{\boldmath $x$}^k; \mbox{\boldmath $z$}^k]}{k+1}\right) \leq \min \left( \frac{1}{2 \rho} \log \vert I_2+\rho \Sigma_x \vert, \frac{2I(\rho)}{\rho}\right) .
\end{equation}


\section{Errors in Linear Smoothing}
In~\cite{taji 19}, we showed that SST Viterbi decoding of QLI codes has some connection with smoothing in linear estimation theory. In this section, we discuss this subject in more detail. As a preparation, we present a result in linear smoothing, which corresponds to the linear minimum variance filter given in Section II-B.

\subsection{Covariance Matrix of the Linear Smoothing Error}
Kailath and Frost~\cite{kai 682} showed that the innovations method~\cite{kai 681} is equally effective for a linear smoothing problem. Suppose that the signal and observation processes are defined by (16) and (17), respectively. Also, we assume the same conditions as those in Section II-B. Let $b>0$ be fixed (i.e., a fixed-interval problem~\cite{saka 72}). For $k<b$, Kailath and Frost~\cite{kai 682} showed the formula
\begin{equation}
\mbox{\boldmath $\hat x$}_{k \vert b}=\mbox{\boldmath $\hat x$}_{k \vert k}+\sum_{l=k+1}^bE[\mbox{\boldmath $x$}_k\mbox{\boldmath $\nu$}_l^T]R_l^{-1}\mbox{\boldmath $\nu$}_l ,
\end{equation}
where $\mbox{\boldmath $\nu$}_l=\mbox{\boldmath $z$}_l-H_l \mbox{\boldmath $\hat x$}_{l \vert l-1}$ is the innovation and $R_l=E[\mbox{\boldmath $\nu$}_l\mbox{\boldmath $\nu$}_l^T]$ is the associated covariance matrix. Then the covariance matrix (denoted by $\Sigma_{k \vert b}$) of the smoothing error $\mbox{\boldmath $x$}_k-\mbox{\boldmath $\hat x$}_{k \vert b}~(k<b)$ is given as follows.
\begin{pro}
For $k=b-1, b-2, \cdots, 1$, we have 
\begin{eqnarray}
\Sigma_{k \vert b} &=& P_k-P(k, k+1)H_{k+1}^TR_{k+1}^{-1}H_{k+1}P(k, k+1)^T \nonumber \\
&& -P(k, k+2)H_{k+2}^TR_{k+2}^{-1}H_{k+2}P(k, k+2)^T \nonumber \\
&& \cdots \nonumber \\
&& -P(k, b)H_b^TR_b^{-1}H_bP(k, b)^T ,
\end{eqnarray}
where
\begin{displaymath}
P_k=E[(\mbox{\boldmath $x$}_k-\mbox{\boldmath $\hat x$}_{k \vert k})(\mbox{\boldmath $x$}_k-\mbox{\boldmath $\hat x$}_{k \vert k})^T]
\end{displaymath}
is the covariance matrix of the filtering error and where $P(k, l)\,(l=k+1, k+2, \cdots, b)$ is defined by
\begin{equation}
P(k, l)=E[(\mbox{\boldmath $x$}_k-\mbox{\boldmath $\hat x$}_{k \vert k})(\mbox{\boldmath $x$}_l-\mbox{\boldmath $\hat x$}_{l \vert l-1})^T] .
\end{equation}
\end{pro}
\begin{IEEEproof}
See Appendix F.
\end{IEEEproof}
\par
The relation between $\Sigma_{k \vert b}$ and $P_k$ implies that the mean-square error in the smoothed estimate is smaller than that in the filtered estimate, and the difference between the two increases as $b-k$ increases. Hence the result is reasonable.
\par
{\it Remark:} Note the matrix
\begin{displaymath}
P(k, l)=E[(\mbox{\boldmath $x$}_k-\mbox{\boldmath $\hat x$}_{k \vert k})(\mbox{\boldmath $x$}_l-\mbox{\boldmath $\hat x$}_{l \vert l-1})^T] .
\end{displaymath}
We see that $\mbox{\boldmath $x$}_k-\mbox{\boldmath $\hat x$}_{k \vert k}$ is the filtering error, whereas $\mbox{\boldmath $x$}_l-\mbox{\boldmath $\hat x$}_{l \vert l-1}$ is the one-step prediction error. That is, $P(k, l)$ consists of two different kinds of errors. It is shown that $P(k, l)$ is reduced to $P_k$ and
\begin{equation}
P(k, k+1), \cdots , P(k, l-2), P(k, l-1) .
\end{equation}
Hence, if $P(k, k+1), \cdots , P(k, l-2), P(k, l-1)$ have been obtained, then $P(k, l)$ is determined. This implies that we can calculate $P(k, l)~(k+1 \leq l \leq b)$ in ascending order of $l$. Notice that the coefficients $F_j$ which appear in the signal equations are used in the above reduction process.

\subsection{Relationship Between General Codes and QLI Codes}
\begin{figure}[tb]
\begin{center}
\includegraphics[width=10.0cm,clip]{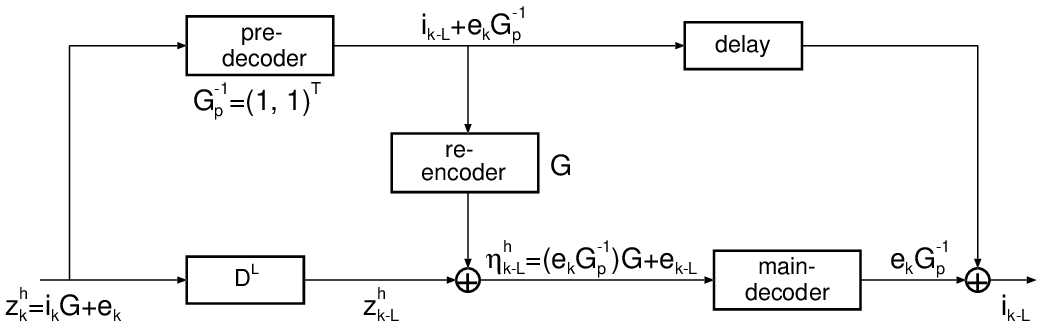}
\end{center}
\caption{The structure of an SST Viterbi decoder for a QLI code (pre-decoder: $G_p^{-1}=(1, 1)^T$).}
\label{Fig.6}
\end{figure}
Fig.6 shows the structure of an SST Viterbi decoder for a QLI code~\cite{mass 71} with generator matrix
\begin{equation}
G(D)=(g_1(D), g_2(D))~(g_1+g_2=D^L,~1 \leq L \leq \nu-1) .
\end{equation}
We observe that the hard-decision input $\mbox{\boldmath $\eta$}_{k-L}^h$ to the main decoder is given by
\begin{eqnarray}
\mbox{\boldmath $\eta$}_{k-L}^h &=& (\mbox{\boldmath $e$}_kG_p^{-1})G+\mbox{\boldmath $e$}_{k-L} \nonumber \\
&=& \mbox{\boldmath $v$}_k+\mbox{\boldmath $e$}_{k-L} ,
\end{eqnarray}
where $G_p^{-1}\stackrel{\triangle}{=}\left(
\begin{array}{c}
1 \\ 
1
\end{array}
\right)$ and where $\mbox{\boldmath $v$}_k=(\mbox{\boldmath $e$}_kG_p^{-1})G$ is the encoded block for the main decoder~\cite{taji 19}. As in the case of $\mbox{\boldmath $r$}_k$, the joint probability density function $p_{\eta}(x, y)$ of $\mbox{\boldmath $\eta$}_{k-L}=\left (\eta_{k-L}^{(1)}, \eta_{k-L}^{(2)}\right)$ is given by
\begin{eqnarray}
p_{\eta}(x, y) &=& \beta_{00}q(x-c)q(y-c)+\beta_{01}q(x-c)q(y+c) \nonumber \\
&& +\beta_{10}q(x+c)q(y-c)+\beta_{11}q(x+c)q(y+c) ,
\end{eqnarray}
where $\beta_{ij}=P(v_k^{(1)}=i, v_k^{(2)}=j)$. The associated covariance matrix (denoted $\Sigma_{\eta}$) becomes
\begin{eqnarray}
\Sigma_{\eta} &=& \left(
\begin{array}{cc}
1+4 \rho \beta_1(1-\beta_1) & 4 \rho \theta_{12}' \\
4 \rho \theta_{12}' & 1+4 \rho \beta_2(1-\beta_2)
\end{array}
\right)  \\
&=& I_2+\rho \Sigma_x' ,
\end{eqnarray}
where
\begin{equation}
\Sigma_x'=\left(
\begin{array}{cc}
4\beta_1(1-\beta_1) & 4 \theta_{12}' \\
4 \theta_{12}' & 4\beta_2(1-\beta_2)
\end{array}
\right)
\end{equation}
and where $\beta_l\,(l=1, 2)$ and $\theta_{12}'$ are defined by
\begin{eqnarray}
\beta_l &=& P(v_k^{(l)}=1),~l=1, 2 \\
\theta_{12}' &=& P(v_k^{(1)}=1, v_k^{(2)}=1)-\beta_1\beta_2 .
\end{eqnarray}
\par
Consider the derived matrix $\Sigma_x'$. 
\par
{\it Remark 1:} We have seen in the previous section that the signal equation (i.e., $F_j$'s) is needed for the calculation of $\Sigma_{k-L \vert k}$. The same is true of $\Sigma_x'$. As stated above, $\Sigma_x'$ is derived based on $p_{\eta}(x, y)$ whose coefficients are $\beta_{ij}=P(v_k^{(1)}=i, v_k^{(2)}=j)$, where $\mbox{\boldmath $v$}_k=(v_k^{(1)}, v_k^{(2)})$ is the encoded block for the main decoder. This implies that coding is equally related to the calculation of $\Sigma_x'$. In our case, convolutional coding corresponds to the signal process. Hence the signal equation is used for $\Sigma_x'$ as well. In fact, $\Sigma_x'$ consists of $\beta_l=P(v_k^{(l)}=1)~(l=1, 2)$ and $\theta_{12}'=P(v_k^{(1)}=1, v_k^{(2)}=1)-\beta_1\beta_2$.
\par
In~\cite{taji 19}, we showed the following:
\begin{itemize}
\item [1)] $\mbox{\boldmath $\eta$}_{k-L}^h$ depends not only on errors $\{\mbox{\boldmath $e$}_s,~s \leq k-L\}$ but also on errors $\{\mbox{\boldmath $e$}_s,~k-L<s \leq k\}$.
\item [2)] $\{\mbox{\boldmath $\eta$}_s^h,~s \leq k-L\}$ and $\{\mbox{\boldmath $z$}_s^h,~s \leq k-L\}$ generate the same syndrome sequence $\{\zeta_s,~s \leq k-L\}$.
\end{itemize}
Then we concluded that $\mbox{\boldmath $\eta$}_{k-L}^h$ has some relation to smoothing in linear estimation theory. Also, we showed that $\mbox{\boldmath $\eta$}_{k-L}^h$ has an expression
\begin{equation}
\mbox{\boldmath $\eta$}_{k-L}^h=(i_{k-L}-\hat i_{k-L \vert k})G(D)+\mbox{\boldmath $e$}_{k-L} .
\end{equation}
Since $i_{k-L}-\hat i_{k-L \vert k}$ is contained in $\mbox{\boldmath $\eta$}_{k-L}^h$, it is natural to think that $\Sigma_x'$ has a connection with the covariance matrix $\Sigma_{k-L \vert k}$ of the smoothing error $\mbox{\boldmath $x$}_{k-L}-\mbox{\boldmath $\hat x$}_{k-L \vert k}$. However, it is difficult to consider that $\Sigma_x'$ corresponds directly to $\Sigma_{k-L \vert k}$.
\par
On the other hand, $\Sigma_x'$ can be compared with $\Sigma_x$, where $\Sigma_x$ is defined by the relation $\Sigma_r=I_2+\rho \Sigma_x$. Since $\Sigma_r$ is the covariance matrix of $\mbox{\boldmath $r$}_k$, whereas $\Sigma_{\eta}$ is the covariance matrix of $\mbox{\boldmath $\eta$}_{k-L}$, we can expect that the inequality $\mbox{tr}(\Sigma_x') \leq \mbox{tr}(\Sigma_x)$ holds, where $\Sigma_x$ is derived by regarding the given QLI code as a general code. As an example, the values of $\mbox{tr}(\Sigma_x')$ for $C_1$ and $C_2$ are shown in Tables VII and VIII, respectively, where these codes are regarded as inherent QLI codes. 
\begin{table}[tb]
\caption{$\frac{1}{2}\mbox{tr}(\Sigma_x')$ for $C_1$ (as a QLI code)}
\label{Table 7}
\begin{center}
\begin{tabular}{c*{5}{|c}}
$E_b/N_0~(\mbox{dB})$ & $\beta_1$ & $4\beta_1(1-\beta_1)$ & $\beta_2$ & $4\beta_2(1-\beta_2)$ & $\frac{1}{2}\mbox{tr}(\Sigma_x')$ \\
\hline
$-10$ & $0.4999$ & $1.0000$ & $0.4981$ & $1.0000$ & $1.0000$ \\
$-9$ & $0.4998$ & $1.0000$ & $0.4970$ & $1.0000$ & $1.0000$ \\
$-8$ & $0.4996$ & $1.0000$ & $0.4954$ & $0.9999$ & $1.0000$ \\
$-7$ & $0.4992$ & $1.0000$ & $0.4929$ & $0.9998$ & $0.9999$ \\
$-6$ & $0.4984$ & $1.0000$ & $0.4892$ & $0.9995$ & $0.9998$ \\
$-5$ & $0.4970$ & $1.0000$ & $0.4835$ & $0.9989$ & $0.9995$ \\
$-4$ & $0.4945$ & $0.9999$ & $0.4752$ & $0.9975$ & $0.9987$ \\
$-3$ & $0.4900$ & $0.9996$ & $0.4632$ & $0.9946$ & $0.9971$ \\
$-2$ & $0.4823$ & $0.9987$ & $0.4461$ & $0.9884$ & $0.9936$ \\
$-1$ & $0.4696$ & $0.9963$ & $0.4226$ & $0.9760$ & $0.9862$ \\
$0$ & $0.4494$ & $0.9898$ & $0.3914$ & $0.9528$ & $0.9713$ \\
$1$ & $0.4191$ & $0.9738$ & $0.3515$ & $0.9118$ & $0.9428$ \\
$2$ & $0.3766$ & $0.9391$ & $0.3033$ & $0.8452$ & $0.8922$ \\
$3$ & $0.3213$ & $0.8723$ & $0.2482$ & $0.7464$ & $0.8094$ \\
$4$ & $0.2565$ & $0.7628$ & $0.1905$ & $0.6168$ & $0.6898$ \\
$5$ & $0.1876$ & $0.6096$ & $0.1346$ & $0.4659$ & $0.5378$ \\
$6$ & $0.1231$ & $0.4318$ & $0.0858$ & $0.3138$ & $0.3728$ \\
$7$ & $0.0710$ & $0.2638$ & $0.0485$ & $0.1846$ & $0.2242$ \\
$8$ & $0.0349$ & $0.1347$ & $0.0236$ & $0.0922$ & $0.1135$ \\
$9$ & $0.0143$ & $0.0564$ & $0.0096$ & $0.0380$ & $0.0472$ \\
$10$ & $0.0047$ & $0.0187$ & $0.0031$ & $0.0124$ & $0.0156$
\end{tabular}
\end{center}
\end{table}
\begin{table}[tb]
\caption{$\frac{1}{2}\mbox{tr}(\Sigma_x')$ for $C_2$ (as a QLI code)}
\label{Table 8}
\begin{center}
\begin{tabular}{c*{5}{|c}}
$E_b/N_0~(\mbox{dB})$ & $\beta_1$ & $4\beta_1(1-\beta_1)$ & $\beta_2$ & $4\beta_2(1-\beta_2)$ & $\frac{1}{2}\mbox{tr}(\Sigma_x')$ \\
\hline
$-10$ & $0.5000$ & $1.0000$ & $0.5000$ & $1.0000$ & $1.0000$ \\
$-9$ & $0.5000$ & $1.0000$ & $0.5000$ & $1.0000$ & $1.0000$ \\
$-8$ & $0.5000$ & $1.0000$ & $0.5000$ & $1.0000$ & $1.0000$ \\
$-7$ & $0.4999$ & $1.0000$ & $0.5000$ & $1.0000$ & $1.0000$ \\
$-6$ & $0.4998$ & $1.0000$ & $0.5000$ & $1.0000$ & $1.0000$ \\
$-5$ & $0.4995$ & $1.0000$ & $0.4999$ & $1.0000$ & $1.0000$ \\
$-4$ & $0.4988$ & $1.0000$ & $0.4997$ & $1.0000$ & $1.0000$ \\
$-3$ & $0.4973$ & $1.0000$ & $0.4993$ & $1.0000$ & $1.0000$ \\
$-2$ & $0.4942$ & $0.9999$ & $0.4981$ & $1.0000$ & $1.0000$ \\
$-1$ & $0.4880$ & $0.9994$ & $0.4953$ & $0.9999$ & $0.9997$ \\
$0$ & $0.4764$ & $0.9978$ & $0.4890$ & $0.9995$ & $0.9987$ \\
$1$ & $0.4559$ & $0.9922$ & $0.4760$ & $0.9977$ & $0.9950$ \\
$2$ & $0.4226$ & $0.9760$ & $0.4515$ & $0.9906$ & $0.9833$ \\
$3$ & $0.3732$ & $0.9357$ & $0.4100$ & $0.9676$ & $0.9517$ \\
$4$ & $0.3084$ & $0.8532$ & $0.3493$ & $0.9092$ & $0.8812$ \\
$5$ & $0.2329$ & $0.7146$ & $0.2717$ & $0.7915$ & $0.7531$ \\
$6$ & $0.1570$ & $0.5294$ & $0.1878$ & $0.6101$ & $0.5698$ \\
$7$ & $0.0923$ & $0.3351$ & $0.1126$ & $0.3997$ & $0.3674$ \\
$8$ & $0.0460$ & $0.1755$ & $0.0569$ & $0.2146$ & $0.1951$ \\
$9$ & $0.0190$ & $0.0746$ & $0.0237$ & $0.0926$ & $0.0836$ \\
$10$ & $0.0062$ & $0.0246$ & $0.0078$ & $0.0310$ & $0.0278$
\end{tabular}
\end{center}
\end{table}
Comparing with the values of $\mbox{tr}(\Sigma_x)$ in Tables I and II, we observe that $\mbox{tr}(\Sigma_x') \leq \mbox{tr}(\Sigma_x)$ actually holds for both $C_1$ and $C_2$. Then the question naturally arises: Does the relation $\mbox{tr}(\Sigma_x') \leq \mbox{tr}(\Sigma_x)$ hold for every QLI code ? In the following, we discuss this subject.
\par
Note that
\begin{eqnarray}
\mbox{tr}(\Sigma_x) &=& 4\alpha_1(1-\alpha_1)+4\alpha_2(1-\alpha_2) \\
\mbox{tr}(\Sigma_x') &=& 4\beta_1(1-\beta_1)+4\beta_2(1-\beta_2) .
\end{eqnarray}
Then a comparison between $\mbox{tr}(\Sigma_x)$ and $\mbox{tr}(\Sigma_x')$ is reduced to that between $\alpha_i(1-\alpha_i)$ and $\beta_j(1-\beta_j)~(i, j=1, 2)$. We have the following.
\begin{lem}
If $\alpha_i \leq \beta_j~(i, j=1, 2)$, then $\alpha_i(1-\alpha_i) \leq \beta_j(1-\beta_j)$ holds.
\end{lem}
\begin{IEEEproof}
We have already shown that $0 \leq \alpha_i \leq \frac{1}{2}$ and $0 \leq \beta_j \leq \frac{1}{2}$ for $0 \leq \epsilon \leq \frac{1}{2}$ (~\cite[Lemma 13, Lemma 15]{taji 19}). Then it suffices to note that $f(x)=x(1-x)$ is an increasing function on the interval $0 \leq x \leq \frac{1}{2}$.
\end{IEEEproof}
\par
From the above lemma, we see that a comparison between $\mbox{tr}(\Sigma_x)$ and $\mbox{tr}(\Sigma_x')$ is further reduced to that between $\alpha_i$ and $\beta_j~(i, j=1, 2)$. Moreover, since $\alpha_l(\beta_l)=P(v_k^{(l)}=1)~(l=1, 2)$, we can compare $\alpha_i$ and $\beta_j$ by considering the encoded block $\mbox{\boldmath $v$}_k=(v_k^{(1)}, v_k^{(2)})$ for the main decoder. Note that $\mbox{\boldmath $v$}_k=\mbox{\boldmath $e$}_kG^{-1}G$ for a general code, whereas $\mbox{\boldmath $v$}_k=\mbox{\boldmath $e$}_kG_p^{-1}G$ for a QLI code. Let us set
\begin{equation}
G^{-1}G=
\left(
\begin{array}{cc}
b_{11} & b_{12} \\
b_{21} & b_{22}
\end{array}
\right) .
\end{equation}
Then we have
\begin{eqnarray}
\mbox{\boldmath $v$}_k &=& (v_k^{(1)}, v_k^{(2)}) \nonumber \\
&=& (e_k^{(1)}, e_k^{(2)})\left(
\begin{array}{cc}
b_{11} & b_{12} \\
b_{21} & b_{22}
\end{array}
\right) \nonumber \\
&=& (e_k^{(1)} b_{11}+e_k^{(2)} b_{21}, e_k^{(1)} b_{12}+e_k^{(2)} b_{22}) .
\end{eqnarray}
Here assume that $b_{1l}(b_{2l})~(l=1, 2)$ contains many $D^j$'s. Then $v_k^{(l)}=e_k^{(1)} b_{1l}+e_k^{(2)} b_{2l}~(l=1, 2)$ consists of many error terms, which results in a large value of $\alpha_l=P(v_k^{(l)}=1)~(l=1, 2)$ (cf.~\cite[Proposition 8]{taji 19}). On the other hand, if $b_{1l}(b_{2l})~(l=1, 2)$ contains a small number of $D^j$'s, then $v_k^{(l)}=e_k^{(1)} b_{1l}+e_k^{(2)} b_{2l}~(l=1, 2)$ consists of a small number of error terms, which results in a small value of $\alpha_l=P(v_k^{(l)}=1)~(l=1, 2)$. Note that the same argument applies for $G_p^{-1}G$ and hence for $\beta_l~(l=1, 2)$. Let $m_l^{\alpha}~(l=1, 2)$ be the number of $D^j$'s contained in the $l$th column of $G^{-1}G$. Similarly, let $m_l^{\beta}~(l=1, 2)$ be the number of $D^j$'s contained in the $l$th column of $G_p^{-1}G$. Then the above argument means that by evaluating $m_i^{\alpha}$ and $m_j^{\beta}$, we can compare the values of $\alpha_i$ and $\beta_j$.
\par
In order to check whether $\mbox{tr}(\Sigma_x') \leq \mbox{tr}(\Sigma_x)$ holds or not for a QLI code, we have applied the above method. Also, we have considered a QLI code with generator matrix~\cite{joha 99}
\begin{eqnarray}
G(D) &=& (g_1(D), g_1(D)+D) \\
&=& (1+Dg'(D), 1+D+Dg'(D)) ,
\end{eqnarray}
where $g'(D)=c_1D+ \cdots +c_{\nu-2}D^{\nu-2}+D^{\nu-1}$ and where $c_i=0~\mbox{or}~1~(1 \leq i \leq \nu-2)$. With respect to the above $G(D)$, we have the following.
\begin{lem}
A right inverse of $G(D)=(1+Dg'(D), 1+D+Dg'(D))$ is given by
\begin{equation}
G^{-1}(D)=\left(
\begin{array}{c}
1+g'(D) \\
g'(D)
\end{array}
\right) .
\end{equation}
\end{lem}
\begin{IEEEproof}
Consider an invariant-factor decomposition~\cite{forn 70} of $G(D)$. Applying elementary column operations (i.e., multiply by the associated matrices on the right), we have
\begin{eqnarray}
\lefteqn{\left(
\begin{array}{cc}
1+Dg'(D) & 1+D+Dg'(D) 
\end{array}
\right)} \nonumber \\
&& \times \left(
\begin{array}{cc}
1+g'(D) & 1+D+Dg'(D) \\
g'(D) & 1+Dg'(D)
\end{array}
\right) \nonumber \\
&& =\left(
\begin{array}{cc}
1 & 0 
\end{array}
\right) .
\end{eqnarray}
\end{IEEEproof}
\par
Since $G^{-1}$ has been given by the above lemma, we can evaluate $m_i^{\alpha}$ and $m_j^{\beta}~(i, j=1, 2)$. We see, however, that $m_i^{\alpha}$ and $m_j^{\beta}~(i, j=1, 2)$ are dependent on the coefficients $c_i~(1 \leq i \leq \nu-2)$ in a complicated way. Hence for the cases that $\nu=5$ and $\nu=6$, we have applied the exhaustive search. The results for $\nu=5$ and $\nu=6$ are shown in Tables IX and X, respectively.
\begin{table}[tb]
\caption{$m_l^{\alpha}$ and $m_l^{\beta}~(l=1, 2)$ ($\nu=5$)}
\label{Table 9}
\begin{center}
\begin{tabular}{c*{6}{|c}}
$c_1$ & $c_2$ & $c_3$ & $m_1^{\alpha}$ & $m_2^{\alpha}$ & $m_1^{\beta}$ & $m_2^{\beta}$ \\
\hline
$0$ & $0$ & $0$ & $6$ & $7$ & $4$ & $6$ \\
$0$ & $0$ & $1$ & $9$ & $10$ & $6$ & $8$ \\
$0$ & $1$ & $0$ & $9$ & $10$ & $6$ & $8$ \\
$0$ & $1$ & $1$ & $10$ & $11$ & $8$ & $10$ \\
$1$ & $0$ & $0$ & $11$ & $10$ & $6$ & $8$\\
$1$ & $0$ & $1$ & $10$ & $9$ & $8$ & $10$ \\
$1$ & $1$ & $0$ & $10$ & $9$ & $8$ & $10$ \\
{\boldmath $1$} & {\boldmath $1$} & {\boldmath $1$} & {\boldmath $11$} & {\boldmath $10$} & {\boldmath $10$} & {\boldmath $12$}
\end{tabular}
\end{center}
\end{table}
\begin{table}[tb]
\caption{$m_l^{\alpha}$ and $m_l^{\beta}~(l=1, 2)$ ($\nu=6$)}
\label{Table 10}
\begin{center}
\begin{tabular}{c*{7}{|c}}
$c_1$ & $c_2$ & $c_3$ & $c_4$ & $m_1^{\alpha}$ & $m_2^{\alpha}$ & $m_1^{\beta}$ & $m_2^{\beta}$ \\
\hline
$0$ & $0$ & $0$ & $0$ & $6$ & $7$ & $4$ & $6$ \\
$0$ & $0$ & $0$ & $1$ & $9$ & $10$ & $6$ & $8$ \\
$0$ & $0$ & $1$ & $0$ & $11$ & $12$ & $6$ & $8$ \\
$0$ & $0$ & $1$ & $1$ & $12$ & $13$ & $8$ & $10$ \\
$0$ & $1$ & $0$ & $0$ & $7$ & $8$ & $6$ & $8$ \\
$0$ & $1$ & $0$ & $1$ & $12$ & $13$ & $8$ & $10$ \\
$0$ & $1$ & $1$ & $0$ & $10$ & $11$ & $8$ & $10$ \\
$0$ & $1$ & $1$ & $1$ & $13$ & $14$ & $10$ & $12$ \\
$1$ & $0$ & $0$ & $0$ & $11$ & $10$ & $6$ & $8$ \\
$1$ & $0$ & $0$ & $1$ & $14$ & $13$ & $8$ & $10$ \\
$1$ & $0$ & $1$ & $0$ & $12$ & $11$ & $8$ & $10$ \\
$1$ & $0$ & $1$ & $1$ & $13$ & $12$ & $10$ & $12$ \\
{\boldmath $1$} & {\boldmath $1$} & {\boldmath $0$} & {\boldmath $0$} & {\boldmath $8$} & {\boldmath $7$} & {\boldmath $8$} & {\boldmath $10$} \\
$1$ & $1$ & $0$ & $1$ & $13$ & $12$ & $10$ & $12$ \\
{\boldmath $1$} & {\boldmath $1$} & {\boldmath $1$} & {\boldmath $0$} & {\boldmath $11$} & {\boldmath $10$} & {\boldmath $10$} & {\boldmath $12$} \\
$1$ & $1$ & $1$ & $1$ & $14$ & $13$ & $12$ & $14$
\end{tabular}
\end{center}
\end{table}
From Tables IX and X, we observe that there are cases (bold-faced) where $\mbox{tr}(\Sigma_x') \leq \mbox{tr}(\Sigma_x)$ does not hold:
\par
1) $\nu=5$, $g'(D)=D+D^2+D^3+D^4$:
\begin{eqnarray}
G(D) &=& (1+D^2+D^3+D^4+D^5, 1+D+D^2+D^3+D^4+D^5) \\
G^{-1}(D) &=& \left(
\begin{array}{c}
1+D+D^2+D^3+D^4 \\
D+D^2+D^3+D^4
\end{array}
\right) .
\end{eqnarray}
We have
\begin{eqnarray}
G^{-1}G &=& \left(
\begin{array}{cc}
1+D+D^3+D^7+D^9 & 1+D^2+D^4+D^5+D^7+D^9 \\
D+D^2+D^4+D^5+D^7+D^9 & D+D^3+D^7+D^9
\end{array}
\right) \\
G_p^{-1}G &=& \left(
\begin{array}{cc}
1+D^2+D^3+D^4+D^5 & 1+D+D^2+D^3+D^4+D^5 \\
1+D^2+D^3+D^4+D^5 & 1+D+D^2+D^3+D^4+D^5
\end{array}
\right) .
\end{eqnarray}
In this case, $m_1^{\alpha}=11$, $m_2^{\alpha}=10$, $m_1^{\beta}=10$, and $m_2^{\beta}=12$. Then it follows from
\begin{equation}
m_1^{\alpha}<m_2^{\beta},~~m_2^{\alpha}=m_1^{\beta}
\end{equation}
that $\mbox{tr}(\Sigma_x) < \mbox{tr}(\Sigma_x')$.
\par
2) $\nu=6$, $g'(D)=D+D^2+D^5$:
\begin{eqnarray}
G(D) &=& (1+D^2+D^3+D^6, 1+D+D^2+D^3+D^6) \\
G^{-1}(D) &=& \left(
\begin{array}{c}
1+D+D^2+D^5 \\
D+D^2+D^5
\end{array}
\right) .
\end{eqnarray}
We have
\begin{eqnarray}
G^{-1}G &=& \left(
\begin{array}{cc}
1+D+D^6+D^{11} & 1+D^2+D^3+D^{11} \\
D+D^2+D^3+D^{11} & D+D^6+D^{11}
\end{array}
\right) \\
G_p^{-1}G &=& \left(
\begin{array}{cc}
1+D^2+D^3+D^6 & 1+D+D^2+D^3+D^6 \\
1+D^2+D^3+D^6 & 1+D+D^2+D^3+D^6
\end{array}
\right) .
\end{eqnarray}
In this case, $m_1^{\alpha}=8$, $m_2^{\alpha}=7$, $m_1^{\beta}=8$, and $m_2^{\beta}=10$. Then it follows from
\begin{equation}
m_1^{\alpha}=m_1^{\beta},~~m_2^{\alpha}<m_2^{\beta}
\end{equation}
that $\mbox{tr}(\Sigma_x) < \mbox{tr}(\Sigma_x')$.
\par
3) $\nu=6$, $g'(D)=D+D^2+D^3+D^5$:
\begin{eqnarray}
G(D) &=& (1+D^2+D^3+D^4+D^6, 1+D+D^2+D^3+D^4+D^6) \\
G^{-1}(D) &=& \left(
\begin{array}{c}
1+D+D^2+D^3+D^5 \\
D+D^2+D^3+D^5
\end{array}
\right) .
\end{eqnarray}
We have
\begin{eqnarray}
G^{-1}G &=& \left(
\begin{array}{cc}
1+D+D^3+D^4+D^6+D^7+D^{11} & 1+D^2+D^7+D^{11} \\
D+D^2+D^7+D^{11} & D+D^3+D^4+D^6+D^7+D^{11}
\end{array}
\right) \\
G_p^{-1}G &=& \left(
\begin{array}{cc}
1+D^2+D^3+D^4+D^6 & 1+D+D^2+D^3+D^4+D^6 \\
1+D^2+D^3+D^4+D^6 & 1+D+D^2+D^3+D^4+D^6
\end{array}
\right) .
\end{eqnarray}
In this case, $m_1^{\alpha}=11$, $m_2^{\alpha}=10$, $m_1^{\beta}=10$, and $m_2^{\beta}=12$. Then it follows from
\begin{equation}
m_1^{\alpha}<m_2^{\beta},~~m_2^{\alpha}=m_1^{\beta}
\end{equation}
that $\mbox{tr}(\Sigma_x) < \mbox{tr}(\Sigma_x')$.
\par
As shown above, we have actually found the cases where $\mbox{tr}(\Sigma_x') \leq \mbox{tr}(\Sigma_x)$ does not hold. Note that if $\Sigma_{k \vert k}$ is independent of $k$ (this is our case), then we have
\begin{equation}
\Sigma_{k-L \vert k}\leq \Sigma_{k-L \vert k-L}=\Sigma_{k \vert k}\leq \Sigma_{k \vert k-1}~(=\Sigma_x) .
\end{equation}
\par
{\it Remark 2:} The notation $\Sigma_{k \vert l}$ is used in connection with linear estimation theory, whereas $\Sigma_x$ is used in coding theory context. Hence an expression $\Sigma_{k \vert k-1}=\Sigma_x$ may be inappropriate. However, we have shown that $\Sigma_x$ corresponds to $M_k=\Sigma_{k \vert k-1}$. Then the above equality is interpreted in this way.
\par
Hence the above result implies that $\mbox{tr}(\Sigma_x')$ is not equal to $\mbox{tr}(\Sigma_{k-L \vert k})$. Moreover, consider the QLI code $C_2$. It follows from the generator matrix $G_2$ that
\begin{equation}
g'(D)=D^2+D^4+D^5 ,
\end{equation}
i.e., $c_1=0,~c_2=1,~c_3=0,~c_4=1$. Then from Table X, we have $\mbox{tr}(\Sigma_x') \leq \mbox{tr}(\Sigma_x)$. On the other hand, comparing the values of $\frac{1}{2}\mbox{tr}(\Sigma_c)$ in Table VI and those of $\frac{1}{2}\mbox{tr}(\Sigma_x')$ in Table VIII, we observe that $\mbox{tr}(\Sigma_c)<\mbox{tr}(\Sigma_x')$ holds. Since $\Sigma_{k-L \vert k}\leq \Sigma_{k-L \vert k-L}=\Sigma_{k \vert k}~(=\Sigma_c)$, this also implies that $\mbox{tr}(\Sigma_x')$ is not equal to $\mbox{tr}(\Sigma_{k-L \vert k})$. Nevertheless, the inequality $\mbox{tr}(\Sigma_x') \leq \mbox{tr}(\Sigma_x)$ holds for almost all QLI codes considered here. Then noting the difference between two equations
\begin{eqnarray}
\mbox{\boldmath $r$}_{k-L}^h &=& (i_{k-L}-\hat i_{k-L \vert k-L})G(D)+\mbox{\boldmath $e$}_{k-L} \nonumber \\
\mbox{\boldmath $\eta$}_{k-L}^h &=& (i_{k-L}-\hat i_{k-L \vert k})G(D)+\mbox{\boldmath $e$}_{k-L} , \nonumber
\end{eqnarray}
we can say that $\mbox{tr}(\Sigma_x')$ is the mean-square error associated with the linear smoothed estimate in a broad sense.

\section{Conclusion}
We have noticed that convolutional coding/Viterbi decoding has the structure of the Kalman filter and hence the innovations method can be applied. In the discrete-time case, if the covariance matrix of the innovation (i.e., the soft-decision input to the main decoder in an SST Viterbi decoder) is calculated, then a matrix corresponding to $M_k$ (covariance matrix of the one-step prediction error) in the Kalman filter can be obtained. Furthermore, a matrix corresponding to $P_k$ (covariance matrix of the filtering error) is derived from that matrix by applying the formula in the Kalman filter. As a result, the input-output mutual information is given using these matrices and the LMMSE is given as the trace of the latter matrix. Thus our goal comes down to the calculation of the covariance matrix of the innovation. In this paper, we have shown that this scenario is really possible. In~\cite{taji 19}, we showed that the hard-decision input to the main decoder in an SST Viterbi decoder is regarded as the innovation of the received data. When the paper was written, however, we did not know how the extracted innovation is used in coding theory. In this paper, we have shown that as in the case of linear filtering theory, the innovation plays an essential role in the derivation of the mutual information and the LMMSE for Viterbi decoding of convolutional codes, which in turn justifies the notion of innovations introduced in our former paper~\cite{taji 19}.


%

\appendices
\section{Proof of Lemma 4}
Let $\rho>0$. By applying the formula related to differentiation of a determinant, we have
\begin{eqnarray}
\frac{d}{d \rho}\log \vert I_{n_0}+\rho M_j \vert &=& \frac{1}{\vert I_{n_0}+\rho M_j \vert}\frac{d}{d \rho} \vert I_{n_0}+\rho M_j \vert \nonumber \\
&=& \mbox{tr} \left((I_{n_0}+\rho M_j)^{-1} \frac{d}{d \rho}(I_{n_0}+\rho M_j) \right) .
\end{eqnarray}
Note that $\frac{d}{d \rho}(I_{n_0}+\rho M_j)=\frac{d R_j}{d \rho}$. It follows from $M_j=P_jR_j$ that
\begin{displaymath}
\frac{d M_j}{d \rho}=\frac{d P_j}{d \rho}R_j+P_j \frac{d R_j}{d \rho} .
\end{displaymath}
Then we have
\begin{eqnarray}
\frac{d R_j}{d \rho} &=& \frac{d}{d \rho}(I_{n_0}+\rho M_j) \nonumber \\
&=& M_j+\rho \frac{d M_j}{d \rho} \nonumber \\
&=& P_jR_j+\rho \frac{d P_j}{d \rho}R_j+\rho P_j \frac{d R_j}{d \rho} . \nonumber
\end{eqnarray}
This is modified as
\begin{equation}
(I_{n_0}-\rho P_j)\frac{d R_j}{d \rho}=\left(P_j+\rho \frac{d P_j}{d \rho}\right)R_j .
\end{equation}
\par
Here consider whether $I_{n_0}-\rho P_j$ has the inverse. Since $\lambda_{max}<\frac{1}{\rho}$ is assumed, the series
\begin{equation}
I_{n_0}+\rho P_j+\rho^2 P_j^2+ \cdots
\end{equation}
is convergent and $(I_{n_0}-\rho P_j)^{-1}$ is given by
\begin{equation}
(I_{n_0}-\rho P_j)^{-1}=I_{n_0}+\rho P_j+\rho^2 P_j^2+ \cdots .
\end{equation}
Hence we obtain
\begin{equation}
\frac{d R_j}{d \rho}=(I_{n_0}-\rho P_j)^{-1}\left(P_j+\rho \frac{d P_j}{d \rho}\right)R_j .
\end{equation}
\par
Let us go back to the evaluation of $\frac{d}{d \rho}\log \vert I_{n_0}+\rho M_j \vert$. We have
\begin{eqnarray}
\frac{d}{d \rho}\log \vert I_{n_0}+\rho M_j \vert &=& \mbox{tr}\left(R_j^{-1}\frac{d R_j}{d \rho} \right) \nonumber \\
&=& \mbox{tr}\left(\frac{d R_j}{d \rho}R_j^{-1} \right) \nonumber \\
&=& \mbox{tr}\left((I_{n_0}-\rho P_j)^{-1}\left(P_j+\rho \frac{d P_j}{d \rho}\right)R_j R_j^{-1} \right) \nonumber \\
&=& \mbox{tr}\left((I_{n_0}-\rho P_j)^{-1}\left(P_j+\rho \frac{d P_j}{d \rho}\right) \right) .
\end{eqnarray}
In the above modifications, we have used the fact that $\mbox{tr}(AB)=\mbox{tr}(BA)$ holds for two square matrices $A$ and $B$~\cite{horn 13}.
\par
Now evaluate the right-hand side. Since the trace of a matrix is invariant under similarity transformations~\cite{horn 13}, we have
\begin{eqnarray}
\lefteqn{\mbox{tr}\left((I_{n_0}-\rho P_j)^{-1}\left(P_j+\rho \frac{d P_j}{d \rho}\right) \right)} \nonumber \\
&& =\mbox{tr}\left(Q^{-1}(I_{n_0}-\rho P_j)^{-1}\left(P_j+\rho \frac{d P_j}{d \rho}\right)Q \right) \nonumber \\
&& =\mbox{tr}\left(Q^{-1}(I_{n_0}-\rho P_j)^{-1}Q \cdot Q^{-1}\left(P_j+\rho \frac{d P_j}{d \rho}\right)Q \right) .
\end{eqnarray}
First, note the term $Q^{-1}(I_{n_0}-\rho P_j)^{-1}Q$. Since the series $I_{n_0}+\rho P_j+\rho^2 P_j^2+ \cdots$ is convergent, we see that the series
\begin{eqnarray}
\lefteqn{I_{n_0}+\rho (Q^{-1}P_jQ)+\rho^2(Q^{-1}P_jQ)^2+\cdots} \nonumber \\
&& =I_{n_0}+\rho \Gamma+\rho^2 \Gamma^2+\cdots
\end{eqnarray}
is also convergent and the sum is equal to
\begin{equation}
Q^{-1}(I_{n_0}+\rho P_j+\rho^2 P_j^2+ \cdots)Q .
\end{equation}
That is, we have
\begin{eqnarray}
\lefteqn{Q^{-1}(I_{n_0}-\rho P_j)^{-1}Q} \nonumber \\
&& =I_{n_0}+\rho \Gamma+\rho^2 \Gamma^2+\cdots \nonumber \\
&& \stackrel{\triangle}{=}\tilde \Gamma .
\end{eqnarray}
Note that $\tilde \Gamma$ is represented using $\lambda_l$. Let
\begin{equation}
\Gamma= \left(
\begin{array}{cccc}
\lambda_1 & 0 & \cdots & 0 \\
0 & \lambda_2 & \cdots & 0 \\
& & \ddots & \\
0 & \cdots & 0 & \lambda_{n_0}
\end{array}
\right) .
\end{equation}
We have
\begin{eqnarray}
\tilde \Gamma &=& \left(
\begin{array}{cccc}
1+\rho \lambda_1+\rho^2 \lambda_1^2+\cdots & 0 & \cdots & 0 \\
0 & 1+\rho \lambda_2+\rho^2 \lambda_2^2+\cdots & \cdots & 0 \\
& & \ddots & \\
0 & \cdots & 0 & 1+\rho \lambda_{n_0}+\rho^2 \lambda_{n_0}^2+\cdots
\end{array}
\right) \nonumber \\
&=& \left(
\begin{array}{cccc}
\frac{1}{1-\rho \lambda_1} & 0 & \cdots & 0 \\
0 & \frac{1}{1-\rho \lambda_2} & \cdots & 0 \\
& & \ddots & \\
0 & \cdots & 0 & \frac{1}{1-\rho \lambda_{n_0}}
\end{array}
\right) \\
&=& \left(
\begin{array}{cccc}
q_1 & 0 & \cdots & 0 \\
0 & q_2 & \cdots & 0 \\
& & \ddots & \\
0 & \cdots & 0 & q_{n_0}
\end{array}
\right) ,
\end{eqnarray}
where
\begin{equation}
q_l\stackrel{\triangle}{=}\frac{1}{1-\rho \lambda_l}>1~(1 \leq l \leq n_0) .
\end{equation}
\par
Hence the right-hand side of
\begin{displaymath}
\frac{d}{d \rho}\log \vert I_{n_0}+\rho M_j \vert=\mbox{tr}\left((I_{n_0}-\rho P_j)^{-1}\left(P_j+\rho \frac{d P_j}{d \rho}\right) \right)
\end{displaymath}
is modified as follows:
\begin{eqnarray}
\lefteqn{\mbox{tr}\left(Q^{-1}(I_{n_0}-\rho P_j)^{-1}\left(P_j+\rho \frac{d P_j}{d \rho}\right)Q \right)} \nonumber \\
&& =\mbox{tr}\left(Q^{-1}(I_{n_0}-\rho P_j)^{-1}Q \cdot Q^{-1}\left(P_j+\rho \frac{d P_j}{d \rho}\right)Q \right) \nonumber \\
&& =\mbox{tr}\left(\tilde \Gamma Q^{-1}\left(P_j+\rho \frac{d P_j}{d \rho}\right)Q \right) \nonumber \\
&& =\mbox{tr}\left(\tilde \Gamma \left(Q^{-1}P_jQ+\rho \,Q^{-1}\frac{d P_j}{d \rho}Q\right) \right) \nonumber \\
&& =\mbox{tr}(\tilde \Gamma(\Gamma+\rho \Phi)) .
\end{eqnarray}
Since $\tilde \Gamma$ is diagonal, $\mbox{tr}(\tilde \Gamma(\Gamma+\rho \Phi))$ is expressed as
\begin{eqnarray}
\lefteqn{\mbox{tr}(\tilde \Gamma(\Gamma+\rho \Phi))} \nonumber \\
&& =q_1(\Gamma+\rho \Phi)_{11}+ \cdots +q_{n_0}(\Gamma+\rho \Phi)_{n_0n_0} .
\end{eqnarray}
If $(\Gamma+\rho \Phi)_{ll}=\frac{d}{d \rho}(\rho \lambda_l)>0~(1 \leq l \leq n_0)$ holds, then noting that $q_l=\frac{1}{1-\rho \lambda_l}>1~(1 \leq l \leq n_0)$, we have
\begin{eqnarray}
\lefteqn{\mbox{tr}(\tilde \Gamma(\Gamma+\rho \Phi))} \nonumber \\
&& >(\Gamma+\rho \Phi)_{11}+ \cdots +(\Gamma+\rho \Phi)_{n_0n_0} \nonumber \\
&& =\mbox{tr}(\Gamma+\rho \Phi) \nonumber \\
&& =\mbox{tr}\left(Q^{-1}P_jQ+\rho Q^{-1}\frac{d P_j}{d \rho}Q \right) \nonumber \\
&& =\mbox{tr}(P_j)+\rho\,\mbox{tr}\left(\frac{dP_j}{d \rho}\right) \nonumber \\
&& =\mbox{tr}(P_j)+\rho\,\frac{d}{d \rho}\mbox{tr}(P_j) \nonumber \\
&& =\frac{d}{d \rho}(\rho\,\mbox{tr}(P_j)) .
\end{eqnarray}
In the above, we have used the fact that differentiation with respect to $\rho$ and the trace operation are commutable. Then we finally have
\begin{equation}
\frac{d}{d \rho}\log \vert I_{n_0}+\rho M_j \vert >\frac{d}{d \rho}(\rho\,\mbox{tr}(P_j)) .
\end{equation}
\par
Now suppose that $\frac{d}{d \gamma}(\gamma \lambda_l)>0~(1 \leq l \leq n_0)$ holds for $\gamma \in (0,~\rho]$. Then by integrating the inequality
\begin{equation}
\frac{d}{d \gamma}\log \vert I_{n_0}+\gamma M_j \vert >\frac{d}{d \gamma}(\gamma\,\mbox{tr}(P_j))
\end{equation}
over the interval $(0, ~\rho]$, we obtain
\begin{equation}
\log \vert I_{n_0}+\rho M_j \vert>\rho\,\mbox{tr}(P_j) .
\end{equation}


\section{Proof of Proposition 10}
Let $\Sigma_r=(\gamma_{ij})$. $\gamma_{ij}~(1 \leq i, j \leq n_0)$ is defined by
\begin{eqnarray}
\gamma_{ij} &=& \int_{-\infty}^{\infty}\cdots \int_{-\infty}^{\infty}(x_i-m_i)(x_j-m_j)p_r(x_1, \cdots, x_{n_0})dx_1 \cdots dx_{n_0} \\
&=& \int_{-\infty}^{\infty}\cdots \int_{-\infty}^{\infty}x_ix_jp_r(x_1, \cdots, x_{n_0})dx_1 \cdots dx_{n_0}-m_i \cdot m_j ,
\end{eqnarray}
where $m_i$ is given by
\begin{equation}
m_i=\int_{-\infty}^{\infty}\cdots \int_{-\infty}^{\infty}x_i p_r(x_1, \cdots, x_{n_0})dx_1 \cdots dx_{n_0} .
\end{equation}
Note that since $p_r(x_1, \cdots, x_{n_0})$ has a particular form, $\gamma_{ij}$ is reduced to
\begin{eqnarray}
\gamma_{ij} &=& \int_{-\infty}^{\infty}\int_{-\infty}^{\infty}(x_i-m_i)(x_j-m_j)p_r(x_i, x_j)dx_idx_j \\
&=& \int_{-\infty}^{\infty}\int_{-\infty}^{\infty}x_ix_jp_r(x_i, x_j)dx_idx_j-m_i \cdot m_j ,
\end{eqnarray}
where $p_r(x_i, x_j)$ is defined by
\begin{eqnarray}
p_r(x_i, x_j) &=& \tilde \alpha_{00}q(x_i-c)q(x_j-c)+\tilde \alpha_{01}q(x_i-c)q(x_j+c) \nonumber \\
&& +\tilde \alpha_{10}q(x_i+c)q(x_j-c)+\tilde \alpha_{11}q(x_i+c)q(x_j+c) .
\end{eqnarray}
\par
First, suppose that $i \neq j$. We have
\begin{eqnarray}
\lefteqn{\int_{-\infty}^{\infty}\int_{-\infty}^{\infty}x_ix_jp_r(x_i, x_j)dx_idx_j} \nonumber \\
&& =\tilde \alpha_{00}\int_{-\infty}^{\infty}\int_{-\infty}^{\infty}x_ix_jq(x_i-c)q(x_j-c)dx_idx_j \nonumber \\
&& +\tilde \alpha_{01}\int_{-\infty}^{\infty}\int_{-\infty}^{\infty}x_ix_jq(x_i-c)q(x_j+c)dx_idx_j \nonumber \\
&& +\tilde \alpha_{10}\int_{-\infty}^{\infty}\int_{-\infty}^{\infty}x_ix_jq(x_i+c)q(x_j-c)dx_idx_j \nonumber \\
&& +\tilde \alpha_{11}\int_{-\infty}^{\infty}\int_{-\infty}^{\infty}x_ix_jq(x_i+c)q(x_j+c)dx_idx_j \nonumber \\
&& = \tilde \alpha_{00}c^2+\tilde \alpha_{01}(-c^2)+\tilde \alpha_{10}(-c^2)+\tilde \alpha_{11}c^2 \nonumber \\
&& = (\tilde \alpha_{00}+\tilde \alpha_{11})c^2-(\tilde \alpha_{01}+\tilde \alpha_{10})c^2 .
\end{eqnarray}
It follows from Lemma 5 that
\begin{eqnarray}
\tilde \alpha_{00}+\tilde \alpha_{11} &=& 1-\alpha_i-\alpha_j+2u \\
\tilde \alpha_{01}+\tilde \alpha_{10} &=& \alpha_i+\alpha_j-2u .
\end{eqnarray}
Then we have
\begin{eqnarray}
\lefteqn{\int_{-\infty}^{\infty}\int_{-\infty}^{\infty}x_ix_jp_r(x_i, x_j)dx_idx_j} \nonumber \\
&& =(1-\alpha_i-\alpha_j+2u)c^2-(\alpha_i+\alpha_j-2u)c^2 \nonumber \\
&& =(1-2\alpha_i-2\alpha_j+4u)c^2 .
\end{eqnarray}
\par
On the other hand, as in the case of $\gamma_{ij}$, $m_i$ is reduced to
\begin{equation}
m_i=\int_{-\infty}^{\infty}x_i p_r(x_i)dx_i ,
\end{equation}
where $p_r(x_i)$ is defined by
\begin{equation}
p_r(x_i)=(1-\alpha_i)q(x_i-c)+\alpha_iq(x_i+c) .
\end{equation}
We have
\begin{eqnarray}
m_i &=& (1-\alpha_i)\int_{-\infty}^{\infty}x_iq(x_i-c)dx_i+\alpha_i\int_{-\infty}^{\infty}x_iq(x_i+c)dx_i \nonumber \\
&=& (1-\alpha_i)c+\alpha_i(-c) \nonumber \\
&=& (1-2\alpha_i)c .
\end{eqnarray}
\par
Hence we obtain
\begin{eqnarray}
\gamma_{ij} &=& \int_{-\infty}^{\infty}\int_{-\infty}^{\infty}x_ix_jp_r(x_i, x_j)dx_idx_j-m_i \cdot m_j \nonumber \\
&=& (1-2\alpha_i-2\alpha_j+4u)c^2-(1-2\alpha_i)(1-2\alpha_j)c^2 \nonumber \\
&=& 4c^2(u-\alpha_i\alpha_j) .
\end{eqnarray}
In the above, since $u=\tilde \alpha_{11}=P(v_k^{(i)}=1, v_k^{(j)}=1)$, we can write
\begin{eqnarray}
u-\alpha_i\alpha_j &=& P(v_k^{(i)}=1, v_k^{(j)}=1)-P(v_k^{(i)}=1)P(v_k^{(j)}=1) \nonumber \\
&=& \theta_{ij} .
\end{eqnarray}
Thus we finally have
\begin{eqnarray}
\gamma_{ij} &=& 4c^2(u-\alpha_i\alpha_j) \nonumber \\
&=& 4c^2 \theta_{ij} \nonumber \\
&=& 4 \rho \theta_{ij} .
\end{eqnarray}
\par
Next, suppose that $i=j$. $\gamma_{ii}~(1 \leq i \leq n_0)$ is given by
\begin{equation}
\gamma_{ii}=\int_{-\infty}^{\infty}x_i^2p_r(x_i)dx_i-(m_i)^2
\end{equation}
We have
\begin{eqnarray}
\lefteqn{\int_{-\infty}^{\infty}x_i^2p_r(x_i)dx_i} \nonumber \\
&&  =(1-\alpha_i)\int_{-\infty}^{\infty}x_i^2q(x_i-c)dx_i+\alpha_i\int_{-\infty}^{\infty}x_i^2q(x_i+c)dx_i \nonumber \\
&& = (1-\alpha_i)(1+c^2)+\alpha_i(1+c^2) \nonumber \\
&& =1+c^2 .
\end{eqnarray}
Hence we obtain
\begin{eqnarray}
\gamma_{ii} &= & \int_{-\infty}^{\infty}x_i^2p_r(x_i)dx_i-(m_i)^2 \nonumber \\
&=& (1+c^2)-(1-2\alpha_i)^2c^2 \nonumber \\
&=& 1+4c^2\alpha_i(1-\alpha_i) \nonumber \\
&=& 1+4 \rho \alpha_i(1-\alpha_i) .
\end{eqnarray}


\section{Proof of Proposition 11}
The $(i, j)~(i \neq j)$ entry of the covariance matrix of $\mbox{\boldmath $v$}_k$ is given by
\begin{eqnarray}
\lefteqn{E[(v_k^{(i)}-E[v_k^{(i)}])(v_k^{(j)}-E[v_k^{(j)}])]} \nonumber \\
&& =E[v_k^{(i)}v_k^{(j)}]-E[v_k^{(i)}]E[v_k^{(j)}] .
\end{eqnarray}
Here, we have
\begin{eqnarray}
E[v_k^{(i)}v_k^{(j)}] &=& P(v_k^{(i)}=1, v_k^{(j)}=1) \times (1 \times 1) \nonumber \\
&=& P(v_k^{(i)}=1, v_k^{(j)}=1) .
\end{eqnarray}
Also, we have
\begin{eqnarray}
E[v_k^{(i)}] &=& P(v_k^{(i)}=1)\times 1 \nonumber \\
&=& P(v_k^{(i)}=1) \nonumber \\
&=& \alpha_i .
\end{eqnarray}
Hence we obtain
\begin{eqnarray}
\lefteqn{E[(v_k^{(i)}-E[v_k^{(i)}])(v_k^{(j)}-E[v_k^{(j)}])]} \nonumber \\
&& =E[v_k^{(i)}v_k^{(j)}]-E[v_k^{(i)}]E[v_k^{(j)}] \nonumber \\
&& =P(v_k^{(i)}=1, v_k^{(j)}=1)-\alpha_i \times \alpha_j \nonumber \\
&& =\theta_{ij} .
\end{eqnarray}
\par
On the other hand, the $(i, i)$ entry is given by
\begin{eqnarray}
\lefteqn{E[(v_k^{(i)}-E[v_k^{(i)}])^2]} \nonumber \\
&& =E[(v_k^{(i)})^2]-(E[v_k^{(i)}])^2 .
\end{eqnarray}
Here, we have
\begin{eqnarray}
E[(v_k^{(i)})^2] &=& P(v_k^{(i)}=1)\times 1^2 \nonumber \\
&=& P(v_k^{(i)}=1) \nonumber \\
&=& \alpha_i .
\end{eqnarray}
Hence we obtain
\begin{eqnarray}
\lefteqn{E[(v_k^{(i)}-E[v_k^{(i)}])^2]} \nonumber \\
&& =E[(v_k^{(i)})^2]-(E[v_k^{(i)}])^2  \nonumber \\
&& =\alpha_i-(\alpha_i)^2 \nonumber \\
&& =\alpha_i(1-\alpha_i) .
\end{eqnarray}


\section{Proof of Proposition 12}
The $(i, j)~(i \neq j)$ entry of the covariance matrix of $\mbox{\boldmath $\tilde x$}_k$ is given by
\begin{eqnarray}
\lefteqn{E[(\tilde x_k^{(i)}-E[\tilde x_k^{(i)}])(\tilde x_k^{(j)}-E[\tilde x_k^{(j)}])]} \nonumber \\
&& =E[\tilde x_k^{(i)}\tilde x_k^{(j)}]-E[\tilde x_k^{(i)}]E[\tilde x_k^{(j)}] .
\end{eqnarray}
Note that
\begin{eqnarray}
\tilde x_k^{(i)}\tilde x_k^{(j)} &=& (1-2v_k^{(i)})(1-2v_k^{(j)}) \nonumber \\
&=& 1-2v_k^{(i)}-2v_k^{(j)}+4v_k^{(i)}v_k^{(j)} .
\end{eqnarray}
Then we have
\begin{eqnarray}
E[\tilde x_k^{(i)}\tilde x_k^{(j)}] &=& 1-2E[v_k^{(i)}]-2E[v_k^{(j)}]+4E[v_k^{(i)}v_k^{(j)}] \nonumber \\
&=& 1-2\alpha_i-2\alpha_j+4P(v_k^{(i)}=1, v_k^{(j)}=1) \nonumber \\
&=& 1-2\alpha_i-2\alpha_j+4\tilde \alpha_{11} .
\end{eqnarray}
Also, we have
\begin{eqnarray}
E[\tilde x_k^{(i)}] &=& 1-2E[v_k^{(i)}] \nonumber \\
&=& 1-2\alpha_i .
\end{eqnarray}
Hence we obtain
\begin{eqnarray}
\lefteqn{E[(\tilde x_k^{(i)}-E[\tilde x_k^{(i)}])(\tilde x_k^{(j)}-E[\tilde x_k^{(j)}])]} \nonumber \\
&& =E[\tilde x_k^{(i)}\tilde x_k^{(j)}]-E[\tilde x_k^{(i)}]E[\tilde x_k^{(j)}] \nonumber \\
&& =(1-2\alpha_i-2\alpha_j+4\tilde \alpha_{11})-(1-2\alpha_i)(1-2\alpha_j) \nonumber \\
&& =4(\tilde \alpha_{11}-\alpha_i \alpha_j) \nonumber \\
&& =4\theta_{ij} .
\end{eqnarray}
\par
On the other hand, the $(i, i)$ entry is given by
\begin{eqnarray}
\lefteqn{E[(\tilde x_k^{(i)}-E[\tilde x_k^{(i)}])^2]} \nonumber \\
&& =E[(\tilde x_k^{(i)})^2]-(E[\tilde x_k^{(i)}])^2 .
\end{eqnarray}
Since
\begin{equation}
(\tilde x_k^{(i)})^2=1-4v_k^{(i)}+4(v_k^{(i)})^2 ,
\end{equation}
we have
\begin{eqnarray}
E[(\tilde x_k^{(i)})^2] &=& 1-4E[v_k^{(i)}]+4E[(v_k^{(i)})^2] \nonumber \\
&=& 1-4\alpha_i+4\alpha_i \nonumber \\
&=& 1 .
\end{eqnarray}
Hence we obtain
\begin{eqnarray}
\lefteqn{E[(\tilde x_k^{(i)}-E[\tilde x_k^{(i)}])^2]} \nonumber \\
&& =E[(\tilde x_k^{(i)})^2]-(E[\tilde x_k^{(i)}])^2 \nonumber \\
&& =1-(1-2\alpha_i)^2 \nonumber \\
&& =4\alpha_i(1-\alpha_i) .
\end{eqnarray}


\section{Proof of Proposition 13}
From the formula in Kalman filter, we have
\begin{equation}
\Sigma_c=\Sigma_x-\rho \Sigma_x (I_2+\rho \Sigma_x)^{-1} \Sigma_x ,
\end{equation}
where
\begin{equation}
I_2+\rho \Sigma_x=\left(
\begin{array}{cc}
1+\rho \sigma_1^2 & \rho \sigma_{12} \\
\rho \sigma_{12} & 1+\rho \sigma_2^2
\end{array}
\right)
\end{equation}
is equal to $\Sigma_r$. $\Sigma_r^{-1}$ is given by
\begin{eqnarray}
\Sigma_r^{-1} &=& \frac{1}{\Delta_r}\left(
\begin{array}{cc}
1+\rho \sigma_2^2 & -\rho \sigma_{12} \\
-\rho \sigma_{12} & 1+\rho \sigma_1^2
\end{array}
\right) \nonumber \\
&=& \frac{1}{\Delta_r}\left[\left(
\begin{array}{cc}
1 & 0 \\
0 & 1
\end{array}
\right)+\rho \left(
\begin{array}{cc}
\sigma_2^2 & -\sigma_{12} \\
-\sigma_{12} & \sigma_1^2
\end{array}
\right)\right] \nonumber \\
&=& \frac{1}{\Delta_r}(I_2+\rho \tilde \Sigma_x) ,
\end{eqnarray}
where
\begin{equation}
\tilde \Sigma_x=\left(
\begin{array}{cc}
\sigma_2^2 & -\sigma_{12} \\
-\sigma_{12} & \sigma_1^2
\end{array}
\right) .
\end{equation}
Then we have
\begin{eqnarray}
\Sigma_c &=& \Sigma_x-\rho \Sigma_x \frac{1}{\Delta_r}(I_2+\rho \tilde \Sigma_x)\Sigma_x \nonumber \\
&=& \Sigma_x-\frac{\rho}{\Delta_r}(\Sigma_x+\rho \Sigma_x \tilde \Sigma_x)\Sigma_x .
\end{eqnarray}
Since
\begin{eqnarray}
\Sigma_x \tilde \Sigma_x &=& \left(
\begin{array}{cc}
\sigma_1^2 & \sigma_{12} \\
\sigma_{12} & \sigma_2^2
\end{array}
\right)\left(
\begin{array}{cc}
\sigma_2^2 & -\sigma_{12} \\
-\sigma_{12} & \sigma_1^2
\end{array}
\right) \nonumber \\
&=& \left(
\begin{array}{cc}
\sigma_1^2\sigma_2^2-(\sigma_{12})^2 & 0 \\
0 & \sigma_1^2\sigma_2^2-(\sigma_{12})^2
\end{array}
\right) \nonumber \\
&=& \Delta_x I_2 ,
\end{eqnarray}
we obtain
\begin{eqnarray}
\Sigma_c &=& \Sigma_x-\frac{\rho}{\Delta_r}(\Sigma_x+\rho \Delta_x I_2)\Sigma_x \nonumber \\
&=& \Sigma_x-\frac{\rho}{\Delta_r}((\Sigma_x)^2+\rho \Delta_x \Sigma_x) .
\end{eqnarray}
Noting that
\begin{eqnarray}
(\Sigma_x)^2 &=& \left(
\begin{array}{cc}
\sigma_1^2 & \sigma_{12} \\
\sigma_{12} & \sigma_2^2
\end{array}
\right)\left(
\begin{array}{cc}
\sigma_1^2 & \sigma_{12} \\
\sigma_{12} & \sigma_2^2
\end{array}
\right) \nonumber \\
&=& \left(
\begin{array}{cc}
\sigma_1^4+(\sigma_{12})^2 & (\sigma_1^2+\sigma_2^2)\sigma_{12} \\
(\sigma_1^2+\sigma_2^2)\sigma_{12} & \sigma_2^4+(\sigma_{12})^2
\end{array}
\right) ,
\end{eqnarray}
we have
\begin{eqnarray}
\Sigma_c &=& \Sigma_x-\frac{\rho}{\Delta_r}\left((\Sigma_x)^2+\rho \Delta_x \Sigma_x \right) \nonumber \\
&=& \left(
\begin{array}{cc}
\sigma_1^2 & \sigma_{12} \\
\sigma_{12} & \sigma_2^2
\end{array}
\right) \nonumber \\
&& -\frac{\rho}{\Delta_r}\left(
\begin{array}{cc}
\sigma_1^2(\sigma_1^2+\rho \Delta_x)+(\sigma_{12})^2 & \sigma_{12}(\sigma_1^2+\sigma_2^2+\rho \Delta_x) \\
\sigma_{12}(\sigma_1^2+\sigma_2^2+\rho \Delta_x) & \sigma_2^2(\sigma_2^2+\rho \Delta_x)+(\sigma_{12})^2
\end{array}
\right) \\
&\stackrel{\triangle}{=}& \left(
\begin{array}{cc}
c_{11} & c_{12} \\
c_{12} & c_{22}
\end{array}
\right) .
\end{eqnarray}
$c_{11}$, $c_{22}$, and $c_{12}$ are modified as follows:
\begin{eqnarray}
c_{11} &=& \sigma_1^2-\frac{\rho}{\Delta_r}\left(\sigma_1^2(\sigma_1^2+\rho \Delta_x)+(\sigma_{12})^2 \right) \nonumber \\
&=& \sigma_1^2 \left (1-\frac{\rho}{\Delta_r}(\sigma_1^2+\sigma_2^2+\rho \Delta_x) \right )+\frac{\rho}{\Delta_r}\left(\sigma_1^2\sigma_2^2-(\sigma_{12})^2 \right) \nonumber \\
&=& \sigma_1^2 \left (1-\frac{\rho}{\Delta_r}(\sigma_1^2+\sigma_2^2+\rho \Delta_x) \right )+\frac{\rho \Delta_x}{\Delta_r}
\end{eqnarray}
\begin{eqnarray}
c_{22} &=& \sigma_2^2-\frac{\rho}{\Delta_r}\left(\sigma_2^2(\sigma_2^2+\rho \Delta_x)+(\sigma_{12})^2 \right) \nonumber \\
&=& \sigma_2^2 \left (1-\frac{\rho}{\Delta_r}(\sigma_1^2+\sigma_2^2+\rho \Delta_x) \right )+\frac{\rho}{\Delta_r}\left(\sigma_1^2\sigma_2^2-(\sigma_{12})^2 \right) \nonumber \\
&=& \sigma_2^2 \left (1-\frac{\rho}{\Delta_r}(\sigma_1^2+\sigma_2^2+\rho \Delta_x) \right )+\frac{\rho \Delta_x}{\Delta_r}
\end{eqnarray}
\begin{eqnarray}
c_{12} &=& \sigma_{12}-\frac{\rho}{\Delta_r}\left(\sigma_{12}(\sigma_1^2+\sigma_2^2+\rho \Delta_x) \right) \nonumber \\
&=& \sigma_{12}\left (1-\frac{\rho}{\Delta_r}(\sigma_1^2+\sigma_2^2+\rho \Delta_x)\right) .
\end{eqnarray}
Let us set
\begin{equation}
\kappa=1-\frac{\rho}{\Delta_r}(\sigma_1^2+\sigma_2^2+\rho \Delta_x) .
\end{equation}
Then we have
\begin{eqnarray}
\Sigma_c &=& \left(
\begin{array}{cc}
c_{11} & c_{12} \\
c_{12} & c_{22}
\end{array}
\right) \nonumber \\
&=& \left(
\begin{array}{cc}
\kappa \sigma_1^2+\frac{\rho \Delta_x}{\Delta_r} & \kappa \sigma_{12} \\
\kappa \sigma_{12} & \kappa \sigma_2^2+\frac{\rho \Delta_x}{\Delta_r}
\end{array}
\right) .
\end{eqnarray}
Here consider the relationship between $\kappa$ and $\Delta_r$. It follows from
\begin{eqnarray}
\Delta_r &=& \vert \Sigma_r \vert \nonumber \\
&=& (1+\rho \sigma_1^2)(1+\rho \sigma_2^2)-\rho^2(\sigma_{12})^2 \nonumber \\
&=& 1+\rho(\sigma_1^2+\sigma_2^2)+\rho^2\left(\sigma_1^2\sigma_2^2-(\sigma_{12})^2 \right) \nonumber \\
&=& 1+\rho(\sigma_1^2+\sigma_2^2)+\rho^2 \Delta_x \nonumber
\end{eqnarray}
that
\begin{equation}
\Delta_r-\rho(\sigma_1^2+\sigma_2^2+\rho \Delta_x)=1 .
\end{equation}
Note that the left-hand side is equal to $\kappa \Delta_r$.  Hence we obtain
\begin{displaymath}
\kappa \Delta_r=\Delta_r-\rho(\sigma_1^2+\sigma_2^2+\rho \Delta_x)=1 ,
\end{displaymath}
i.e.,
\begin{equation}
\kappa=\frac{1}{\Delta_r} .
\end{equation}
Thus we finally have
\begin{eqnarray}
\Sigma_c &=& \left(
\begin{array}{cc}
\kappa \sigma_1^2+\kappa \rho \Delta_x & \kappa \sigma_{12} \\
\kappa \sigma_{12} & \kappa \sigma_2^2+\kappa \rho \Delta_x
\end{array}
\right) \nonumber \\
&=& \kappa \left(
\begin{array}{cc}
\sigma_1^2+\rho \Delta_x & \sigma_{12} \\
\sigma_{12} & \sigma_2^2+\rho \Delta_x
\end{array}
\right) \nonumber \\
&=& \frac{1}{\Delta_r}\left(
\begin{array}{cc}
\sigma_1^2+\rho \Delta_x & \sigma_{12} \\
\sigma_{12} & \sigma_2^2+\rho \Delta_x
\end{array}
\right) .
\end{eqnarray}


\section{Proof of Proposition 15}
For the expression
\begin{eqnarray}
\mbox{\boldmath $\hat x$}_{k \vert b} &=& \mbox{\boldmath $\hat x$}_{k \vert k}+E[\mbox{\boldmath $x$}_{k}\mbox{\boldmath $\nu$}_{k+1}^T]R_{k+1}^{-1}\mbox{\boldmath $\nu$}_{k+1} \nonumber \\
&& +E[\mbox{\boldmath $x$}_{k}\mbox{\boldmath $\nu$}_{k+2}^T]R_{k+2}^{-1}\mbox{\boldmath $\nu$}_{k+2} \nonumber \\
&& \cdots \nonumber \\
&& +E[\mbox{\boldmath $x$}_{k}\mbox{\boldmath $\nu$}_b^T]R_b^{-1}\mbox{\boldmath $\nu$}_b ,
\end{eqnarray}
note the term $E[\mbox{\boldmath $x$}_k\mbox{\boldmath $\nu$}_{k+1}^T]$. Using the equation
\begin{equation}
\mbox{\boldmath $\nu$}_{k+1}=H_{k+1}(\mbox{\boldmath $x$}_{k+1}-\mbox{\boldmath $\hat x$}_{k+1 \vert k})+\mbox{\boldmath $w$}_{k+1} ,
\end{equation}
we have
\begin{eqnarray}
E[\mbox{\boldmath $x$}_k\mbox{\boldmath $\nu$}_{k+1}^T] &=& E[\mbox{\boldmath $x$}_k(\mbox{\boldmath $x$}_{k+1}-\mbox{\boldmath $\hat x$}_{k+1 \vert k})^T]H_{k+1}^T+E[\mbox{\boldmath $x$}_k\mbox{\boldmath $w$}_{k+1}^T] \nonumber \\
&=& E[\mbox{\boldmath $x$}_k(\mbox{\boldmath $x$}_{k+1}-\mbox{\boldmath $\hat x$}_{k+1 \vert k})^T]H_{k+1}^T+0 \nonumber \\
&=& E[(\mbox{\boldmath $x$}_k-\mbox{\boldmath $\hat x$}_{k \vert k}+\mbox{\boldmath $\hat x$}_{k \vert k})(\mbox{\boldmath $x$}_{k+1}-\mbox{\boldmath $\hat x$}_{k+1 \vert k})^T]H_{k+1}^T \nonumber \\
&=& E[(\mbox{\boldmath $x$}_k-\mbox{\boldmath $\hat x$}_{k \vert k})(\mbox{\boldmath $x$}_{k+1}-\mbox{\boldmath $\hat x$}_{k+1 \vert k})^T]H_{k+1}^T \nonumber \\
&& +E[\mbox{\boldmath $\hat x$}_{k \vert k}(\mbox{\boldmath $x$}_{k+1}-\mbox{\boldmath $\hat x$}_{k+1 \vert k})^T]H_{k+1}^T \nonumber \\
&=& E[(\mbox{\boldmath $x$}_k-\mbox{\boldmath $\hat x$}_{k \vert k})(\mbox{\boldmath $x$}_{k+1}-\mbox{\boldmath $\hat x$}_{k+1 \vert k})^T]H_{k+1}^T+0 \nonumber \\
&=& P(k, k+1)H_{k+1}^T .
\end{eqnarray}
In the above modifications, we have used the relations
\begin{equation}
E[\mbox{\boldmath $x$}_k\mbox{\boldmath $w$}_{k+1}^T]=0
\end{equation}
\begin{equation}
E[\mbox{\boldmath $\hat x$}_{k \vert k}(\mbox{\boldmath $x$}_{k+1}-\mbox{\boldmath $\hat x$}_{k+1 \vert k})^T]=0 ,
\end{equation}
where the former is obtained from the assumption that the future noise is uncorrelated with the past signal (see (19)), whereas the latter is justified by the fact that $\mbox{\boldmath $\hat x$}_{k \vert k}$ is orthogonal to $\mbox{\boldmath $x$}_{k+1}-\mbox{\boldmath $\hat x$}_{k+1 \vert k}$. Similarly, we have
\begin{eqnarray}
E[\mbox{\boldmath $x$}_k\mbox{\boldmath $\nu$}_{k+2}^T] &=& P(k, k+2)H_{k+2}^T \\
&\cdots&  \nonumber \\
E[\mbox{\boldmath $x$}_k\mbox{\boldmath $\nu$}_b^T] &=& P(k, b)H_b^T .
\end{eqnarray}
Hence we obtain
\begin{eqnarray}
\mbox{\boldmath $\hat x$}_{k \vert b} &=& \mbox{\boldmath $\hat x$}_{k \vert k}+P(k, k+1)H_{k+1}^TR_{k+1}^{-1}\mbox{\boldmath $\nu$}_{k+1} \nonumber \\
&& +P(k, k+2)H_{k+2}^TR_{k+2}^{-1}\mbox{\boldmath $\nu$}_{k+2} \nonumber \\
&& \cdots \nonumber \\
&& +P(k, b)H_b^TR_b^{-1}\mbox{\boldmath $\nu$}_b .
\end{eqnarray}
As a result, we have
\begin{eqnarray}
\mbox{\boldmath $x$}_k-\mbox{\boldmath $\hat x$}_{k \vert b} &=& (\mbox{\boldmath $x$}_k-\mbox{\boldmath $\hat x$}_{k \vert k})-P(k, k+1)H_{k+1}^TR_{k+1}^{-1}\mbox{\boldmath $\nu$}_{k+1} \nonumber \\
&& -P(k, k+2)H_{k+2}^TR_{k+2}^{-1}\mbox{\boldmath $\nu$}_{k+2} \nonumber \\
&& \cdots \nonumber \\
&& -P(k, b)H_b^TR_b^{-1}\mbox{\boldmath $\nu$}_b .
\end{eqnarray}
\par
Using this equation, let us derive the covariance matrix of the smoothing error, i.e.,
\begin{equation}
\Sigma_{k \vert b}=E[(\mbox{\boldmath $x$}_k-\mbox{\boldmath $\hat x$}_{k \vert b})(\mbox{\boldmath $x$}_k-\mbox{\boldmath $\hat x$}_{k \vert b})^T] .
\end{equation}
We have
\begin{eqnarray}
\Sigma_{k \vert b} &=& E[(\mbox{\boldmath $x$}_k-\mbox{\boldmath $\hat x$}_{k \vert k})(\mbox{\boldmath $x$}_k-\mbox{\boldmath $\hat x$}_{k \vert k})^T] \nonumber \\
&& -E[(\mbox{\boldmath $x$}_k-\mbox{\boldmath $\hat x$}_{k \vert k})\mbox{\boldmath $\nu$}_{k+1}^T]R_{k+1}^{-1}H_{k+1}P(k, k+1)^T \nonumber \\
&& -E[(\mbox{\boldmath $x$}_k-\mbox{\boldmath $\hat x$}_{k \vert k})\mbox{\boldmath $\nu$}_{k+2}^T]R_{k+2}^{-1}H_{k+2}P(k, k+2)^T \nonumber \\
&& \cdots \nonumber \\
&& -E[(\mbox{\boldmath $x$}_k-\mbox{\boldmath $\hat x$}_{k \vert k})\mbox{\boldmath $\nu$}_b^T]R_b^{-1}H_bP(k, b)^T \nonumber \\
&& -P(k, k+1)H_{k+1}^TR_{k+1}^{-1}E[\mbox{\boldmath $\nu$}_{k+1}(\mbox{\boldmath $x$}_k-\mbox{\boldmath $\hat x$}_{k \vert k})^T] \nonumber \\
&& +P(k, k+1)H_{k+1}^TR_{k+1}^{-1}E[\mbox{\boldmath $\nu$}_{k+1}\mbox{\boldmath $\nu$}_{k+1}^T]R_{k+1}^{-1}H_{k+1}P(k, k+1)^T \nonumber \\
&& -P(k, k+2)H_{k+2}^TR_{k+2}^{-1}E[\mbox{\boldmath $\nu$}_{k+2}(\mbox{\boldmath $x$}_k-\mbox{\boldmath $\hat x$}_{k \vert k})^T] \nonumber \\
&& +P(k, k+2)H_{k+2}^TR_{k+2}^{-1}E[\mbox{\boldmath $\nu$}_{k+2}\mbox{\boldmath $\nu$}_{k+2}^T]R_{k+2}^{-1}H_{k+2}P(k, k+2)^T \nonumber \\
&& \cdots \nonumber \\
&& -P(k, b)H_b^TR_b^{-1}E[\mbox{\boldmath $\nu$}_b(\mbox{\boldmath $x$}_k-\mbox{\boldmath $\hat x$}_{k \vert k})^T] \nonumber \\
&& +P(k, b)H_b^TR_b^{-1}E[\mbox{\boldmath $\nu$}_b\mbox{\boldmath $\nu$}_b^T]R_b^{-1}H_bP(k, b)^T ,
\end{eqnarray}
where we have used the property of the innovations:
\begin{equation}
E[\mbox{\boldmath $\nu$}_l\mbox{\boldmath $\nu$}_l^T]=R_l,~~E[\mbox{\boldmath $\nu$}_l\mbox{\boldmath $\nu$}_m^T]=0~(l \neq m) .
\end{equation}
In (290), we have $E[\mbox{\boldmath $\nu$}_l\mbox{\boldmath $\nu$}_l^T]=R_l$. Also, we have
\begin{eqnarray}
\lefteqn{E[(\mbox{\boldmath $x$}_k-\mbox{\boldmath $\hat x$}_{k \vert k})\mbox{\boldmath $\nu$}_{k+1}^T]} \nonumber \\
&& =E[(\mbox{\boldmath $x$}_k-\mbox{\boldmath $\hat x$}_{k \vert k})\left((\mbox{\boldmath $x$}_{k+1}-\mbox{\boldmath $\hat x$}_{k+1 \vert k})^TH_{k+1}^T+\mbox{\boldmath $w$}_{k+1}^T \right)] \nonumber \\
&& =E[(\mbox{\boldmath $x$}_k-\mbox{\boldmath $\hat x$}_{k \vert k})(\mbox{\boldmath $x$}_{k+1}-\mbox{\boldmath $\hat x$}_{k+1 \vert k})^T]H_{k+1}^T+0 \nonumber \\
&& =P(k, k+1)H_{k+1}^T .
\end{eqnarray}
Similarly, we have
\begin{eqnarray}
E[(\mbox{\boldmath $x$}_k-\mbox{\boldmath $\hat x$}_{k \vert k})\mbox{\boldmath $\nu$}_{k+2}^T] &=& P(k, k+2)H_{k+2}^T \\
&\cdots& \nonumber \\
E[(\mbox{\boldmath $x$}_k-\mbox{\boldmath $\hat x$}_{k \vert k})\mbox{\boldmath $\nu$}_b^T] &=& P(k, b)H_b^T .
\end{eqnarray}
In the above modifications, we have used the relations
\begin{equation}
E[(\mbox{\boldmath $x$}_k-\mbox{\boldmath $\hat x$}_{k \vert k})\mbox{\boldmath $w$}_l^T]=0~(k+1 \leq l \leq b) .
\end{equation}
Notice that $\mbox{\boldmath $\hat x$}_{k \vert k} \in Z^k$, where $Z^k$ the space spanned by $\{\mbox{\boldmath $z$}_j=H_j\mbox{\boldmath $x$}_j+\mbox{\boldmath $w$}_j,~0 \leq j \leq k\}$. Then the above is justified by the assumption that the future noise is uncorrelated with the past signal (see (19)) and by the fact that $\mbox{\boldmath $w$}_l$ are orthogonal to each other.
\par
Thus we finally obtain
\begin{eqnarray}
\Sigma_{k \vert b} &=& P_k-P(k, k+1)H_{k+1}^TR_{k+1}^{-1}H_{k+1}P(k, k+1)^T \nonumber \\
&& -P(k, k+2)H_{k+2}^TR_{k+2}^{-1}H_{k+2}P(k, k+2)^T \nonumber \\
&& \cdots \nonumber \\
&& -P(k, b)H_b^TR_b^{-1}H_bP(k, b)^T .
\end{eqnarray}




\ifCLASSOPTIONcaptionsoff
  \newpage
\fi

\end{document}